\documentclass[reprint, prd, amsmath, amssymb, superscriptaddress, nofootinbib]{revtex4-1}
\usepackage{graphicx,color,float}
\usepackage[caption=false]{subfig}
\usepackage{placeins}
\usepackage{rviewport}
\definecolor{colorLink} {rgb}{0,0,0.8} 
\definecolor{colorCite} {rgb}{0,0,0.8} 
\definecolor{colorURL} {rgb}{0,0,0.8} 
\usepackage[colorlinks=true,linktocpage=true,linkcolor=colorLink,citecolor=colorCite,urlcolor=colorURL]{hyperref}

\usepackage[sort&compress]{natbib}

\DeclareMathOperator{\sgn}{sgn}

\begin{document}

\title{Progress towards characterizing \\ultrahigh energy cosmic ray sources}

\author{Marco Stein Muzio}
\email{msm659@nyu.edu}
\affiliation{Center for Cosmology and Particle Physics, Department of Physics, New York University, 726 Broadway, New York, New York, USA}
\author{Michael Unger}
\email{Michael.Unger@kit.edu}
\affiliation{Institut f\"ur Kernphysik, Karlsruher Institut f\"ur Technologie, 76021 Karlsruhe, Germany}
\author{Glennys R. Farrar}
\email{gf25@nyu.edu}
\affiliation{Center for Cosmology and Particle Physics, Department of Physics, New York University, 726 Broadway, New York, New York, USA}

\date{\today}

\begin{abstract}	
We use a multimessenger approach to constrain realistic mixed composition models of ultrahigh energy cosmic ray sources using the latest cosmic ray, neutrino, and gamma-ray data. We build on the successful Unger-Farrar-Anchordoqui 2015 (UFA15) model which explains the shape of the spectrum and its complex composition evolution via photodisintegration of accelerated nuclei in the photon field surrounding the source. We explore the constraints which can currently be placed on the redshift evolution of sources and the temperature of the photon field surrounding the sources. We show that a good fit is obtained to all data either with a source which accelerates a narrow range of nuclear masses or a Milky Way-like mix of nuclear compositions, but in the latter case the nearest source should be 30-50 Mpc away from the Milky Way in order to fit observations from the Pierre Auger Observatory. We also ask whether the data allow for a subdominant purely protonic component at UHE in addition to the primary UFA15 mixed composition component. We find that such a two-component model can significantly improve the fit to cosmic ray data while being compatible with current multimessenger data. 
\end{abstract}

\maketitle

\section{Introduction}

\par
The origin of ultrahigh energy cosmic rays (UHECRs), $E \gtrsim 10^{18}$ eV, is a long standing problem in high-energy astrophysics. While the magnetic field of the Milky Way and extragalactic magnetic fields make direct detection of sources very challenging, a multimessenger approach can help narrow the range of candidate sources. Signals from UHECR sources include CRs, neutrinos, and gamma-rays. Neutrinos result from the decay of neutrons and pions produced in the source environment when UHECRs interact with gas and photon fields surrounding the source. They are also produced during extragalactic propagation, due to interactions with the cosmic microwave background (CMB) and extragalactic background light (EBL). High energy gamma-rays are produced by neutral pion decay both in the source environment and in propagation, but gamma-rays are more commonly produced at lower energies through electromagnetic (EM) cascades initiated by Bethe-Heitler pair-production off the CMB and EBL during propagation. 

\par
The Unger, Farrar, and Anchordoqui (UFA15,~\cite{UFA15}) framework provides a way to characterize UHECR sources by basic parameters of the CR accelerator and its surroundings. The parameters are adjusted to fit the UHECR spectrum and composition. As shown in~\cite{UFA15} the model naturally explains the origin of the ankle in the CR spectrum, the light composition below the ankle, and increasingly heavy composition above the ankle. This is achieved by considering an accelerator, characterized by a spectral index, rigidity cutoff, and composition, emitting UHECRs into a surrounding environment. The environment is envisaged as containing both photons and a turbulent magnetic field. The photon field is characterized by its temperature (or peak energy if a broken power-law). A detailed description of the magnetic properties and size of the surrounding environment is not needed; one only needs the ratio of a UHECR's interaction time with a photon to its escape time from the source environment, as a function of UHECR energy, to fully characterize the source environment.Inspired by the diffusion of charged particles in a turbulent magnetic field, we model the energy-dependence of the escape time as a power law in rigidity. The energy dependence of the interaction times are based on the known cross sections of photonuclear interactions. Once a UHECR escapes from the environment it propagates to Earth, losing energy through Bethe-Heitler, photopion production, and photodisintegration interactions with the CMB and EBL. Fitting the observed UHECR spectrum and composition then fixes the parameters of the model, such as the escape-to-interaction time ratio at a reference energy and its energy dependence. Once these parameters are known empirically, sources whose properties are not compatible with their values can be excluded. Besides providing a very general framework for describing cosmic ray sources, the UFA15 model can be efficiently implemented numerically as discussed in~\cite{UFA15}, allowing precise fitting of the model to data.

\par
In the UFA15 model, the ankle is formed via photodisintegration of nuclei, leading to a pile-up of protons at energies below the ankle and a spread of nuclear compositions. The UFA15 mechanism yields good fits to both the entire UHECR spectrum and the composition dependence on energy~\cite{UFA15}. This mechanism for generating the ankle has also been studied in~\cite{Globus+15, Fang+17, Kachelriess+17, Supanitsky+18, Boncioli+18}. Studies investigating the implications of fitting both the CR spectrum and composition on UHECR source parameters have also been conducted without considering interactions in the source environment \cite{Aab+16e, Heinze+19}. 

\par
Our purpose here is two-fold. First, we investigate what constraints can be placed on UHECR sources using the most up-to-date multimessenger data. Second, we ask whether UHECR data suggest any elaboration or improvement of the basic UFA15 model ~\cite{UFA15}.

Previous multimessenger studies~\cite{Berezinsky+16, Gavish+16, Liu+16, Supanitsky16, vanVliet16, Aloisio+17, Biehl+17, Fang+17, Globus+17, Kachelriess+17, AlvesBatista+18, Supanitsky+18, Heinze+19} used gamma-ray data from the \textit{Fermi}-Large Area Telescope (LAT)~\cite{Ackermann+14, FermiLAT+15} and IceCube~\cite{Aartsen+13} to constrain neutrino signals. The IceCube neutrino bounds~\cite{Aartsen+13} were not strong enough to be particularly constraining and gamma-rays proved to only be significantly constraining for pure-proton models~\cite{Heinze+15, Berezinsky+16, Liu+16, Supanitsky16, vanVliet16, Muzio+17}. 

\par
This paper is organized as follows. In Section~\ref{sec:multimessengerdata}, we present the multimessenger data we will use to constrain UHECR source models. In Section~\ref{sec:models+variations}, we discuss the degree to which parameters of the UFA15 model, specifically redshift evolution of UHECR production and temperature of the source environment, can be constrained by multimessenger data. In Section~\ref{sec:refinements}, we discuss possible refinements of the UFA15 model. We summarize our results in Section~\ref{sec:summary}. In addition, Appendix~\ref{app:EMcascades} details the methods we used to simulate EM cascades and Appendix~\ref{app:supplfigs} provides some supplementary figures. 

\section{Multimessenger Data} \label{sec:multimessengerdata}

\begin{figure*}[htp]
	\centering
	\begin{minipage}{\linewidth}
		\centering
	\subfloat[\label{fig:UFA_benchmark_CR}]{\includegraphics[width=\textwidth]{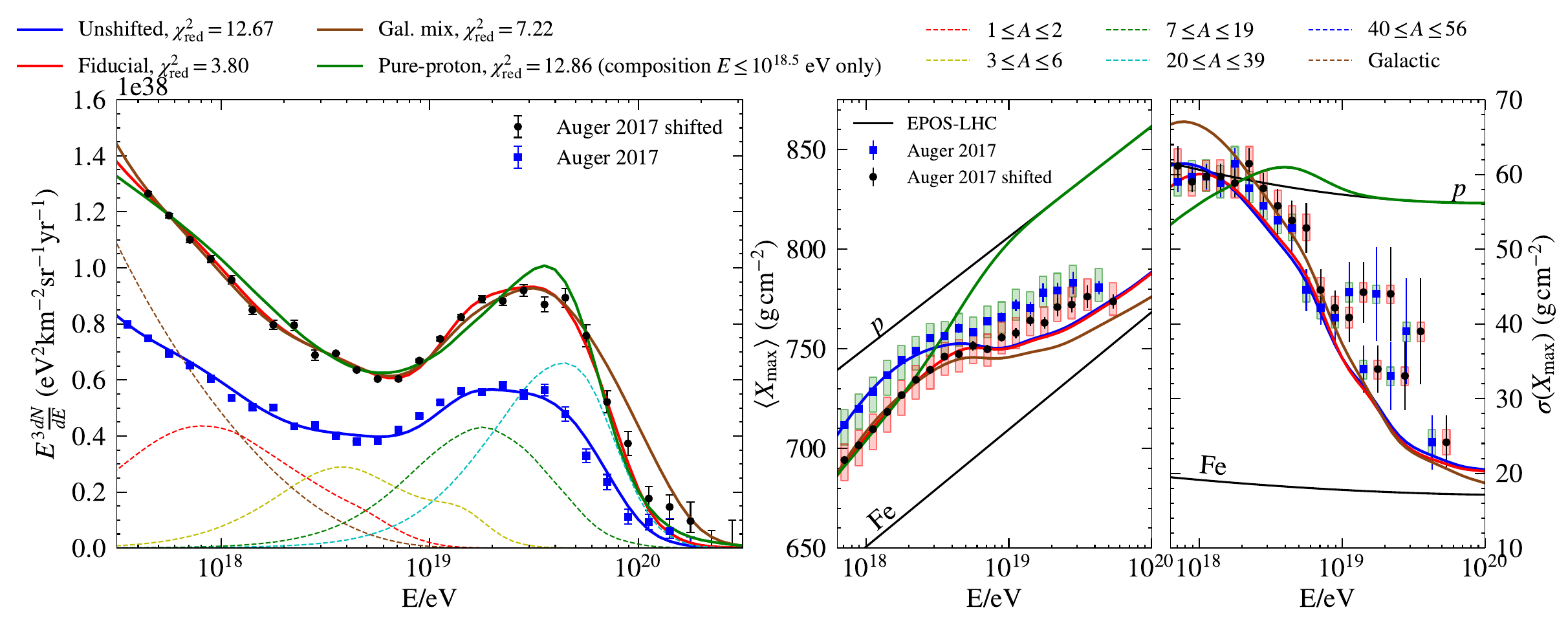}}
	\end{minipage}
	\begin{minipage}{\linewidth}
		\centering
		\subfloat[\label{fig:UFA_benchmark_secondaries}]{\includegraphics[width=\textwidth]{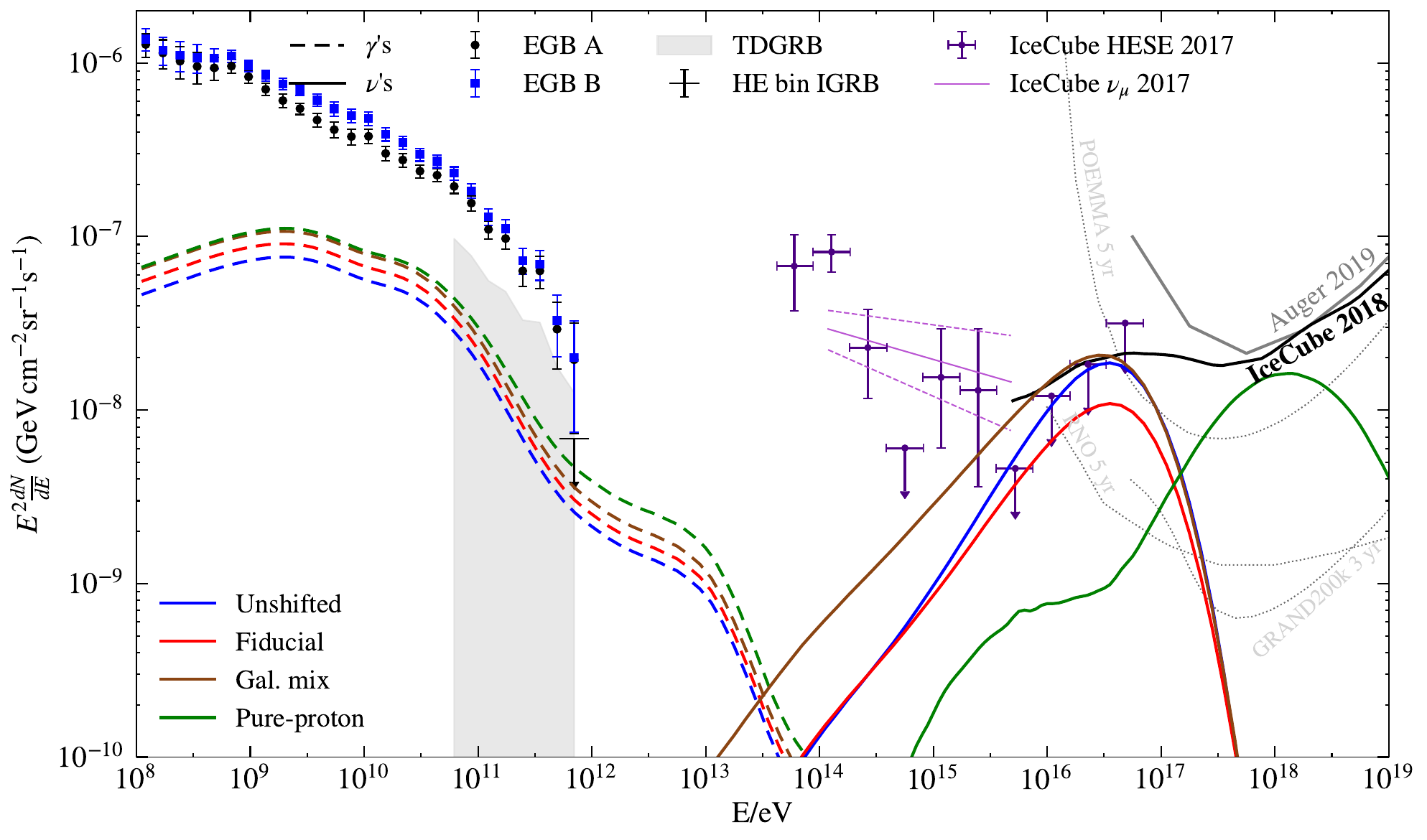}}
	\end{minipage}
	\caption{Predictions for (a) CR and (b) multimessenger signals in benchmark models from UFA15 updated to latest Auger data -- unshifted (blue), fiducial (red), and galactic mix (brown) -- along with a pure-proton model as detailed in the text (green). \textbf{Top:} The CR spectrum (left) and composition (right) at Earth. For illustration, a breakdown of the spectrum at Earth by mass group is shown for the fiducial model (dashed colored lines). Data points are the 2017 Auger spectrum and composition, shifted (black circles, see text for details) and unshifted (blue squares). $X_\mathrm{max}$ data is plotted with error bars and shaded boxes denoting statistical and systematic uncertainties. \textsc{EPOS-LHC} is used to infer the $X_\mathrm{max}$ distribution moments from composition predictions. For reference, the predictions of \textsc{EPOS-LHC} for pure-proton and pure-iron compositions are shown (solid black lines). The reduced $\chi^2$ quoted for the pure-proton model only includes composition data up to $E = 10^{18.5}$ eV. \textbf{Bottom:} The predicted gamma-ray (dashed lines) and neutrino fluxes (solid lines) are plotted along with the EGB in model A (black circles) and model B (blue squares), which differ in their assumptions used to calculate the galactic foregrounds~\cite{Ackermann+14}. The upper-bound on the TDGRB (grey band) and the highest energy bin of the IGRB (black upper-bound) are also plotted. The astrophysical neutrino flux measured by IceCube (data points and inferred spectrum) are shown in purple (dark and light, respectively). Upper-bounds from IceCube (black) and Auger (grey) on the EHE cosmic neutrino flux (solid lines) are shown along with projected sensitivities (dashed grey lines) for the GRAND200k, POEMMA, and RNO experiments~\cite{Alvarez-Muniz+18, Krizmanic19, Connolly19}.}
	\label{fig:UFA_benchmark_summary}
\end{figure*}

\par
In this section we review the data products used in our study. Any viable UHECR source model must, of course, be able to explain both the observed UHECR spectrum and the observed composition. We focus on the UHECR data taken by the Pierre Auger Observatory (Auger) because of its higher statistics~\cite{Aab+14a, Aab+14b} relative to the Telescope Array (TA). The UHECR composition measurements can be estimated most reliably from measurements of the depth of shower maximum, $X_\mathrm{max}$. These measurements are consistent between the Pierre Auger and TA collaborations~\cite{deSouza17}, but the much greater statistics of the Pierre Auger Observatory enables it to make geometric cuts to obtain an essentially uniform acceptance in $X_\mathrm{max}$ and thus an $X_\mathrm{max}$ distribution with minimal detector or acceptance bias. The spectra measured by the two observatories are consistent within their respective systematic energy scale uncertainties over a wide range of energies, though there are hints of larger discrepancies at ultrahigh energies $> 10$ EeV~\cite{Tsunesada+17}. Following UFA15, we allow for a systematic shift in the Auger energy and $X_\mathrm{max}$ scales. 

\par
It is important to stress that, for both observatories, inferences on composition from the data depend on extrapolating models of UHE hadronic physics, which introduces a significant uncertainty. The $X_\mathrm{max}$ of events detected by the fluorescence detector (FD) can be used to infer the composition. Auger measures the distributions of $X_\mathrm{max}$ over many extensive air showers and publishes these distributions along with their first two moments (mean and standard deviation)~\cite{Bellido17, Aab+14a, Aab+14b}. Here we consider the first two moments of these distributions. In order to relate these moments to the composition one must interpret them via a hadronic event generator (HEG) to obtain the mean and variance of the logarithmic mass distribution ($\langle \ln{A} \rangle$ and $\mathrm{V}(\ln{A})$) of UHECRs at Earth ~\cite{Abreu+13}. Commonly used LHC-tuned HEGs include \textsc{EPOS-LHC}~\cite{Pierog+13}, \textsc{Sibyll2.3c}~\cite{Riehn+15}, and \textsc{QGSJet-II-04}~\cite{Ostapchenko10}. Throughout this paper we use \textsc{EPOS-LHC} and \textsc{Sibyll2.3c}. We do not employ \textsc{QGSJet-II-04} because it has been found to give considerably inconsistent interpretations of air shower data~\cite{Abreu+13, Bellido17}. 

\par
UHECRs give rise to neutrinos produced via photohadronic interactions with the CMB and EBL during UHECR propagation and also within the UHECR source environment via photohadronic and hadronic interactions with ambient photon fields and gas. Intergalactic space is practically transparent to neutrinos allowing them to propagate unimpeded from their source to their point of observation, though they do undergo both redshift energy loss and flavor oscillations. IceCube has detected an astrophysical neutrino flux in the $\sim 50$ TeV to $\sim 10$ PeV energy range~\cite{Kopper17}. Bounds on neutrinos beyond $\sim 10$ PeV energies, hereafter \textit{extremely-high energy (EHE) cosmic neutrinos}, are provided by the IceCube~\cite{Aartsen+18}, Auger~\cite{Aab+19}, and ANITA~\cite{Allison+18, Gorham+19} experiments. IceCube's latest bound from 9 years of observation provides the strongest constraint in the $\sim 10$ PeV to $\sim 10$ EeV energy range, while the combined bound from four ANITA balloon flights provides the strongest constraint above $\sim 10$ EeV. We can use these limits to exclude UHECR source models which predict $N_\nu > 2.44$ $(3.09)$ at the 90\% (95\%) CL in the given energy range~\cite{Feldman+97}.

\par
UHECRs also give rise to gamma-rays. These are produced via neutral pion decay, both in the source environment and during UHECR propagation, and via EM cascades initiated by UHECR interactions with the CMB and EBL. However, since the universe is opaque to gamma-rays with energies $\gtrsim 0.1-1$ TeV~\cite{DeAngelis+18}, it is not possible to observe gamma-rays directly produced in UHECR interactions except very nearby. Instead, gamma-rays which have cascaded down to energies $\lesssim$ TeV are observed. In the $100$ MeV to $820$ GeV energy range, LAT reports the total extragalactic gamma-ray spectrum, the \textit{extragalactic gamma-ray background} (EGB), and the \textit{isotropic gamma-ray background} (IGRB), defined to be the EGB with resolved point sources subtracted~\cite{Ackermann+14}. Many source types are expected to contribute to both the EGB and IGRB, the most significant among them being blazars, mis-aligned active galactic nuclei (AGNs), star-forming galaxies, and millisecond pulsars~\cite{Fornasa+15}. Estimates have been made of the \textit{truly} diffuse gamma-ray background (TDGRB) --- the gamma-ray background not coming from gamma-ray point sources --- based on the fraction of the EGB above $50$ GeV produced by both resolved and unresolved point sources~\cite{FermiLAT+15}. The estimate in ~\cite{FermiLAT+15} provides a bound on the \textit{integral} TDGRB flux above $50$ GeV, which we present throughout this paper as an approximate \textit{differential} bound. This approximate differential bound was obtained by subtracting the most conservative estimate of the point source contribution to the EGB estimated in~\cite{FermiLAT+15} --- about $62\%$, when subtracting $1\sigma$ from the central value, of the total EGB model B integrated flux --- from the EGB in each bin above $50$ GeV. The IGRB also acts as an upper-bound on the diffuse gamma-ray flux arising from UHECRs, but in practice only its highest energy bin is more constraining than the TDGRB estimate. This highest energy IGRB bin places a limit of $\sim 7\times 10^{-9}$ GeV cm$^{-2}$ s$^{-1}$ sr$^{-1}$ on the flux in the $\sim 820$ GeV energy bin. As shown in~\cite{Muzio+17}, and will be further developed below, mixed composition UHECR source models are only weakly constrained by current gamma-ray data. 

\par
Each of these components --- the CR spectrum, its composition, neutrinos, and gamma-rays --- help constrain UHECR source models. In this paper we identify what are the most constraining data for different source model parameters. We also extend and elaborate the UFA15 modeling in several ways and update the CR data used. The $\chi^2$ values reported throughout this paper are the result of a combined fit to both the Auger spectrum above $E =10^{17.5}$ eV and the Auger composition for $E\geq10^{17.8}$ eV. CR interactions within the source environment are calculated as described in Appendix C of~\cite{UFA15}, which takes into account only photohadronic interactions. Here we use the approximate treatment of~\cite{UFA15} and assign an average elasticity of $0.8$ for photopion production. We performed further studies with \textsc{SOPHIA}~\cite{Mucke+99} and found that a full treatment of pion production, including the distribution of elasticities for single- and multi-pion production, leads to an increase in the predicted number of neutrinos by a factor of $\lesssim 2$ compared to our approximate treatment, allowing us to place conservative constraints here. Extragalactic CR propagation and secondary production are based on simulations using \textsc{CRPropa3}~\cite{Batista+16} and EM cascades are simulated using \textsc{ELMAG}~\cite{Kachelriess+11} (see Appendix~\ref{app:EMcascades} for details). Throughout all of our calculations we adopt the Gilmore12 EBL~\cite{Gilmore+11} and a star formation rate (SFR) evolution~\cite{Robertson+15} unless stated otherwise. An extragalactic magnetic field (EGMF) strength of $1$ nG was used throughout all EM cascade simulations, but our results are insensitive to its value in the $10^{-17}-10^{-8}$ G range (see Appendix~\ref{app:EMcascades}). All spectral indices are defined via a power-law with \textit{positive} exponent, e.g. $\Phi^\mathrm{inj} \sim E^\gamma_\mathrm{inj}$. Lastly, we alert the reader that the term ``mixed composition'' appears in two different contexts. Generally, we assume the composition of \textit{observed} UHECRs to be an admixture of nuclear masses, in accordance with standard interpretation of the air shower development using LHC-tuned HEGs. However, occasionally we consider the possibility that the HEGs are not valid and the observed composition is actually pure-proton. Thus we sometimes have occasion to distinguish between pure and mixed composition as observed at Earth. Usually, however, when we refer to mixed or pure composition, we are referring to the nature of the composition emerging from the accelerator, before processing through interactions in the photon field of the source environment.  

\section{Models and Variations} \label{sec:models+variations}

\subsection{UFA15 Benchmark Models} \label{sec:benchmarks}

\par
Figure~\ref{fig:UFA_benchmark_summary} displays the multimessenger signals from three benchmark source scenarios considered in the UFA15 paper~\cite{UFA15} as well as the predictions of a pure-proton scenario with a galactic component. For the purposes of Fig.~\ref{fig:UFA_benchmark_summary}, EPOS-LHC is used following UFA15. In the first two benchmark scenarios, the accelerator injects a single mass into the source environment. The injected mass is a fit parameter and turns out to be near Si. In the third scenario, the accelerator injects a mixed composition based on the flux fractions of nuclei observed in the Milky Way at $1$ TeV~\cite{PDG}; we call this the \textit{galactic mix} scenario. The first \textit{unshifted} single-mass scenario is fit directly to the published Auger spectrum and composition~\cite{Fenu17, Bellido17, Aab+14a, Aab+14b}. The \textit{fiducial} scenario fits the model to the Auger spectrum shifted by one bin upwards in the energy scale and to the Auger $X_\mathrm{max}$ values shifted down by $-1\sigma_\mathrm{syst. X_\mathrm{max}}$. (These shifts were obtained in UFA15 by finding the combination of $-1/0/+1$ bin of energy and $-1/0/+1\times\sigma_\mathrm{syst. X_\mathrm{max}}$ in $X_\mathrm{max}$ scale which gave the best-fit to the data.) The shift by one bin in the logarithm of energy ($0.1$ decades) corresponds to slightly more than a $+1\sigma_\mathrm{syst.E}$ ($\approx 14$\%) shift. Throughout this paper we adopt these same shifts when fitting to the Auger data, except in the case of the unshifted model. Model parameters have been updated from UFA15 to fit the latest Auger dataset. 

\par
Composition measurements can put strong constraints on the types of source models which adequately describe the Auger data. The benchmark UFA15 scenarios provide good fits to the CR spectrum and composition data, whereas interpreted using \textsc{EPOS-LHC} or \textsc{Sibyll2.3c} the data rules out pure-proton source models~\cite{Aab+16a, Bellido17}. However, the LHC-tuned HEGs are known to not properly describe the hadronic component of UHECR air showers~\cite{Aab+16d}. These shortcomings in the modeling of the hadronic part of air showers do not neccessarily falsify the fidelity of the description of the electromagnetic component (and thus $X_\mathrm{max}$), but they make it plausible that variation of composition inferences with different HEGs do not cover the full range of theoretical uncertainties. Indeed it has been argued that pure-proton composition can fit the $X_\mathrm{max}$ and $\sigma(X_\mathrm{max})$ measurements if one allows for accelerator-compatible modifications of the HEGs above $E \approx 10^{18.5}$ eV~\cite{Allen+13, Farrar+13, Farrar14, Stodolsky18}.\footnote{But see~\cite{Aab+16a} adducing independent evidence for a mixed composition in the $10^{18.5}-10^{19}$ eV range.} Thus it is of interest to investigate whether a pure-proton model can give a reasonable accounting of the full UHECR spectrum and the $X_\mathrm{max}$ observables up to $E = 10^{18.5}$ eV and at the same time satisfy multimessenger constraints. Figure~\ref{fig:UFA_benchmark_summary} therefore shows not only the UFA15 benchmarks, but also the best-fit scenario when the extragalactic component is taken to be pure-proton and the Auger data is shifted as described above. For consistency with the consideration of systematic uncertainties in~\cite{UFA15}, we also investigated the effect of shifting the data by $\pm1$ bin in energy and $\pm 1 \sigma_\mathrm{syst. X_\mathrm{max}}$ in $X_\mathrm{max}$. We checked if the pure-proton fit can be improved by using different systematic shifts of the energy and $X_\mathrm{max}$ scale, but no significantly better fit resulted from this study (best reduced $\chi^2 = 12.19$). Thus, a pure-proton model does not describe the Auger data well, even when allowing for the additional degrees of freedom from a low-energy Galactic component and ignoring the UHE $X_\mathrm{max}$ data.

\par
The qualitative differences in multimessenger signals between mixed composition and pure-proton models are interesting. Firstly, pure-proton scenarios generally produce a higher gamma-ray flux compared to mixed composition scenarios. This is due to pure-proton scenarios producing CRs with higher energy-per-nucleon on average, which allows energy to be more efficiently transferred into EM cascades via pair-production off of the CMB. This effect is strong enough that it leaves the pure-proton model in possible tension with the upper-bound provided by the highest energy bin of the IGRB (due to the fact that some portion of this bin's flux is due to unresolved point sources), even if some modified UHE particle physics were to bring the UHECR composition data into adequate compatibility. In the future, refined limits on the gamma-rays that can be attributed to UHECR sources could allow for the exclusion of pure-proton scenarios independently of UHE particle physics, unless the predicted gamma-rays are observed.\footnote{We disagree with Liu et al.~\cite{Liu+16} who claim that gamma-rays already exclude the pure-proton scenario. This difference is due to their choice of a more constraining galactic foreground model, LAT's galactic foreground model A \cite{Ackermann+14}. However, LAT considers several galactic foreground models equally~\cite{Ackermann+14}. Therefore, for the purposes of constraining UHECR source models, we have chosen the galactic foreground model which assigns the least flux to diffuse galactic sources, LAT's galactic foreground model B. This is appropriate in order to place conservative limits.}

\par
In neutrinos, the pure-proton scenario produces a characteristic UHE peak in the neutrino spectrum, as can be seen in Fig.~\ref{fig:UFA_benchmark_secondaries}. The pure-proton model predicts $1.79$ events above $\sim 10$ PeV and thus evades the 90\% CL bound. Future neutrino data --- with sufficient sensitivity to probe the higher-energy peak --- will provide strong constraints on the amount of UHE protons compatible with data, since any scenario producing considerable amounts of protons above $\sim 10^{19.7}$ eV would produce this UHE peak feature. This peak is due to photopion production off of the CMB. By contrast, such a UHE neutrino peak is absent in mixed composition scenarios which do not produce considerable amounts of protons of sufficiently high energies. Instead, mixed composition scenarios produce neutrino spectra with peaks in the $\sim 0.1 - 10$ PeV energy range. While cosmogenic neutrinos (those produced enroute from source environment to Earth) are produced in all scenarios, the neutrino fluxes at Earth are dominated by neutrinos produced in the source environment via photopion production. Since the neutrinos in mixed composition scenarios are dominated by those produced in the source environment, the number of peaks in the neutrino spectrum is sensitive to the assumptions made about the number of peaks in the photon field in the source environment.  

The gamma-ray signals predicted by the UFA15 benchmark models are well below the EGB flux, as one sees from Fig.~\ref{fig:UFA_benchmark_secondaries}. However, the neutrino flux for the best-fit unshifted and galactic mix models yield $2.75$ and $4.21$ events, respectively, in the current IC2018 exposure and are thus excluded at 90\% and 95\% CL. The best-fit fiducial model remains unconstrained by the current IceCube limits, producing only $2.12$ events in the current exposure. 

\par
In the next sections we explore how these multimessenger constraints suggest refinements and elaborations of the basic UFA15 framework, giving further insight into UHECR sources. 

\subsection{Source Evolution}
\label{sec:evo}
\par
We begin by considering the constraints which can be placed on the evolution of UHECR sources. As we shall see, the strongest constraints come from the UHECR spectrum and composition. We consider two parametrizations of the source evolution, $\xi(z)$, the source comoving CR power density at redshift $z$ relative to its value today: a SFR evolution~\cite{Robertson+15},

\begin{equation}\label{eq:SFR}
	\xi_\mathrm{SFR}(z) \propto \frac{(1+z)^a}{1+[(1+z)/b]^c}~~,
\end{equation}

\noindent
where $a=3.26\pm 0.21$, $b=2.59\pm0.14$ and $c=5.68\pm 0.19$; and, a single power-law with an exponential cutoff,

\begin{equation}\label{eq:source_evols}
	\xi(z) =
	\begin{cases}
		(1+z)^m & \quad z<z_0 \\
		(1+z_0)^m e^{-(z-z_0)} & \quad z\ge z_0 ~~.
	\end{cases}
\end{equation}

\begin{figure}
	\centering
	\includegraphics[width=\linewidth]{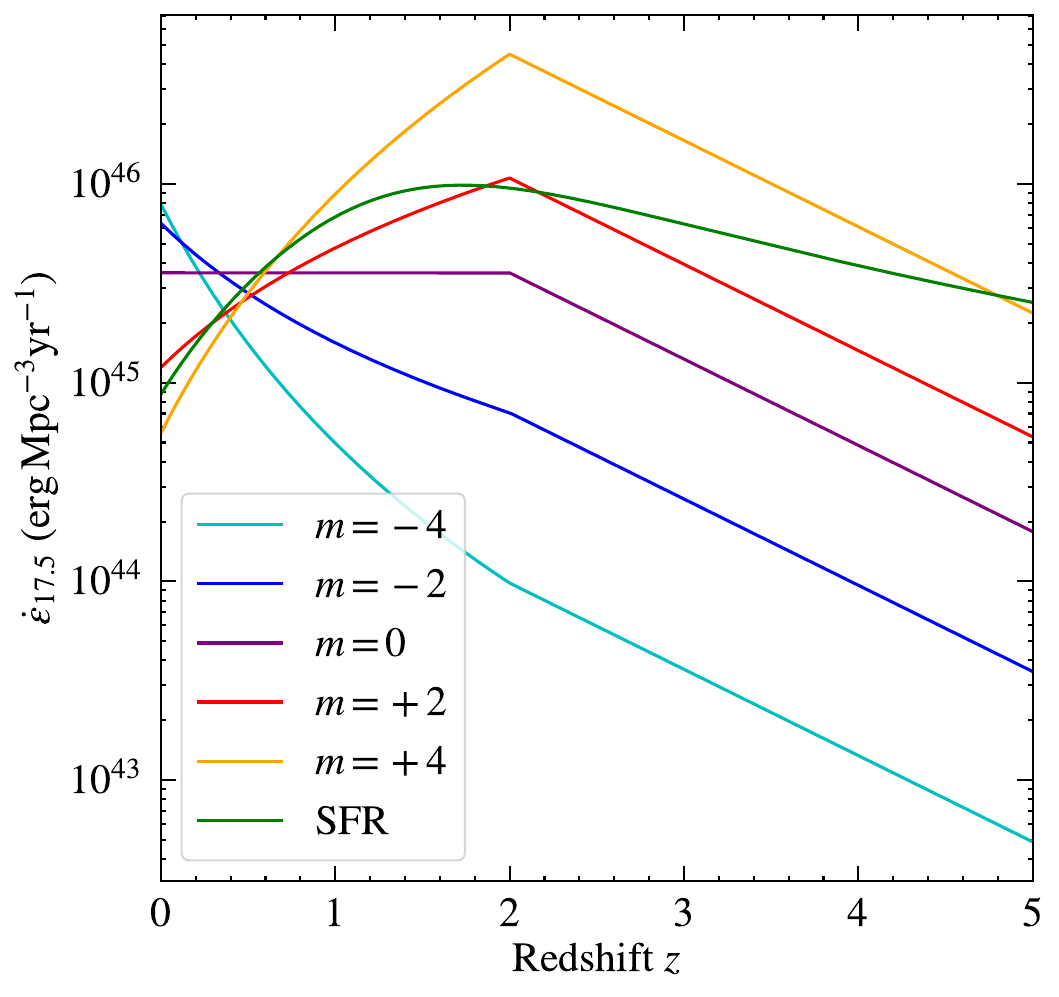}
	\caption{Comoving CR power density above $10^{17.5}$ eV as a function of redshift in several source evolution parameterizations. }
	\label{fig:source_evolutions}
\end{figure}

\noindent
We fix $z_0 = 2$ following UFA15. Source models with positive (negative) $m$ represent models with comoving CR power density increasing (decreasing) as redshift increases for $z\lesssim 1$. 
See Fig.~\ref{fig:source_evolutions} for plots of the evolution in illustrative cases, with normalization scaled according to the UHECR power density injected above $10^{17.5}$ eV in the given best-fit single-mass model, using \textsc{EPOS-LHC} and shifted Auger data for definiteness.

\begin{figure}
	\centering
	\includegraphics[width=\linewidth]{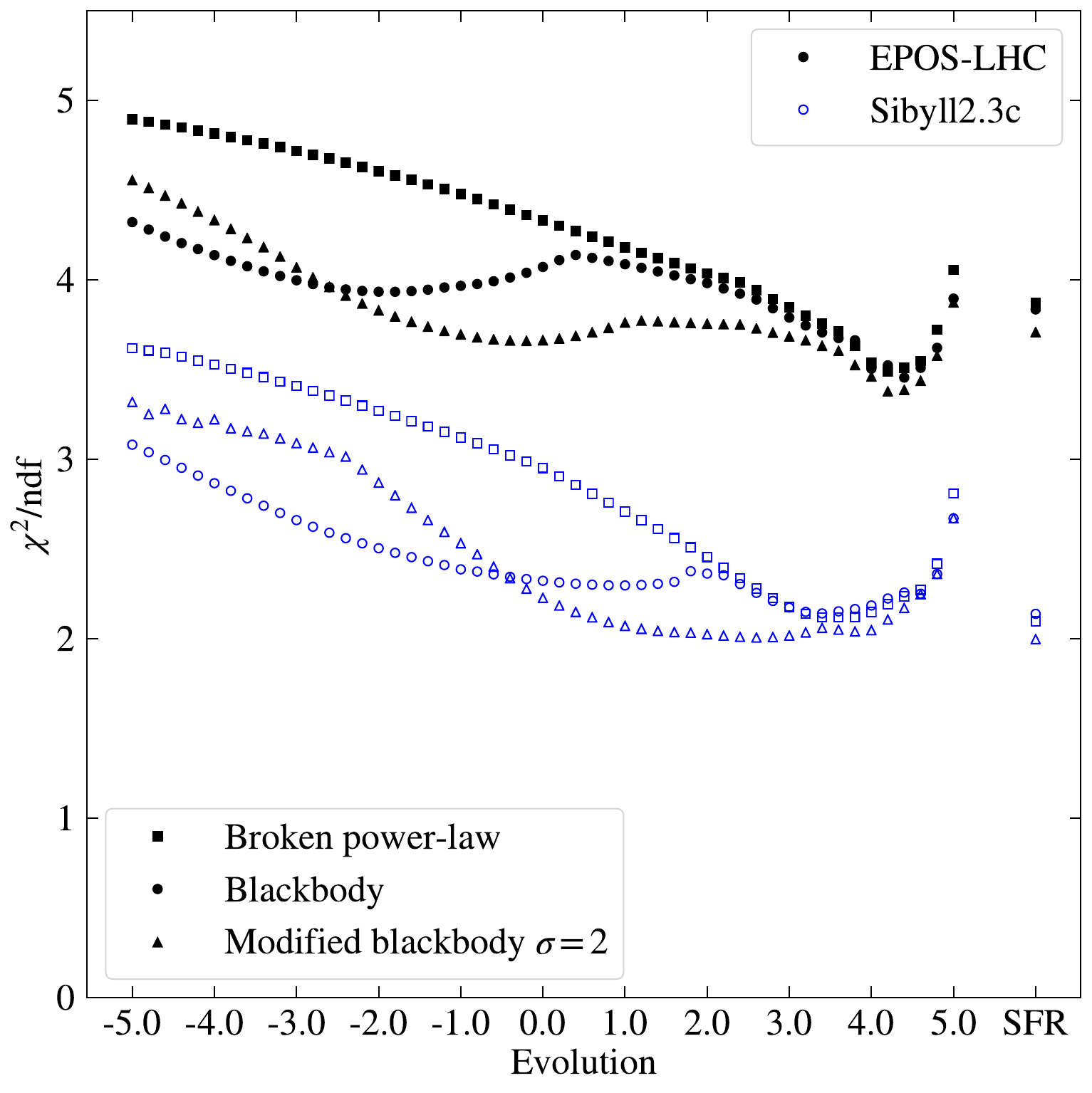}
	\caption{\label{fig:evochi2}Reduced $\chi^2$ of best-fit models as a function of source evolution index $m$ for various parameterizations of the photon field and hadronic interaction models.}
\end{figure}

\begin{figure*}
	\centering
	\begin{minipage}{\linewidth}
		\centering
		\subfloat[]{\label{fig:UFA_evo_CR} \includegraphics[width=\textwidth]{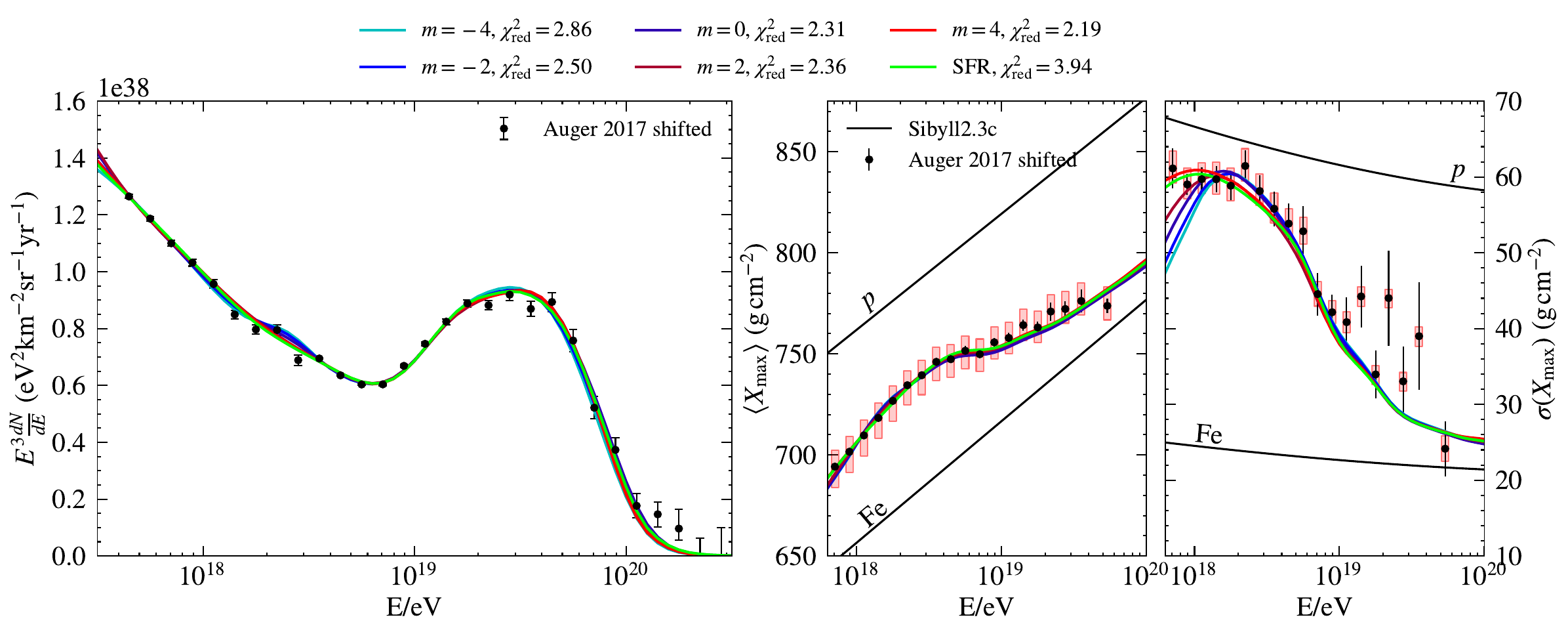} }
	\end{minipage}
	\begin{minipage}{\linewidth}
		\centering
		\subfloat[]{\label{fig:UFA_evo_secondaries} \includegraphics[width=\textwidth]{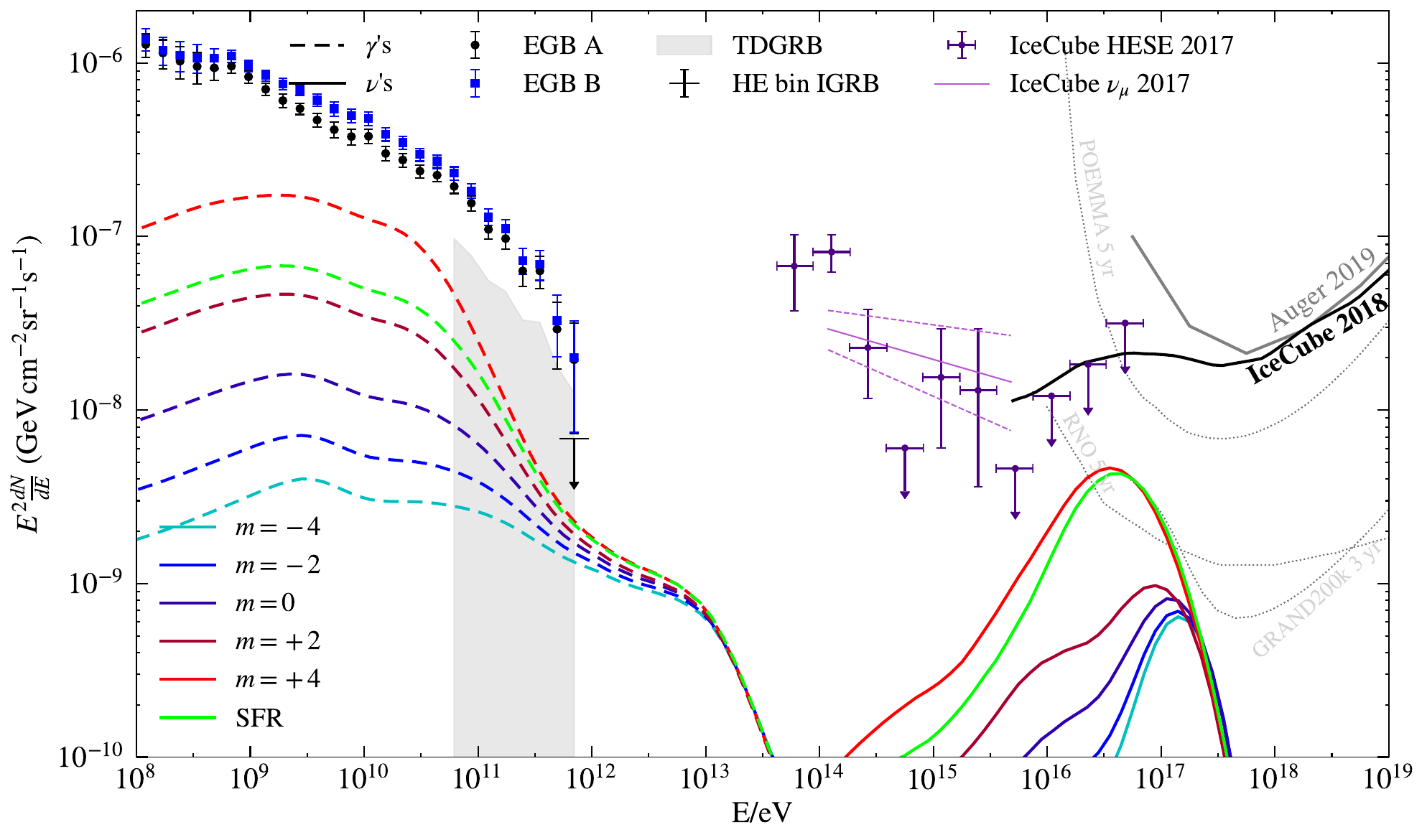} }
	\end{minipage}
	\caption{Impact of source evolution on the predicted CR and multimessenger signals. All power-law evolutions have an exponential cutoff fixed at $z_0=2$, c.f. equation~\eqref{eq:source_evols} and Fig.~\ref{fig:source_evolutions}. \textbf{Top:} The CR spectrum (left) and composition (right) at Earth. Data points are the 2017 Auger spectrum and composition shifted as described in \S~\ref{sec:benchmarks}. \textsc{Sibyll2.3c} is used to infer the $X_\mathrm{max}$ distribution moments from composition predictions. For reference, the predictions of \textsc{Sibyll2.3c} for pure-proton and pure-iron compositions are shown (solid black lines). \textbf{Bottom:} Gamma-ray and neutrino signals at Earth. Data points same as in Fig.~\ref{fig:UFA_benchmark_summary}. }
	\label{fig:UFA_evo_summary}
\end{figure*}

\par

To quantify the constraint, Fig.~\ref{fig:evochi2} (and Fig. 10(a) of~\cite{UFA15}) shows the quality of the fit to UHECR spectrum and composition for SFR evolution and evolutions with $-5\leq m \leq 5$. Independently of the HEG, positive source evolutions are favored, with SFR or models having $2 \lesssim m \lesssim 4$ giving the best-fits. The preferred range of evolution indices is similar to some models of the SFR and medium low luminosity AGNs~\cite{Gavish+16}. The best-fitting model of all considered up to this point has SFR evolution when air shower data is interpreted using \textsc{Sibyll2.3c} and $m=4.2$ when interpreted with \textsc{EPOS-LHC}, giving a reduced $\chi^2$ of $2.00$ and $3.38$, respectively. 

\par
We confirm the positive correlation between the evolution index $m$ and the preferred spectral index of CRs injected into the source, reported in~\cite{UFA15} (cf. their Fig. 10(c)). In general, the Auger data is best reproduced using an injected spectral index harder than $\gamma_\mathrm{inj}=-2$, while the best-fitting evolutions (SFR or $m\gtrsim 2$) prefer $\gamma_\mathrm{inj}\gtrsim-1$. 

\par
Figure~\ref{fig:UFA_evo_summary} shows the UHECR and multimessenger predictions for SFR and some reference evolution indices $m$ using \textsc{Sibyll2.3c}. For illustrative purposes, \textsc{Sibyll2.3c} is used due to its systematically better ability to describe air shower data (c.f. Fig.~\ref{fig:evochi2}). The low energy UHECR composition data is most constraining on source evolution, as is apparent visually in the right panel of Fig.~\ref{fig:UFA_evo_CR}. The source evolution at redshifts $z\gtrsim 1$ only impacts UHECR predictions at energies $\lesssim 10^{17.8}$ eV, where the extragalactic component of the spectrum is dominated by UHECRs from high redshifts. We do not fit such low energies because the Galactic contribution is not well constrained. Thus the quality of the UHECR fit is sensitive to the behavior of $\xi(z)$ only for $z \lesssim 1$. However, the inferred comoving CR power density of UHECR sources does depend on the high redshift behavior of the source evolution.

\par
While the source evolution at high redshifts is not constrained by UHECRs alone, it can, in principle, be constrained by other messengers. However, the gamma-ray and neutrino signals produced by these scenarios are well below current bounds, and so, multimessenger data are not constraining. However, it is worth noting that the shape of the gamma-ray flux is very sensitive to the source evolution. The strong increase in overall gamma-ray flux with the evolution index $m$ is due to the increase in the average CR propagation distance as the evolution becomes more positive. This results in CRs undergoing more Bethe-Heitler pair-production interactions transferring more energy into EM cascades which populate the LAT energy band. To quantify the future constraining power of neutrino data on source evolution, we use the number of IC86-years to $90\%$ CL exclusion.\footnote{Here, IC86-years to $90\%$ CL exclusion refers to the number of years of exposure IceCube must have in its 86-string configuration in order to exclude a source model.} In order for neutrino data to be strongly constraining, in either the evolution index $m$ or the redshift cutoff $z_0$, another $10$ IC86-years of exposure are needed regardless of the HEG. Beyond that, future neutrino constraints will be strongly sensitive to the HEG used to interpret Auger data. Details and plots can be found in Appendix~\ref{app:supplfigs}. 

\subsection{Peak Photon Energy around the Source}

\par
\begin{figure*}
	\centering
	\begin{minipage}{\linewidth}
		\centering
		\subfloat[]{\label{fig:UFA_temp_CR}\includegraphics[width=\textwidth]{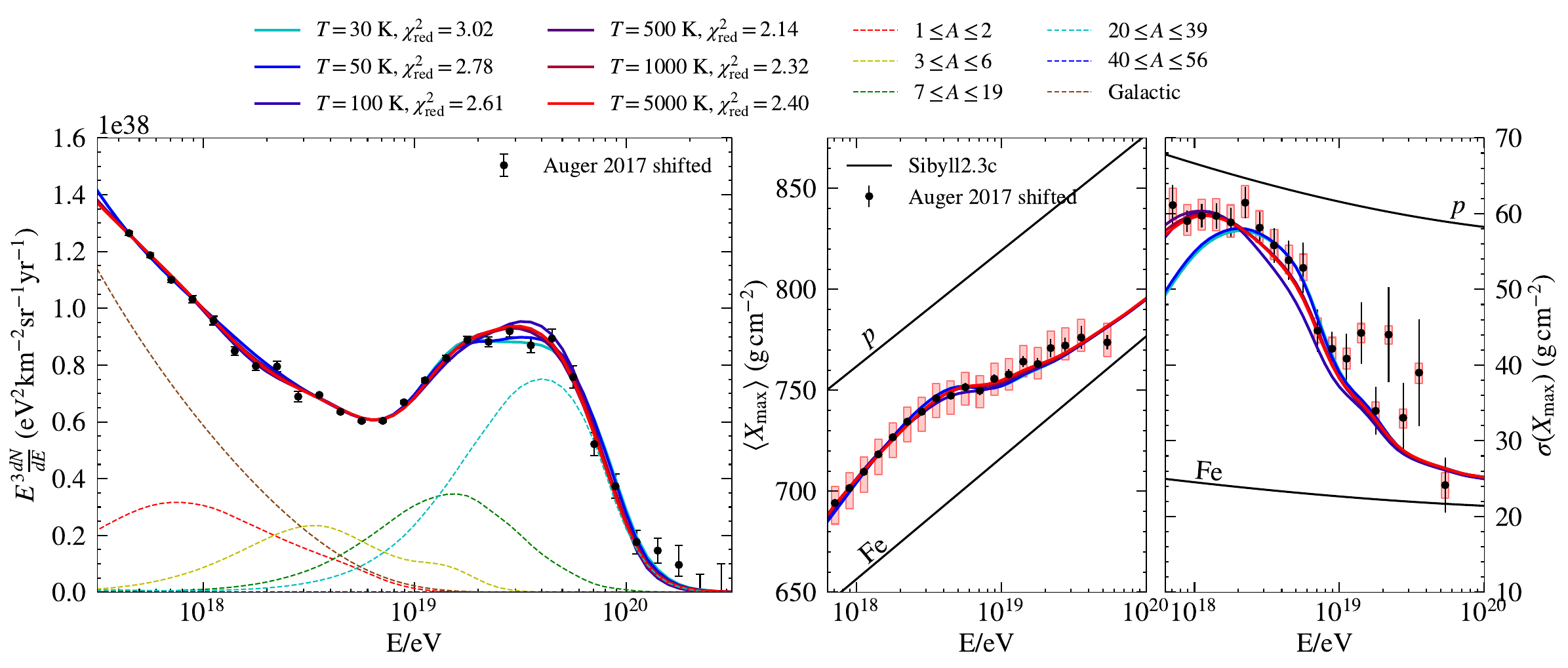}}
	\end{minipage}
	\begin{minipage}{\linewidth}
		\centering
		\subfloat[]{\label{fig:UFA_temp_secondaries}\includegraphics[width=\textwidth]{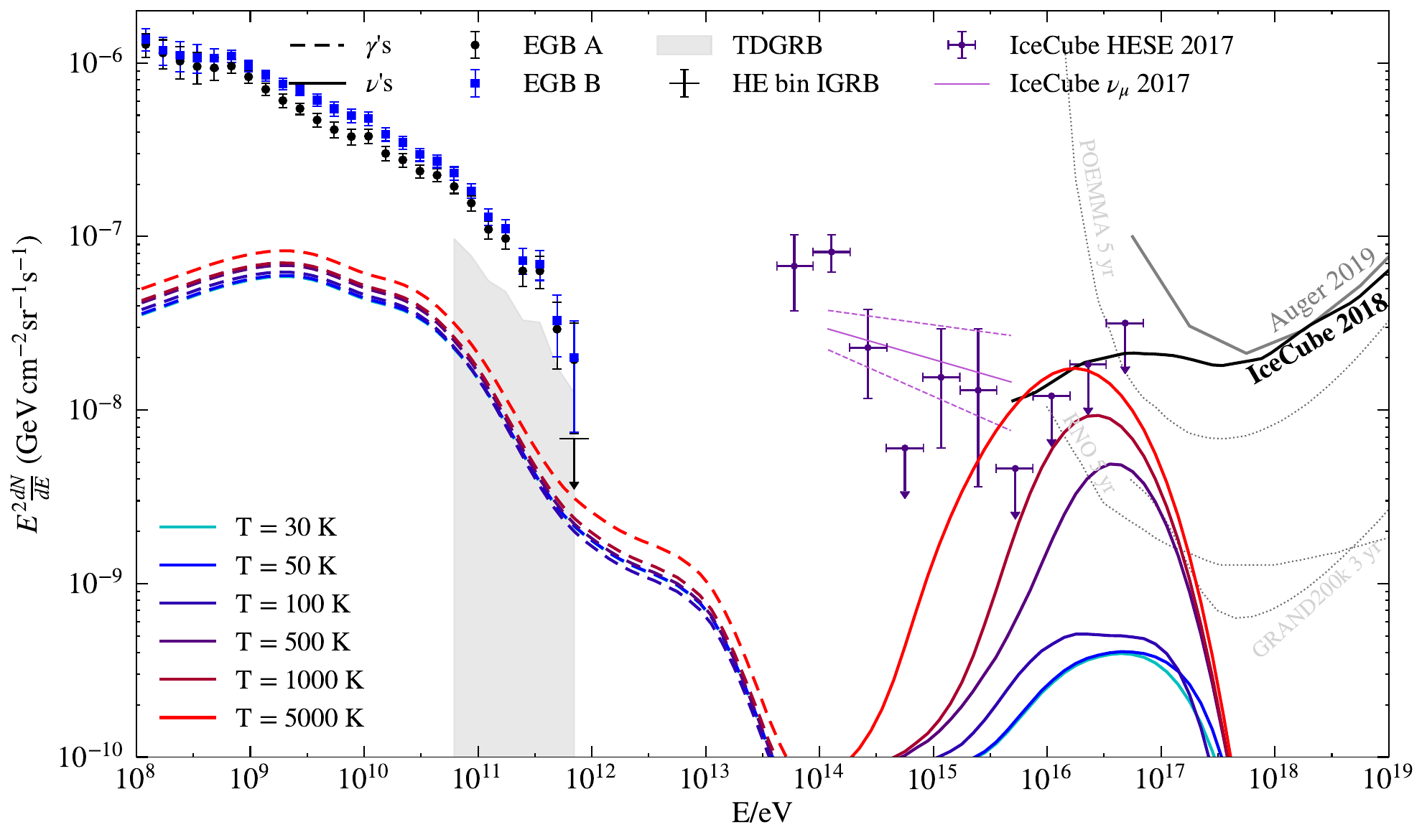}}
	\end{minipage}
	\caption{Impact of source temperature on the predicted CR and multimessenger signals. Ambient photon fields inside the source are described by a blackbody spectrum. Data points are the same as in Fig.~\ref{fig:UFA_evo_summary}. \textbf{Top:} The CR spectrum (left) and composition (right) at Earth. For illustration, a breakdown of the spectrum by mass group is shown for the $500$ K case (dashed colored lines). \textbf{Bottom:} Gamma-ray and neutrino signals at Earth. Evidently, the neutrino signal at $\sim10$ PeV is a sensitive probe of the peak photon energy in the source environment.}
	\label{fig:UFA_temp_summary}
\end{figure*}

The UHECR spectrum and composition are relatively insensitive by themselves to the peak photon energy of photons surrounding the source. By adjusting other source parameters, nearly identical CR spectra and compositions can be obtained for different peak photon energies. This is evident in Fig.~\ref{fig:UFA_temp_summary} which shows the best-fit source model with ambient photon fields described by a blackbody spectrum at different temperatures. Not surprisingly, the gamma-ray flux produced by these models is only weakly sensitive to the photon field peak energy since the gamma-ray flux normalization depends mostly on the total energy-per-nucleon leaving the source environment and this must be roughly the same in all cases in order to fit the Auger spectrum.\footnote{Of course, the gamma-ray flux normalization also depends on the source evolution but we take this to be fixed by considerations of the previous section, setting it to SFR for concreteness in this section.} Moreover, gamma-rays produced in the source environment are subdominant to those produced in propagation except for in the most negative source evolutions.

\begin{figure}[!htpb]
	\centering
	\includegraphics[width=\linewidth]{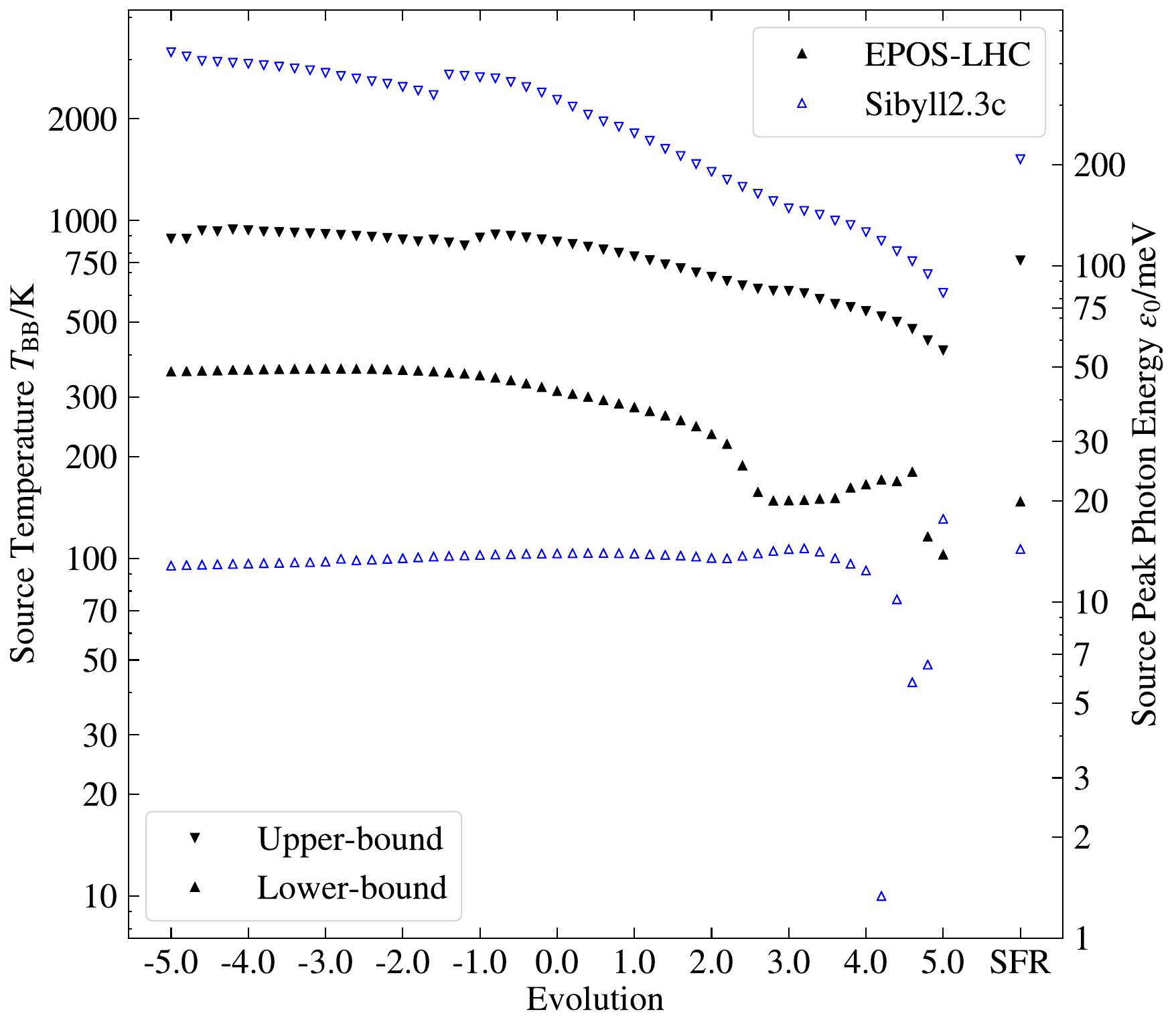}
	\caption{Upper- and lower-bounds on the blackbody temperature of the source photon field from IceCube and data from Auger, respectively. Values were obtained by linear interpolation, so fine structure may not be significant. For ease, these constraints are converted to constraints on the peak photon energy $\varepsilon_0$ of a broken power-law photon field (right y-axis).}
	\label{fig:maxTplt}
\end{figure}

\par
By contrast to the UHECRs themselves, neutrinos are a sensitive probe of the temperature or the peak photon energy of the photon field surrounding the source. The reason for this is two-fold. First, the hotter photon fields lower the threshold energy necessary for nucleons to produce photopions. Second, hotter photon fields more efficiently photodisintegrate nuclei producing more secondary neutrons, which escape the source, and more protons, which have lower rigidity than their parent nucleus and therefore undergo more interactions on average before escaping. But the neutrino flux normalization is sensitive to other UHECR source model parameters besides the peak photon energy, namely the injected spectral index and the source evolution. Softer spectral indices result in relatively more CRs at lower energies. These low energy CRs have a longer escape time and so undergo more interactions within the source environment producing more neutrinos. On the other hand, more positive source evolutions produce more neutrinos in propagation, due to longer UHECR propagation times. 

\par
To obtain conservative constraints on the source temperature from neutrinos, we minimize the number of neutrinos produced in the source environment by fixing the injected spectral index to $\gamma_\mathrm{inj}=-1$ and injecting a single-mass composition. Further, we choose a blackbody rather than broken power-law photon energy distribution, since it minimizes the number of neutrinos produced in the source environment as shown in UFA15 (see their Fig. $12$). The resulting upper limits on the temperature of the photon field surrounding the accelerator are shown in Fig.~\ref{fig:maxTplt} for various source evolutions and both HEGs. The corresponding constraint on the peak photon energy/temperature for other photon field parameterizations can be obtained using equation (A3) in~\cite{UFA15}. Across all evolutions, ambient photon fields with blackbody temperatures $T_\mathrm{BB} \geq 4000$ K are excluded at $90\%$ CL based on neutrino data alone. For SFR and source evolutions with $m \gtrsim 2$, as favored by UHECR data, we exclude photon fields hotter than $T_\mathrm{BB} \simeq 2000$ K, at $90\%$ CL. For a broken power-law photon field these bounds correspond to an upper-bound on the peak photon energy of $\varepsilon_0 \leq 500$ meV across all evolutions and $\varepsilon_0 \lesssim 250$ meV for SFR and $m \gtrsim 2$ source evolutions.

\par
A lower bound can also be placed on the source temperature by considering the goodness-of-fit to the Auger spectrum and composition. Compared to the best-fit, the $\chi^2$ for the CR data dramatically worsens as the source photon field temperature goes to zero (see Fig. $12$ in~\cite{UFA15}). This is because cool photon fields do not produce enough photodisintegration in the source environment to reproduce the Auger composition or the population of subankle protons. We consider temperatures producing fits which are $3\sigma$ worse than the best-fit to be excluded by Auger data. This constrains CR sources to have photon fields hotter than $T_\mathrm{BB} = 10$ K at the $90\%$ CL, regardless of evolution or HEG. This lower-bound corresponds to a minimal peak photon energy of $\varepsilon_0 = 1$ meV for a broken power-law photon field. 

\section{Refinements to UFA15} \label{sec:refinements}

\subsection{Distance to Nearest Source}

\par
Depending on how particles are entrained in the accelerator, the UHECRs emerging from the accelerator can have a relatively pure composition or a mixed composition. The benchmark model comparisons of Fig.~\ref{fig:UFA_benchmark_CR} show that the fiducial model, with single-mass injected, has $\chi^2_{\rm red} = 3.69$ while the corresponding galactic mix model gives a poorer fit, $\chi^2_{\rm red} = 7.14$. In this subsection, we investigate whether this contrast is an indication that the source accelerates a relatively pure rather than a broadly mixed composition. 

\par
Examining the galactic mix curve in Fig.~\ref{fig:UFA_benchmark_CR}, it is evident that the poor fit is partly caused by the predicted spectrum extending too high in energy relative to the data. This cannot be fixed by reducing the maximum energy of the accelerator, for a given composition, because proper placement of the ankle fixes the rigidity cutoff $R_\mathrm{max}$ of UHECRs from the accelerator. This is due to the fact that nucleons freed in photodisintegration interactions obey the relation $E^{\rm max}_{p,\rm PD} = A^{-1} E^{\rm max}_A = (Z/A) R^{\rm max} = 1/2 R^{\rm max}$, where $E^{\rm max}_{p,\rm PD}$ is fixed by the relative position of the ankle and the observed spectral cutoff. So insisting on a composition with heavy components, such as Fe, forces the spectrum to extend to higher energy than in the well-fitting case that Si is the heaviest component. 

\par
However there is a potential cure for a spectrum which extends too high in energy. Due to the Greisen-Zatsepin-Kuzmin (GZK) cutoff~\cite{Greisen66, Zatsepin+66}, the CR spectrum at the highest energies is very sensitive to the distance of the nearest source (see e.g.~\cite{Taylor+11}). To study this sensitivity, we employ a model using a galactic mixture of injected CRs and SFR source evolution, but with extragalactic propagation which has a minimum comoving source distance from Earth, $D_\mathrm{min}$. 

\par
\begin{figure*}[!htp]
	\centering
	\begin{minipage}{\linewidth}
		\centering
		\subfloat[]{\label{fig:UFA_mindist_CR}\includegraphics[clip,rviewport=0 0 1 1.1, width=\textwidth]{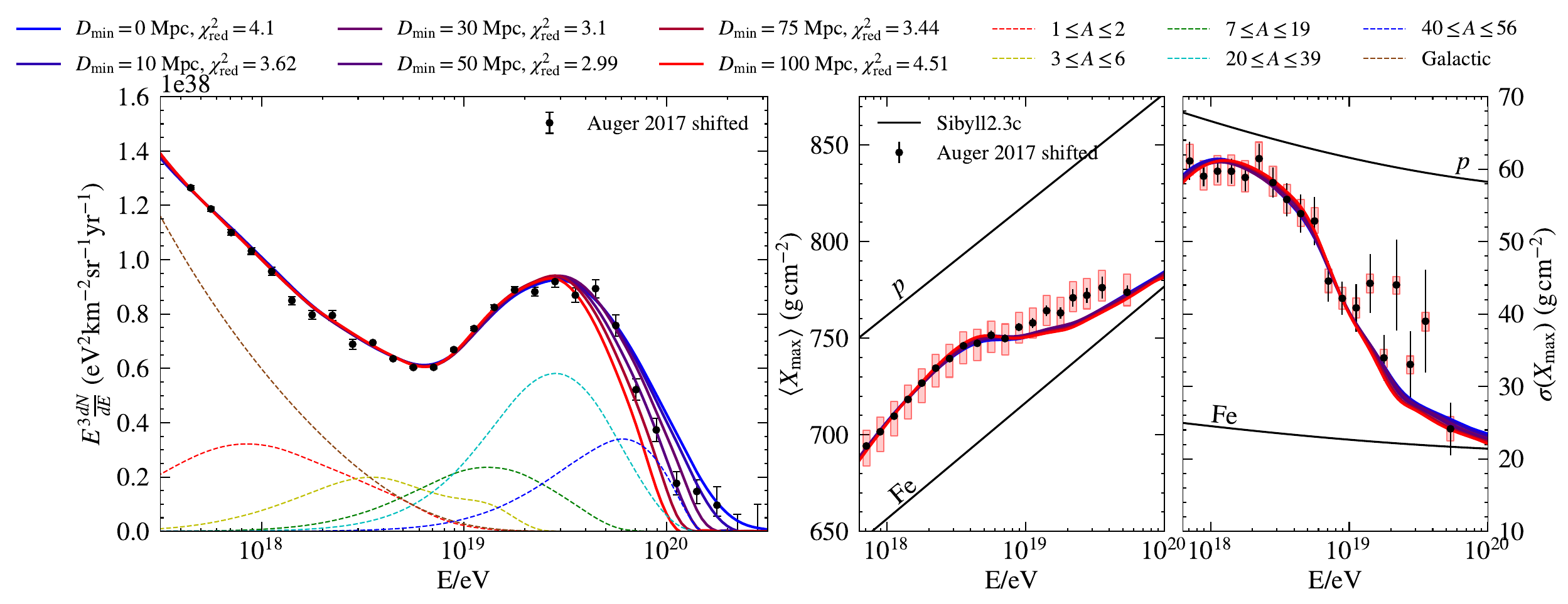}}
	\end{minipage}
	\begin{minipage}{\linewidth}
		\centering
		\subfloat[]{\label{fig:UFA_mindist_secondaries}\includegraphics[width=\textwidth]{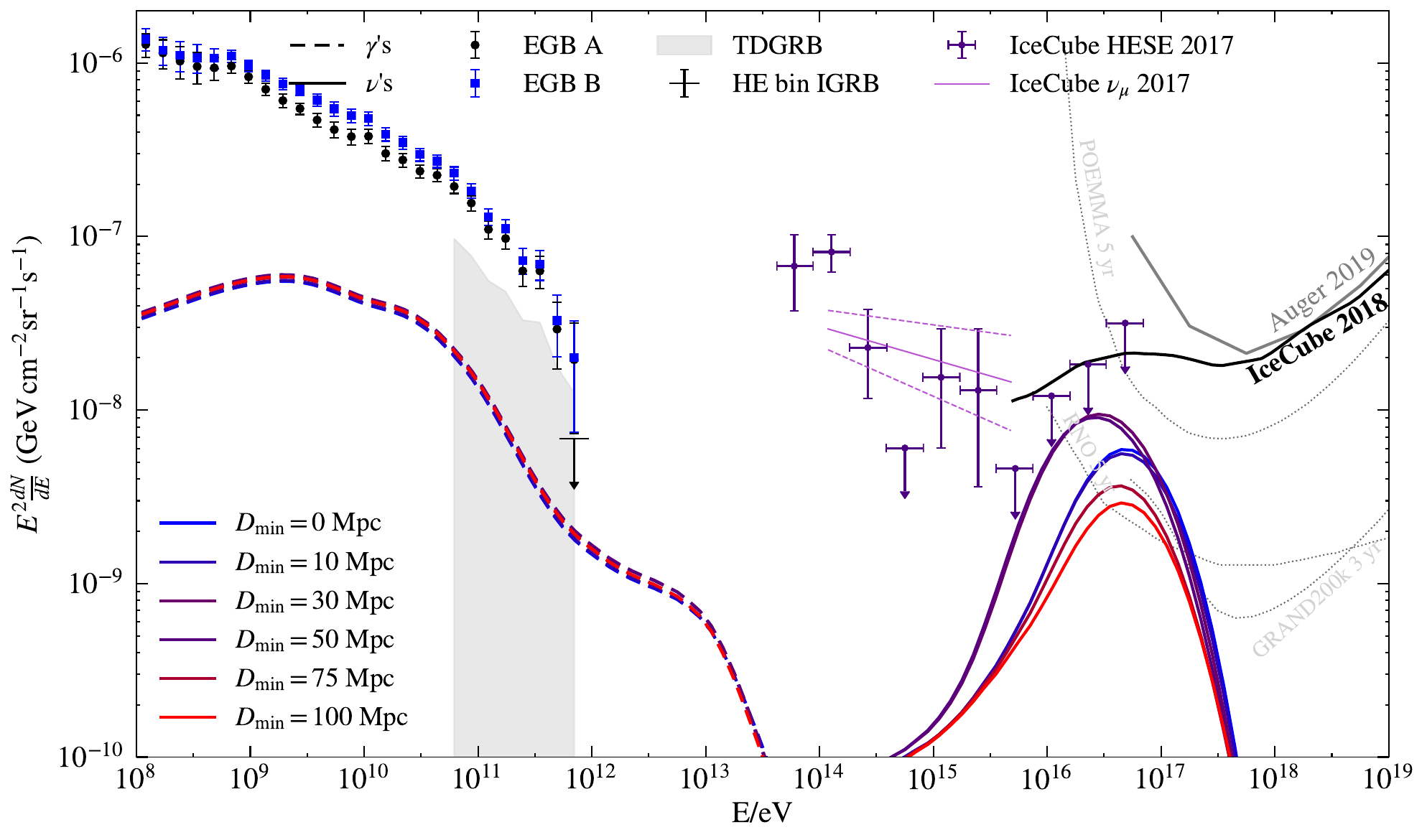}}
	\end{minipage}
	\caption{Impact of nearest source distance $D_\mathrm{min}$ on the predicted CR and multimessenger signals. Spectra were obtained using a SFR evolution beyond $D_\mathrm{min}$ and zero sources for $D < D_\mathrm{min}$. Data points are the same as in Fig.~\ref{fig:UFA_evo_summary}. \textbf{Top:} The CR spectrum (left) and composition (right) at Earth. For illustration, a breakdown of the spectrum by mass group is shown for nearest source distance of $50$ Mpc case (dashed colored lines). \textbf{Bottom:} Gamma-ray and neutrino signals at Earth. Gamma-rays are insensitive to the value of $D_\mathrm{min}$, while the neutrino signal weakens for larger values of $D_\mathrm{min}$.}
	\label{fig:UFA_mindist_summary}
\end{figure*}

Figure~\ref{fig:UFA_mindist_summary} shows that the fit to the Auger spectrum at the highest energies can be substantially improved for galactic mix models by introducing a non-zero distance to the nearest source. For the shifted 2017 Auger spectrum, the best-fit is obtained for a nearest source distance of $D_\mathrm{min} = 30-50$ Mpc, independent of the HEG used to interpret air shower data. However, it is important to keep in mind several points.

\par
Firstly, while the fit to the spectral cutoff is improved by $D_\mathrm{min} = 30-50$ Mpc, the fit to $X_\mathrm{max}$ at high energy remains worse than for our single-composition scenarios.

\begin{figure*}[!htpb]
	\centering
	\includegraphics[width=\linewidth]{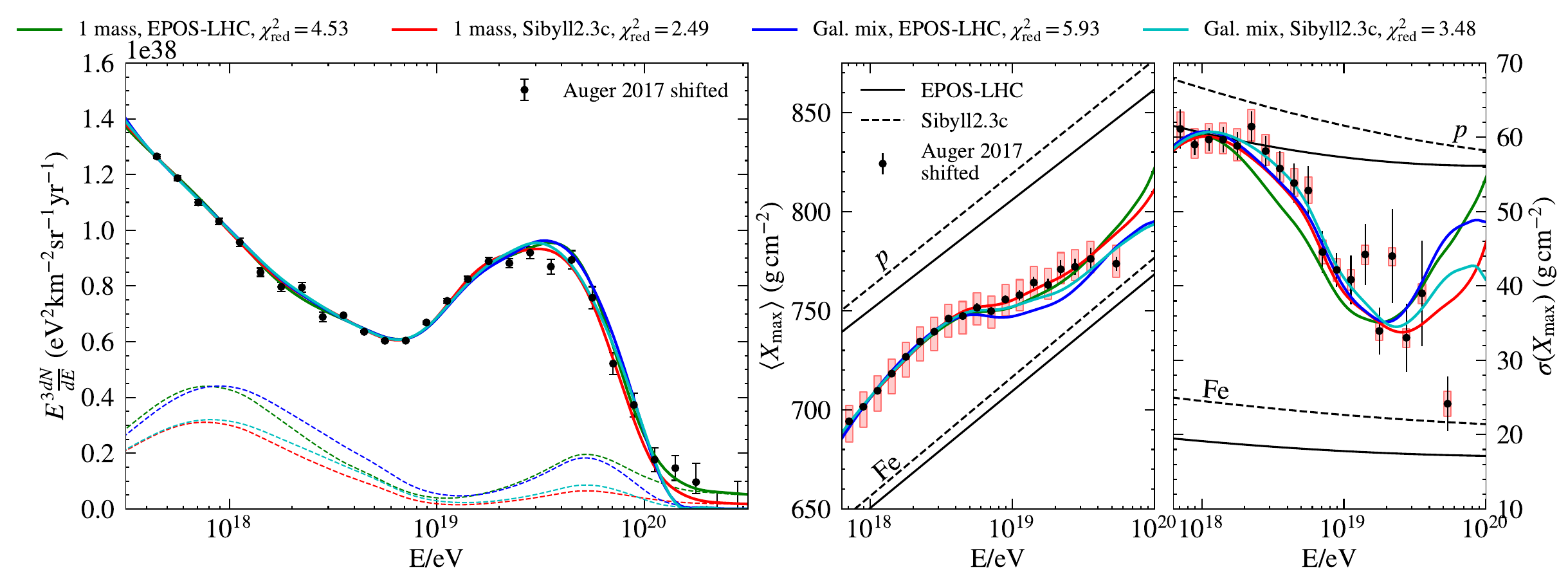}
	\caption{CR signals in the subdominant pure-proton component model which gives the maximal observed proton fraction above $50$ EeV compatible with current multimessenger data. The CR spectrum (left) and composition (right) at Earth. The proton spectrum at Earth is shown for each model (colored dashed lines).}
	\label{fig:UHEproton_maxobsp}
\end{figure*}

\par
Secondly, the best-fit $D_\mathrm{min}$ depends strongly on both the heaviest nucleus injected into the source environment and the systematic shifts in energy and $X_\mathrm{max}$ we have used. For instance, if we left the Auger data unshifted the best-fit distance to the nearest source would be farther from Earth, while if instead we used a lighter composition injected into the source environment the best-fit distance to the nearest source would be nearer Earth. Therefore, mild adjustments in the injected composition might allow the fit to $X_\mathrm{max}$ at highest energies to be improved while also fitting the spectral cutoff; we do not pursue such adjustments here.

\par
Thirdly, although one might think that increasing the source temperature could equally well solve the problem of too much flux at the highest energies, allowing $D_\mathrm{min}=0$ Mpc by producing comparable amounts of photodisintegration as would be created during propagation, this scenario is excluded. Source temperatures high enough to produce the required amount of photodisintegration are excluded by the excess of neutrinos that are produced in the source as a result.

\par
Finally, multimessenger signals cannot be used to constrain $D_\mathrm{min}$ since both the gamma-ray flux within the LAT energy range and the EHE cosmic neutrino flux are fairly insensitive to this parameter. This is because these fluxes are mostly due to cumulative contributions throughout the history of the universe, so that high redshifts make a significant contribution to the observed fluxes. However, a nonzero distance to the nearest source increases the minimum amount of photodisintegration experienced by CRs in propagation reducing the need for pre-escape attenuation of the highest energy CRs. Therefore, a relatively distant nearest source can describe the Auger spectrum with a cooler ambient photon field and so we expect the neutrino flux to be smaller in this case. 

\par
In a single-mass scenario, where a pure composition emerges from the accelerator, data prefers a relatively nearby, $\lesssim 10$ Mpc, closest source as can be seen in Fig.~\ref{fig:UFA_mindist_singlemass} of Appendix~\ref{app:supplfigs}. However, even in the single-mass case this preference is strongly sensitive to the injected composition and systematic shifts in energy and $X_\mathrm{max}$ which are chosen. 

\par
This leads us to conclude that current data is not able to discriminate between a mixed and pure composition injected by the accelerator, in the absence of knowledge of the distance to the nearest source.

\subsection{Subdominant Pure-Proton Component} \label{sec:UHEp}

\par
Here we consider the possibility of a second, subdominant component of UHECRs in addition to CRs originating from sources described within the UFA15 framework. Specifically, we consider that some population of sources produces a pure-proton component escaping the source environment which extends to energies $\gtrsim 10$ EeV. Similar studies were conducted in \cite{vanVliet+19,Moller+18}, but without constraining the proton fraction to CR data. A pure-proton composition was originally expected for GRBs~\cite{Waxman95}, due to the extreme temperatures in the collapsar which dissolve nuclei depending on the radius at which they are injected into the jet. The possibility that either pure-proton or mixed compositions can follow from different conditions has been discussed by multiple authors, e.g., see~\cite{Horiuchi+12,Zhang+17}. Phenomenologically, it is interesting to investigate whether such an additional component may better describe the Auger data and whether such a component can be ruled out by multimessenger data. Further motivation to study an additional UHE light component comes from recent observational evidence using Auger’s surface detector (SD), which shows that the rate of increase of the average nuclear mass with energy is slowing at the highest energies~\cite{Aab+17}. That analysis makes use of the fact that the SD rise-time is sensitive to composition and that the order-of-magnitude larger SD dataset provides adequate statistics to study trends to higher energies than is possible with the fluorescence detector’s measurements of $X_\mathrm{max}$.

\begin{figure*}
	\centering
	\begin{minipage}{\linewidth}
		\centering
		\subfloat[]{\label{fig:UHEproton_CR}\includegraphics[width=\textwidth]{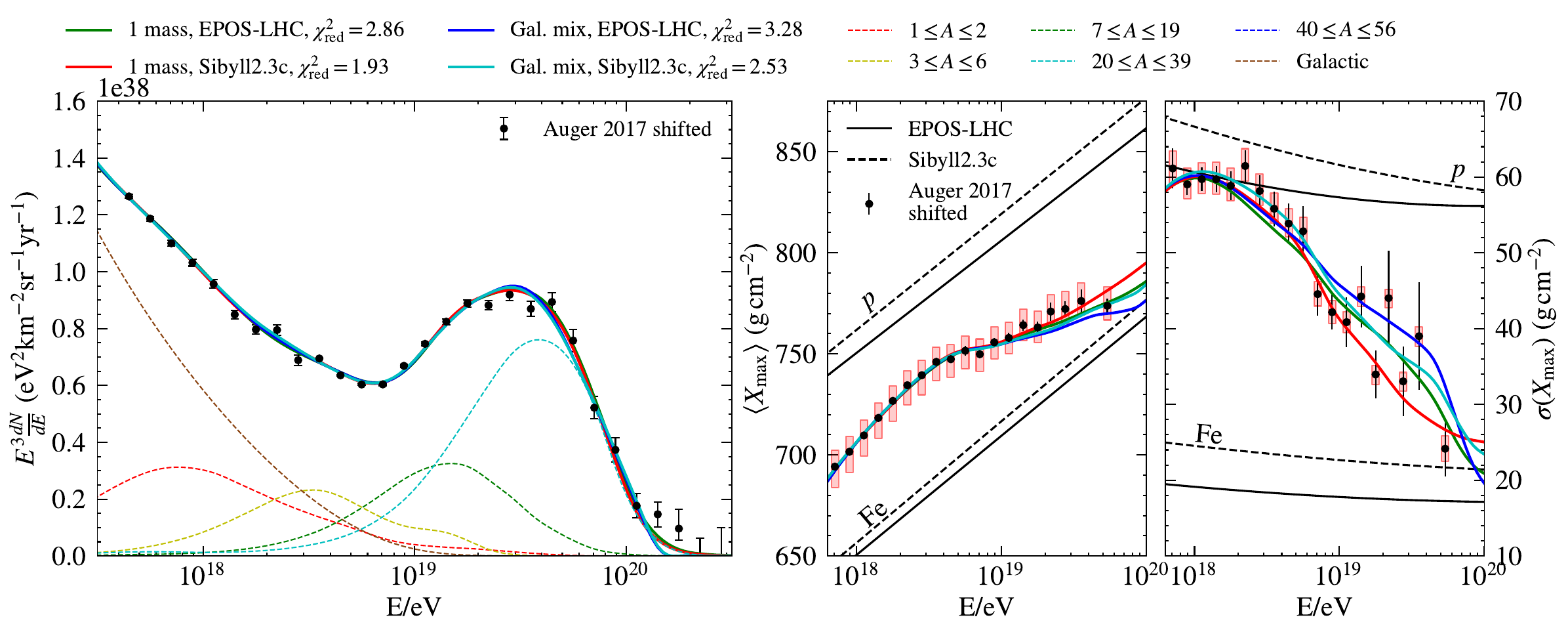}}
	\end{minipage}
	\begin{minipage}{\linewidth}
		\centering
		\subfloat[]{\label{fig:UHEproton_secondaries}\includegraphics[width=\textwidth]{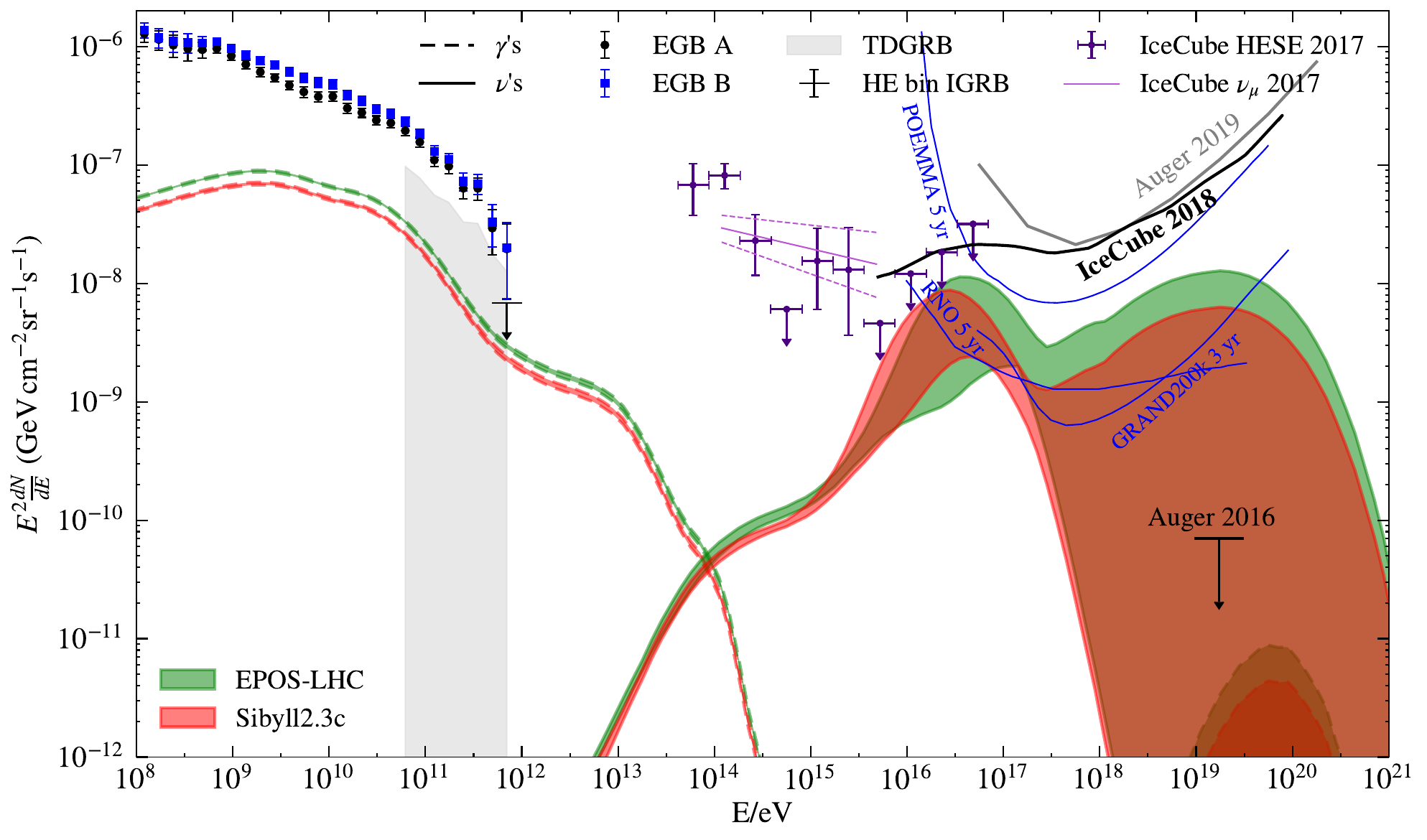}}
	\end{minipage}
	\caption{Predictions of the best-fit UFA15 models including a subdominant pure-proton component for CRs and the range of multimessenger signals across all viable subdominant pure-proton component models. Data points are the same as in Fig.~\ref{fig:UFA_evo_summary}. \textbf{Top:} The CR spectrum (left) and composition (right) at Earth. The breakdown of the observed UHECR spectrum by mass group is shown for the best-fitting model, i.e. the subdominant pure-proton component in addition to a single-mass UFA15 component using \textsc{Sibyll2.3c} (dashed colored lines). \textbf{Bottom:} The range of gamma-ray and neutrino signals possible with the additional subdominant pure-proton component. Projected sensitivities for future neutrino detectors are highlighted in blue. The Auger 2016 upper-bound on UHE gamma-rays~\cite{Aab+16c} is plotted in black.}
	\label{fig:UHEproton_summary}
\end{figure*}

\begin{figure*}[!htpb]
	\centering
	\begin{minipage}{0.49\linewidth}
		\centering
		\subfloat[]{\label{fig:UHEproton_map_epos}\includegraphics[width=\textwidth]{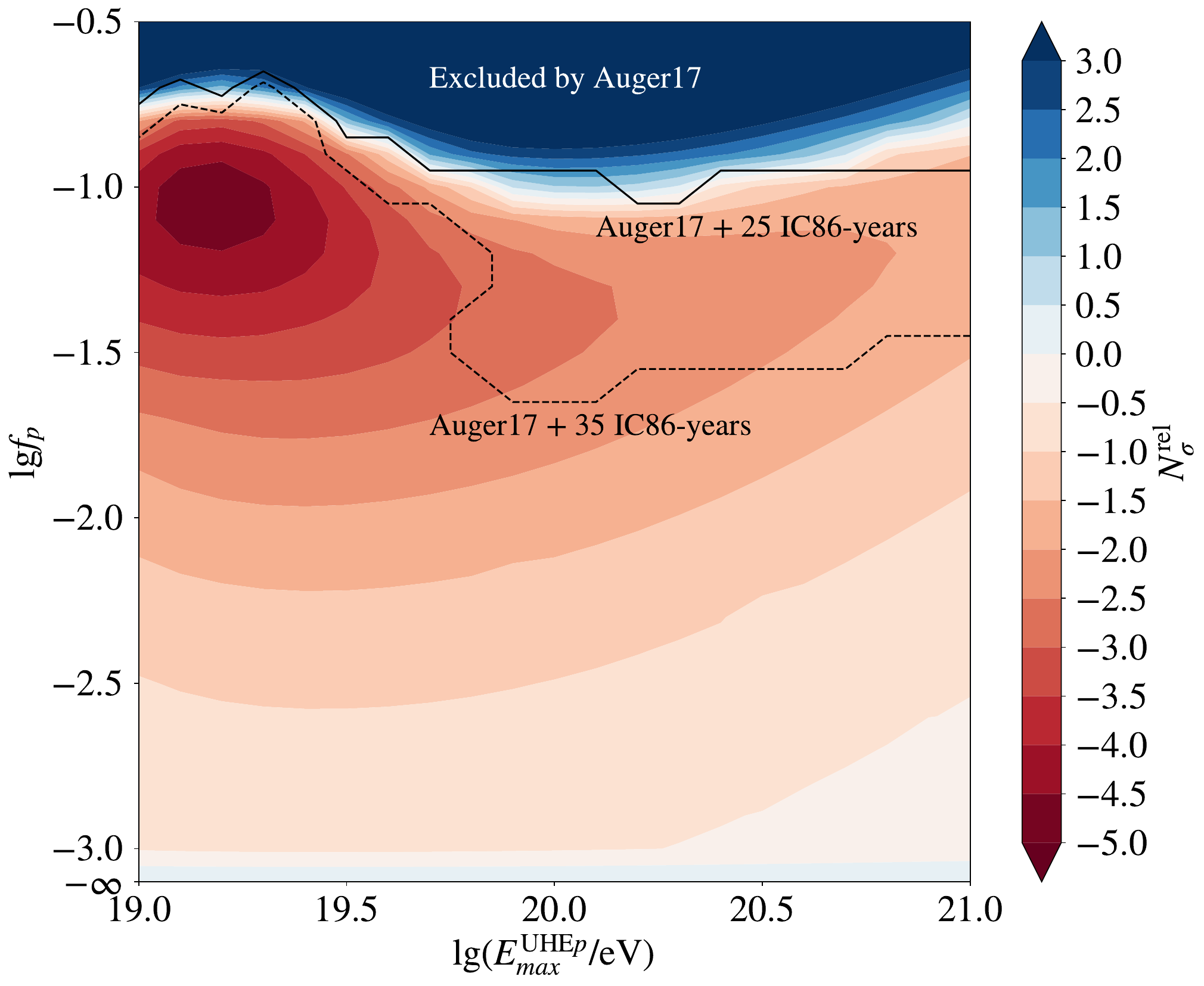}}
	\end{minipage}
	\begin{minipage}{0.49\linewidth}
		\centering
		\subfloat[]{\label{fig:UHEproton_map_sibyll}\includegraphics[width=\textwidth]{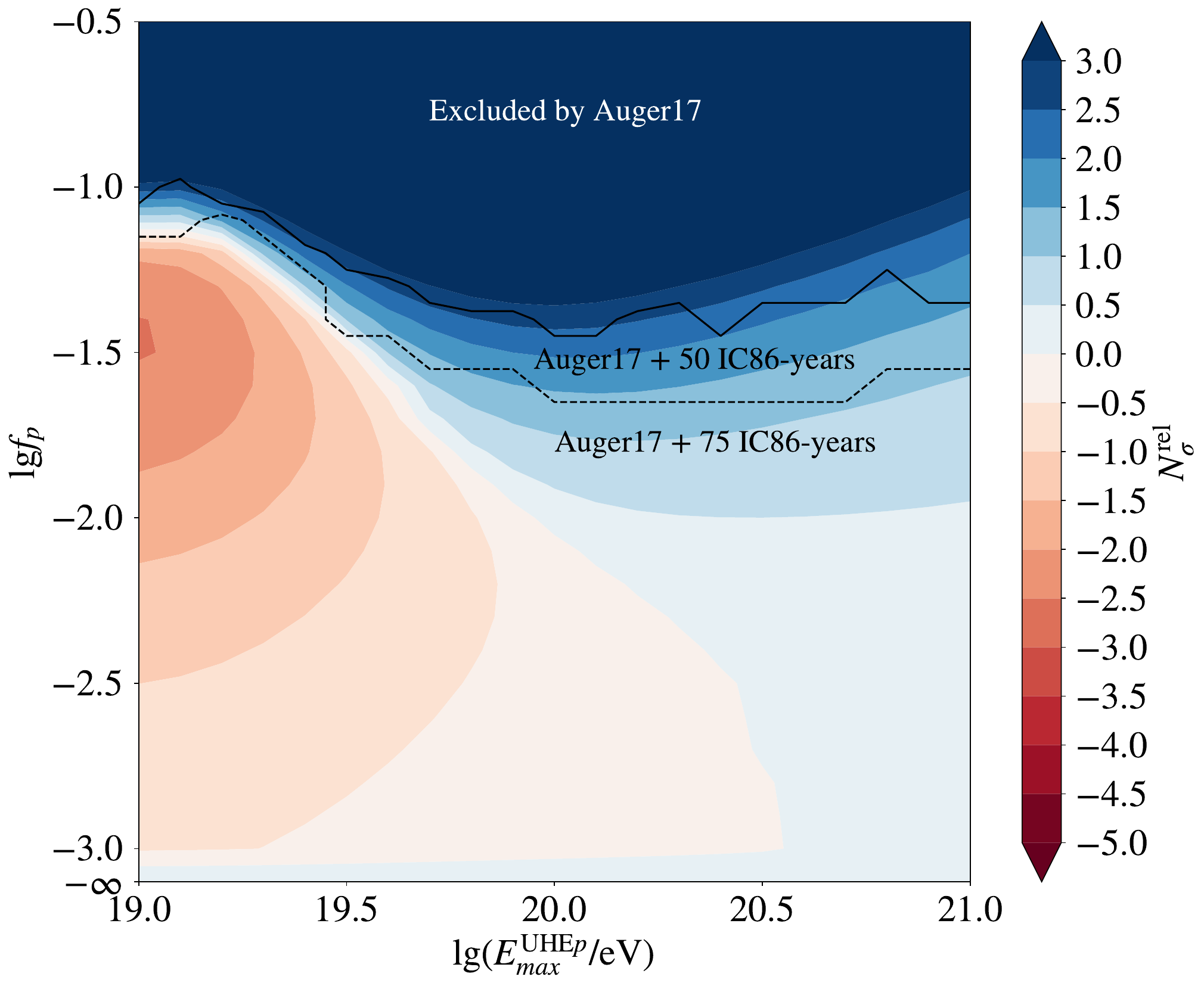}}
	\end{minipage}\\
	\begin{minipage}{0.49\linewidth}
		\centering
		\subfloat[]{\label{fig:UHEproton_map_galmix_epos}\includegraphics[width=\textwidth]{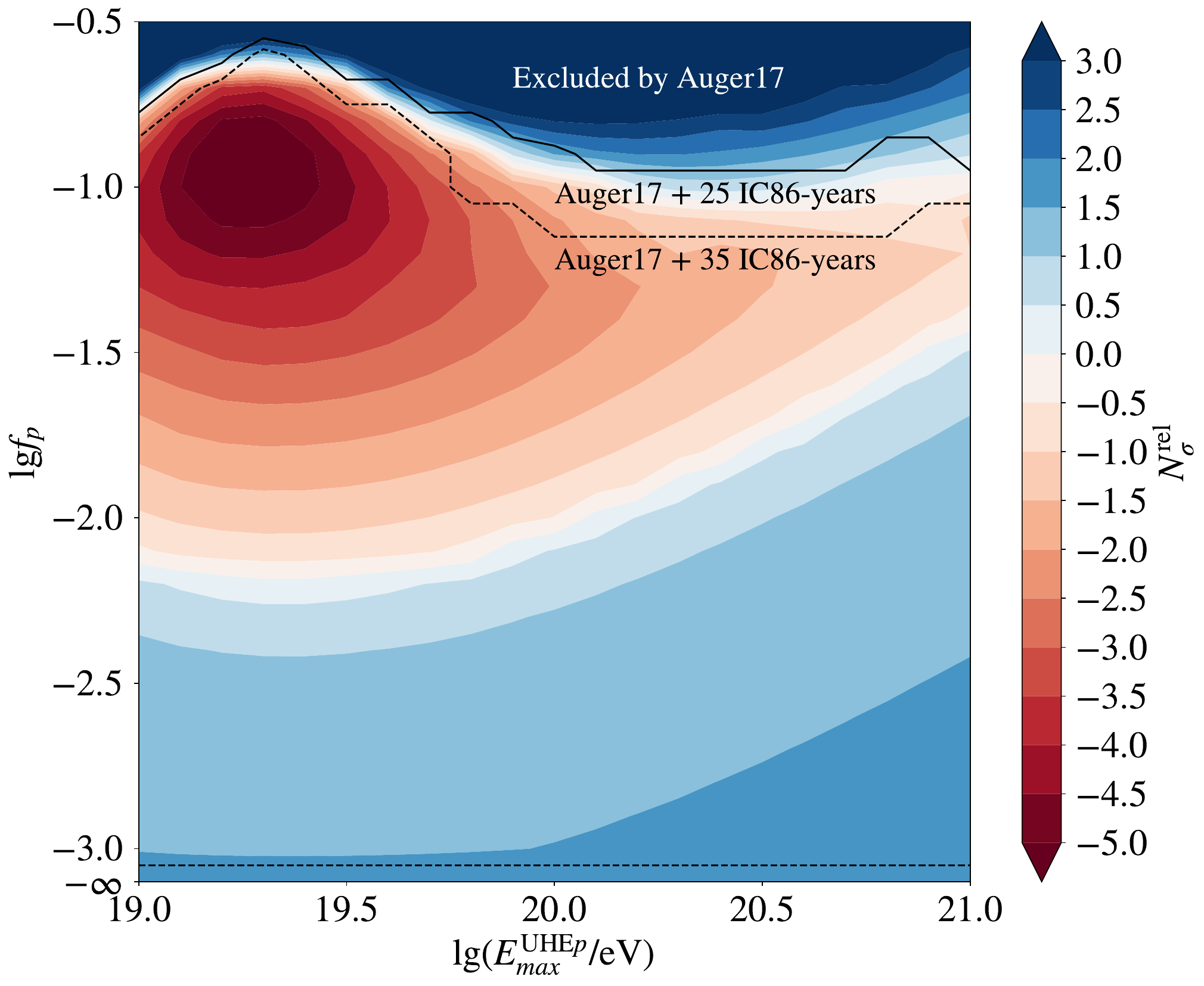}}
	\end{minipage}
	\begin{minipage}{0.49\linewidth}
		\centering
		\subfloat[]{\label{fig:UHEproton_map_galmix_sibyll}\includegraphics[width=\textwidth]{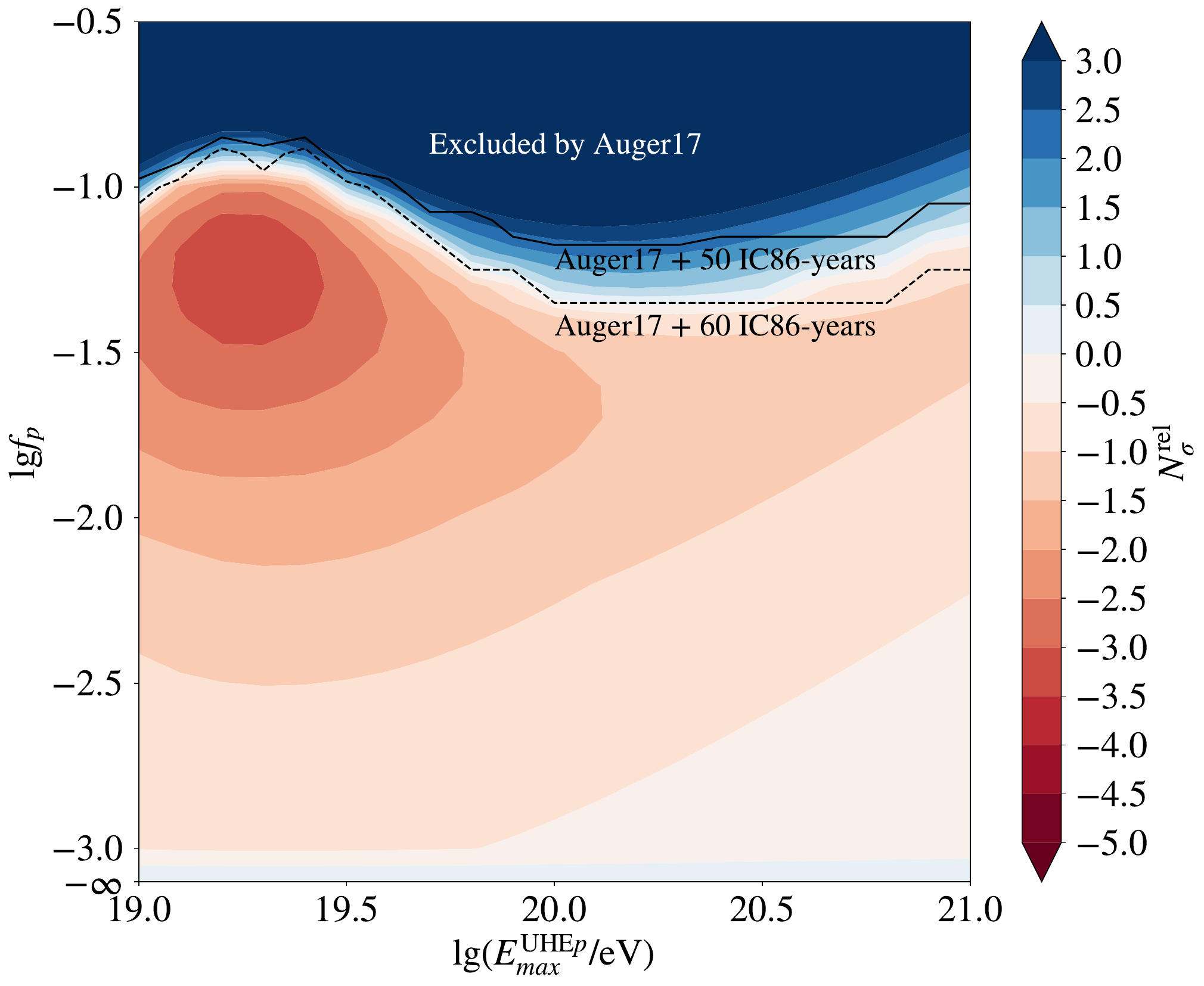}}
	\end{minipage}
	\caption{Effect on fit due to an additional pure-proton component with (a) single-mass \textsc{EPOS-LHC}, (b) single-mass \textsc{Sibyll2.3c}, (c) galactic mix \textsc{EPOS-LHC}, and (d) galactic mix \textsc{Sibyll2.3c} models. The color indicates the number of standard deviations from the corresponding best-fit UFA15 model without the extra component. Black contours indicate the combined CR-neutrino bounds with the dark blue region excluded by current Auger data at the $90\%$ CL.}
	\label{fig:UHEproton_maps}
\end{figure*}

\par
To study this question we introduce an additional component characterized by only three parameters: the spectral index of the escaping protons, $\gamma_p$, the cutoff energy of their spectrum, $E_\mathrm{max}^{\mathrm{UHE}p}$, and the fraction $f_p$ of energy carried by these protons relative to the total energy of all CRs escaping their source with $E > E_\mathrm{ref} = 10^{19}$ eV. Quantitatively, $f_p$ and $\Phi_p$ are defined as

\begin{align} \label{eq:fp_def}
	f_p = \frac{\int_{E_\mathrm{ref}}^\infty E\Phi_p \mathrm{d}E}{\int_{E_\mathrm{ref}}^\infty E(\Phi_p + \Phi_\mathrm{mix}) \mathrm{d}E} ~~,
\end{align}

\noindent 
and

\begin{align}
	\Phi_p \sim E^{\gamma_p} e^{-E/E^{\mathrm{UHE}p}_\mathrm{max}} ~~,
\end{align}

\noindent
where $\Phi_p$ is the spectrum of escaping second-component protons and $\Phi_\mathrm{mix}$ is that for the mixed composition (i.e. UFA15) component.

\par
We perform our analysis for four cases: injecting a single-mass or galactic mix for the UFA15 component while using \textsc{EPOS-LHC} or \textsc{Sibyll2.3c} as the HEG. We fix both injected spectral indices to $\gamma_\mathrm{inj} = \gamma_p = -1$, the former because it is typical of the best-fits of UFA15 models with no extra component and the latter because we are interested in the possibility of a protonic component which can be visible to high energies. For $f_p$ and $E_\mathrm{max}^{\mathrm{UHE}p}$ we step through a grid of values. In the case of an injected galactic mix, we allow the distance to the nearest source to float between $0-100$ Mpc. The fits of these four cases with the maximum proton fraction and best-fit proton fraction are shown in Figs.~\ref{fig:UHEproton_maxobsp} and~\ref{fig:UHEproton_summary} respectively and the $\chi^2$ contours are displayed in Fig.~\ref{fig:UHEproton_maps} in units of $N^\mathrm{rel}_\sigma$, the number of standard deviations the resulting model is from the best-fit with $f_p=0$.\footnote
{Here we follow the PDG~\cite{PDG, Rosenfeld75} and define the number of standard deviations from the best-fit as $N'_\sigma = S^{-1} \sqrt{\chi^2_\mathrm{model}-\chi^2_\mathrm{min}}$, where $S=\sqrt{\chi^2_\mathrm{min}/N_\mathrm{dof}}$ is a scale factor introduced in~\cite{Rosenfeld75} to enlarge the uncertainties in account of a poor $\chi^2$ at the minimum, $N_\mathrm{dof}$ is the number of degrees of freedom, $\chi^2_\mathrm{min}$ is the $\chi^2$ for the global minimum, and $\chi^2_\mathrm{model}$ is the $\chi^2$ for a given model. $N_\sigma^\mathrm{rel}$ is defined as $N_\sigma^\mathrm{rel}=\sgn(\chi^2-\chi^2|_{f_p=0}) S^{-1} \sqrt{\chi^2_\mathrm{max}-\chi^2_\mathrm{min}}$, where $\chi^2_\mathrm{max}=\max(\chi^2, \chi^2|_{f_p=0})$ and $\chi^2_\mathrm{min}=\min(\chi^2, \chi^2|_{f_p=0})$.} In all four cases, the global minimum has a non-zero value of $f_p$, indicating that the addition of a light component of UHECRs at high energies can better describe the data. Interpreting the Auger $X_\mathrm{max}$ data using \textsc{EPOS-LHC}, the fit improves by more than $5\sigma$ relative to the global minimum for both the single-mass and galactic mix cases. These improvements in the fit quality are most strongly driven by an improved description of $\langle \ln{A}\rangle$. High values of $f_p$ are most strongly prevented by a degrading description of $V(\ln{A})$ (see Appendix~\ref{app:supplfigs} for more details). The contour lines in Fig.~\ref{fig:UHEproton_maps} show that currently only high values of $f_p$ are excluded by data from Auger and IceCube. (Presently, the bound is mostly driven by the Auger data.) Figure~\ref{fig:UHEproton_summary} shows the CR spectra and composition and multimessenger signals predicted for the global best-fits in each of the four cases. The second peak at higher energy in the neutrino flux is characteristic of a non-negligible amount of trans-GZK protons escaping the source environment, which is strongly dependent on $E_\mathrm{max}^{\mathrm{UHE}p}$. Similarly, there is a UHE peak in the gamma-ray flux due to GZK production of photons. However, this peak is not constrained by current bounds on UHE photons from Auger~\cite{Aab+16c}.  

\par
\begin{figure*}[!htpb]
	\centering
	\begin{minipage}{0.49\linewidth}
		\centering
		\subfloat[]{\label{fig:obsfp_map_epos}\includegraphics[width=\textwidth]{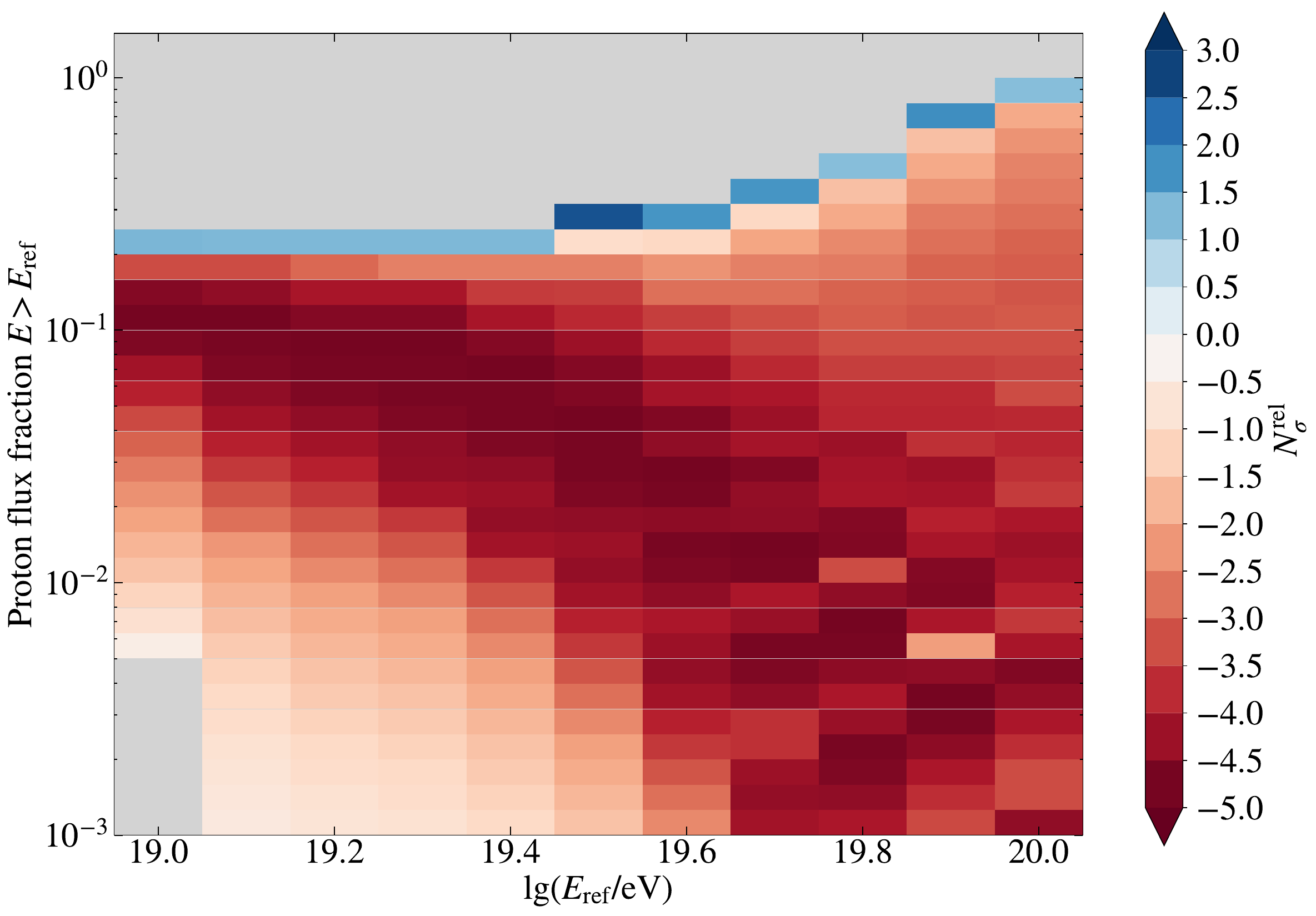}}
	\end{minipage}
	\begin{minipage}{0.49\linewidth}
		\centering
		\subfloat[]{\label{fig:obsfp_map_sibyll}\includegraphics[width=\textwidth]{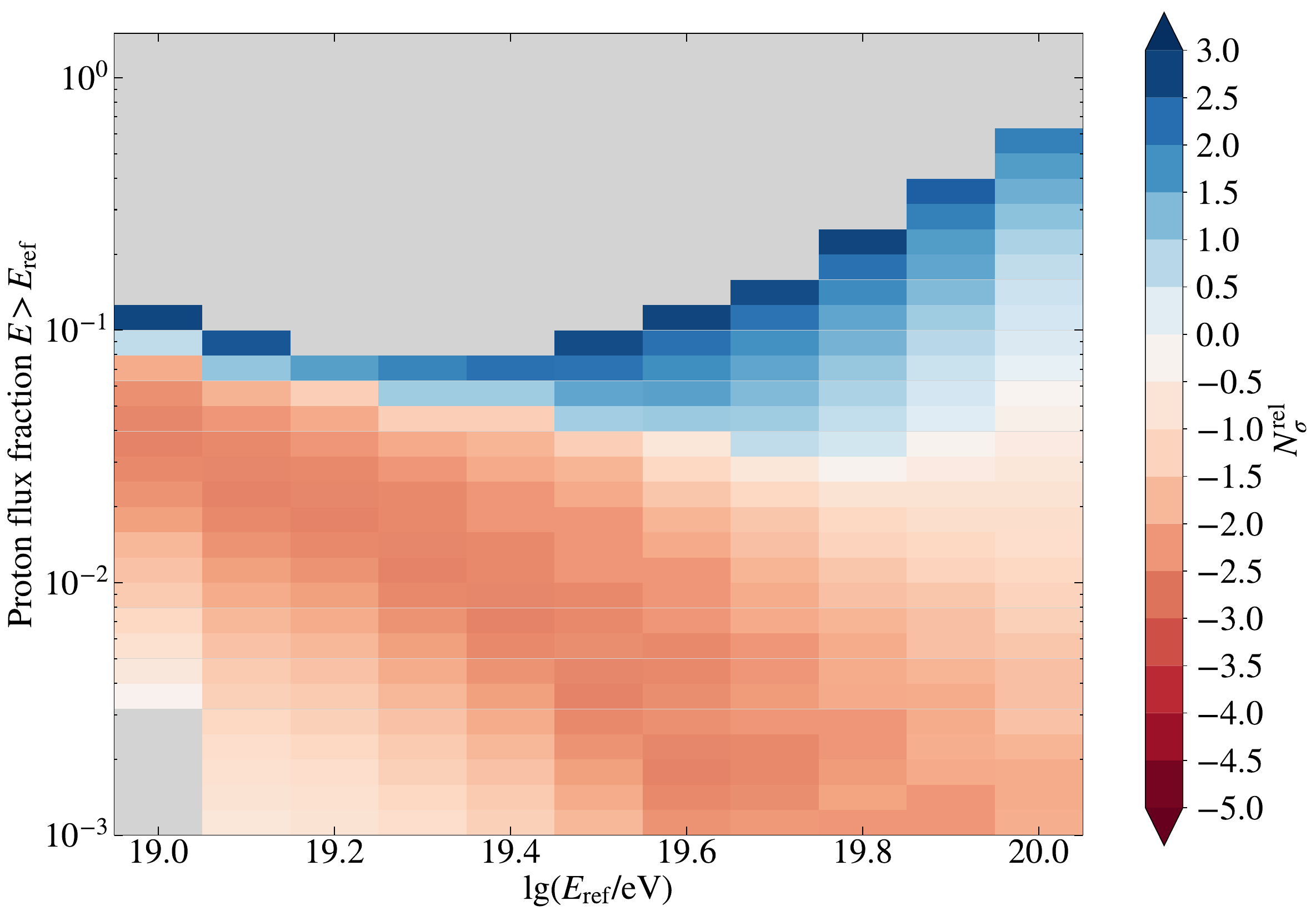}}
	\end{minipage}\\
	\begin{minipage}{0.49\linewidth}
		\centering
		\subfloat[]{\label{fig:obsfp_map_galmix_epos}\includegraphics[width=\textwidth]{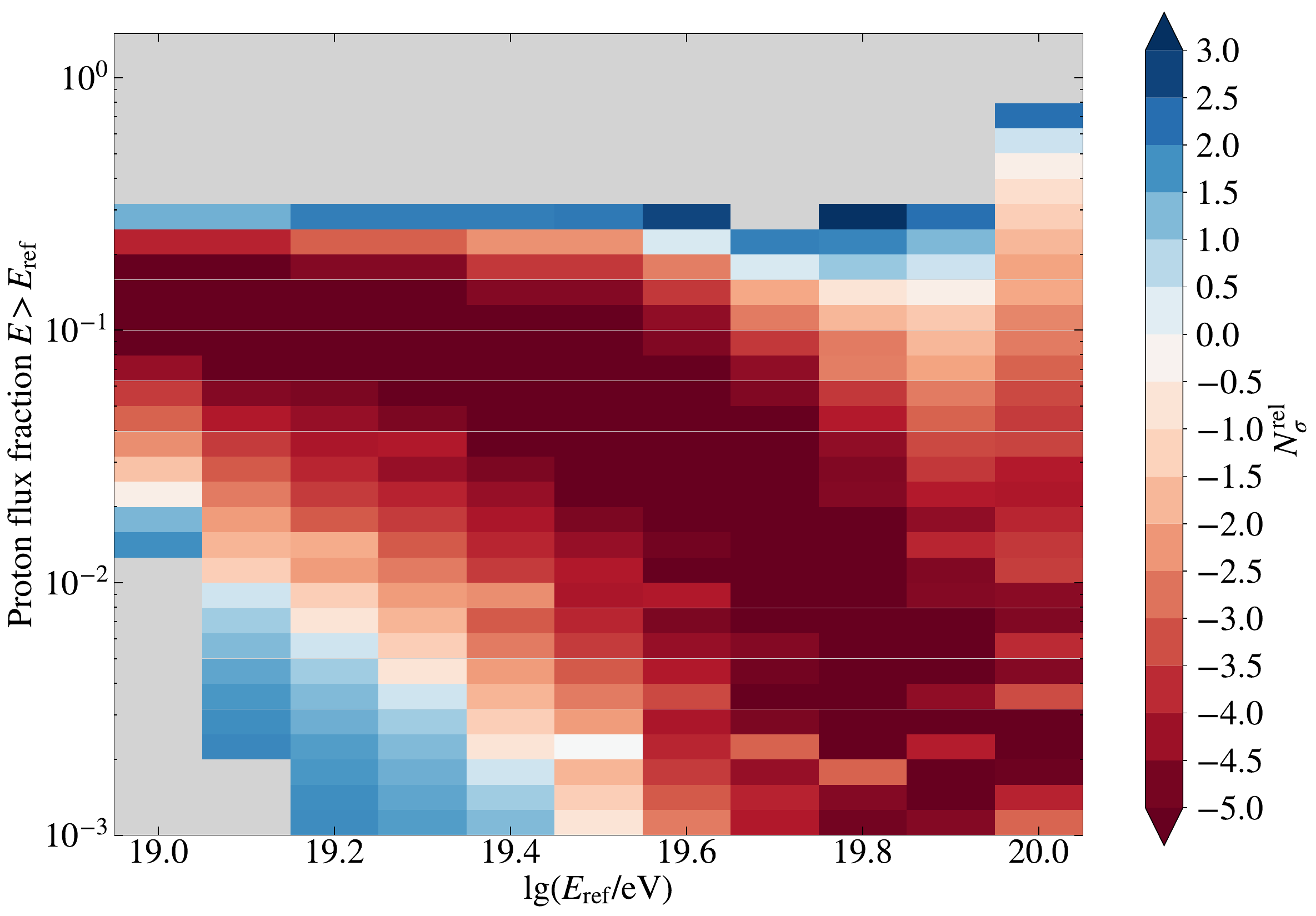}}
	\end{minipage}
	\begin{minipage}{0.49\linewidth}
		\centering
		\subfloat[]{\label{fig:obsfp_map_galmix_sibyll}\includegraphics[width=\textwidth]{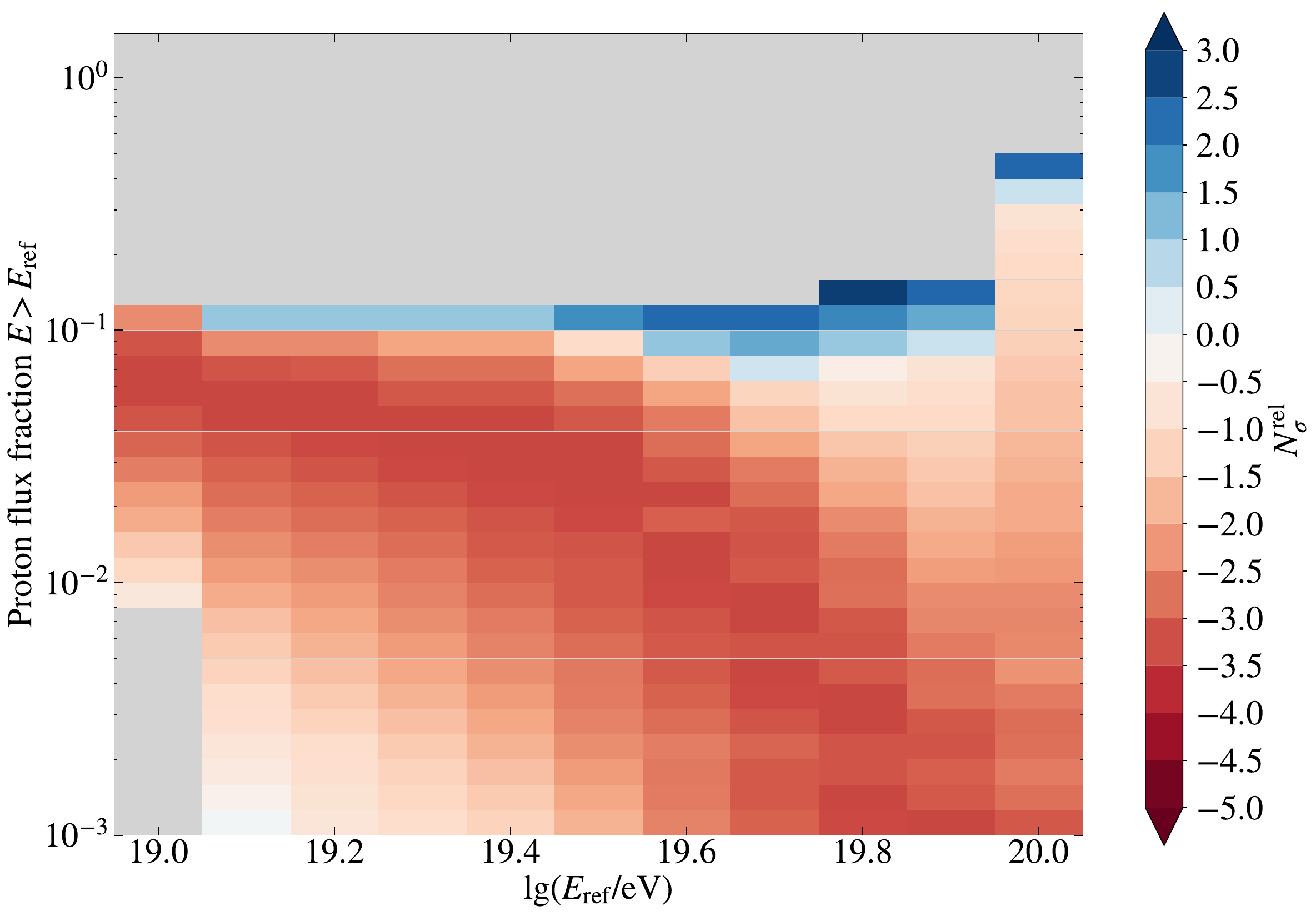}}
	\end{minipage}
	\caption{Maximal fit improvement (color scale) as a function of reference energy $E_\mathrm{ref}$ and observed proton fraction above that $E_\mathrm{ref}$ when a subdominant pure-proton component is included along with a (a) single-mass \textsc{EPOS-LHC}, (b) single-mass \textsc{Sibyll2.3c}, (c) galactic mix \textsc{EPOS-LHC}, and (d) galactic mix \textsc{Sibyll2.3c} UFA15 component. Color indicates the number of standard deviations from the best-fit model without a subdominant pure-proton component for the best-fit model allowed by current data and able to produce the specified observed proton fraction. Grey regions indicated observed proton flux fractions not realized by the models we considered. Only models consistent with current Auger data are plotted.}
	\label{fig:obsfp_maps}
\end{figure*}

In order to understand the observational implications of these fits with a secondary purely protonic component to UHECRs, we need to map from the model parameters $f_p$ and $E_\mathrm{max}^{\mathrm{UHE}p}$ to the observed proton (number) flux fraction above some specified reference energy $E_\mathrm{ref}$. This mapping is shown in Fig.~\ref{fig:obsfp_maps}, in which the color scale indicates $N_\sigma^\mathrm{rel}$ for the best-fit model able to produce that fraction. A more elaborate version of Fig.~\ref{fig:obsfp_maps}, detailing the dependence of the observed proton flux fraction on the source cutoff energy $E_\mathrm{max}^{\mathrm{UHE}p}$, is shown in Fig.~\ref{fig:obsfp_detailed_maps} of Appendix~\ref{app:supplfigs}. Figure~\ref{fig:obsfp_maps} shows there are viable scenarios which produce an observed CR flux above $50$ EeV which is at least $10\%$ protons in all the composition-HEG combinations considered. At the extreme, more the $35\%$ of the observed CR flux above $50$ EeV could be protons. If these protons exist and can be identified on an event-by-event basis, they will potentially allow for a new era of CR astronomy. The UHECR predictions of the models of each type having maximal protonic contributions are shown in Fig.~\ref{fig:UHEproton_maxobsp}. As can be seen, both spectrum and $X_\mathrm{max}$ are described very well, but the last data point on $\sigma(X_\mathrm{max})$ is much lower than the model predictions. We estimate the chance probability to observe a $\sigma(X_\mathrm{max})$ less than or equal to the observed value given the composition fractions of our prediction by drawing realizations of $\sigma(X_\mathrm{max})$ for the observed number of events ($N=62$) using the parametrization of $X_\mathrm{max}$ distributions from \cite{Arbeletche+19}. This yields a chance probability of $P=1.55\%$ for the single-mass \textsc{Sibyll2.3c} model and it can be concluded that our maximum proton model is only in mild tension with the currently available low-statistics measurements at UHE.

\par
While Fig.~\ref{fig:UHEproton_maps} shows that continued running of current neutrino experiments will not be able to constrain the remaining parameter space, but future neutrino detectors should be able to put strong constraints on $f_p$, as can be seen in Fig.~\ref{fig:UHEproton_secondaries}. Future mass-sensitive UHECR detectors, such as AugerPrime~\cite{Aab+16b} and POEMMA~\cite{Olinto+17}, should also be able to constrain $f_p$ considerably. 

\par
For this study we considered an additional pure-proton component but any scenario which lightens the observed composition at energies $\gtrsim 10$ EeV will enable an improvement in the fit quality to the Auger data relative to the UFA15 component alone. Given that a pure-proton scenario should be the most constrained, due to the larger EHE cosmic neutrino and gamma-ray fluxes pure-proton models produce compared to mixed composition models, our choice shows that there is broadly room for an additional light component above $10$ EeV.

\section{Summary} \label{sec:summary}

\par
In this paper we have examined the degree to which UHECR sources and their environments can be constrained by current CR and multimessenger data. The combination of CR composition and neutrino data are the most strongly constraining, once a good fit to the UHECR spectrum is demanded. We find that any realistic UHECR source model not excluded by CR or neutrino data is unconstrained by gamma-ray data.

\par
Current UHECR data mildly favors positive source evolutions (CR production increasing with redshift for $z\lesssim 1$), regardless of our parametrization of the ambient photon field inside the source or the hadronic event generator (HEG) used to interpret air shower data (see Fig.~\ref{fig:evochi2}). Current neutrino experiments do not strongly constrain the source evolution. For this purpose, exposures of at least a decade longer than are currently available are needed (see Fig.~\ref{fig:mz0_map}). Such exposures will be possible with future neutrino experiments. Constraints on the source evolution are strongly dependent on the HEG used to interpret air shower data, so reducing the particle physics uncertainties in air shower modeling will be necessary to fully exploit the power of future neutrino experiments to constrain UHECR source models.

\par
Current neutrino data is already constraining for the temperature of the ambient photon field inside the source (see Fig.~\ref{fig:maxTplt}). Ambient photon fields whose blackbody temperature is $T_\mathrm{BB} \geq 4000$ K are excluded, regardless of the source evolution assumed and the HEG used to interpret air shower data. For preferred source evolutions (SFR and those with $m\gtrsim 2$), sources are constrained to temperatures $T_\mathrm{BB} < 2000$ K. Complementarily, Auger data constrains the source to have temperature hotter than $T_\mathrm{BB} = 10$ K. These bounds correspond to the peak photon energy of a broken power-law photon field being constrained to be $1 \leq \varepsilon_0 \leq 500$ meV. 

\par
We also investigated the compatibility of pure-proton models with current CR and multimessenger data, setting aside CR composition data given the particle physics uncertainties that exist at UHEs. We find that, in fact, pure-proton models are compatible with both gamma-ray and neutrino data. Rather it is the shape of the resulting CR spectrum which rules out these models (see Fig.~\ref{fig:UFA_benchmark_CR}). 

\par
Finally, we considered two important refinements of the original UFA15 modeling, in addition to updating to the latest Auger data: allowing for the nearest source to be at some minimum distance and allowing for a subdominant pure-proton component at UHE. The best-fits for a simple UFA15 scenario are when a narrow mass range near Si is injected by the accelerator and the distance to the nearest source is small. However if the nearest source in the Auger field-of-view is $30-50$ Mpc away, a galactic mix of masses gives an acceptable fit, for either HEG (see Fig.~\ref{fig:UFA_mindist_CR}). 
\par
An additional UHECR component consisting of protons escaping the source with energies $\gtrsim 10$ EeV can strongly improve fits to Auger data. Allowing for a subdominant pure-proton component results in more than $5\sigma$ improvement in the quality of fit for some scenarios (see Fig.~\ref{fig:UHEproton_maps}). Such an additional component is largely unconstrained by gamma-ray and neutrino data and may produce more than $10\%$ of the observed CR flux above $50$ EeV. If high energy protons are present in the spectrum and tagged using event-by-event composition indicators, accessible thanks to the AugerPrime upgrade, a subset of high rigidity events can be identified. Their deflections in the galactic magnetic field would be smaller than those of the predominantly lower-rigidity main component, potentially permitting CR astronomy. The prominence of such a subdominant pure-proton component will be strongly constrained by future neutrino experiments like GRAND200k and RNO, as well as by future high-exposure mass-sensitive UHECR observatories such as AugerPrime and POEMMA. 

\acknowledgments

\par
We thank Luis Anchordoqui for many helpful suggestions and comments, Foteini Oikonomou for her considerable help resolving simulation inaccuracies, and Noemie Globus for reviewing our manuscript and providing valuable feedback. We also thank David Walz for his support regarding questions about \textsc{CRPropa3}, Michael Kachelrie\ss{} for his support regarding questions about \textsc{ELMAG}, and Sergio Petrera for providing an update of the model parameters~\cite{Abreu+13} for \textsc{Sibyll2.3c}. This work was supported in part through the NYU IT High Performance Computing resources, services, and staff expertise. MSM would like to thank the NYU Physics Department for support throughout this work. The research of GRF has been supported by NSF-PHY-1212538 and NSF-AST-1517319. MU would like to thank the hospitality of CCPP NYU, where part of this research was carried out.

\newpage

\bibliographystyle{apsrev4-1}
\bibliography{MUF19_PRDv2}

\appendix

\section{Calculation of EM Cascades} \label{app:EMcascades}

\par
In order to efficiently calculate the predicted flux of gamma-rays (and electron-positrons) over a wide variety of source models we have taken advantage of the following observation: for both CRs and EM cascades, the average observed flux on Earth from a given particle depends only on the primary particle's initial energy, distance from Earth, and particle type.\footnote{From a computational perspective, the average observed flux also depends on the extragalactic background light (EBL) model and cosmology chosen. For this purpose we have used the Gilmore12 EBL~\cite{Gilmore+11} and a flat FRW cosmology.} The observed EM spectrum also depends on the extragalactic magnetic field (EGMF) strength, but we will show that for the LAT energy band the observed EM cascade is insensitive to the parameter as long as the EGMF is within the range observationally allowed values. 

\begin{figure*}[!htpb]
	\centering
	\includegraphics[width=\linewidth]{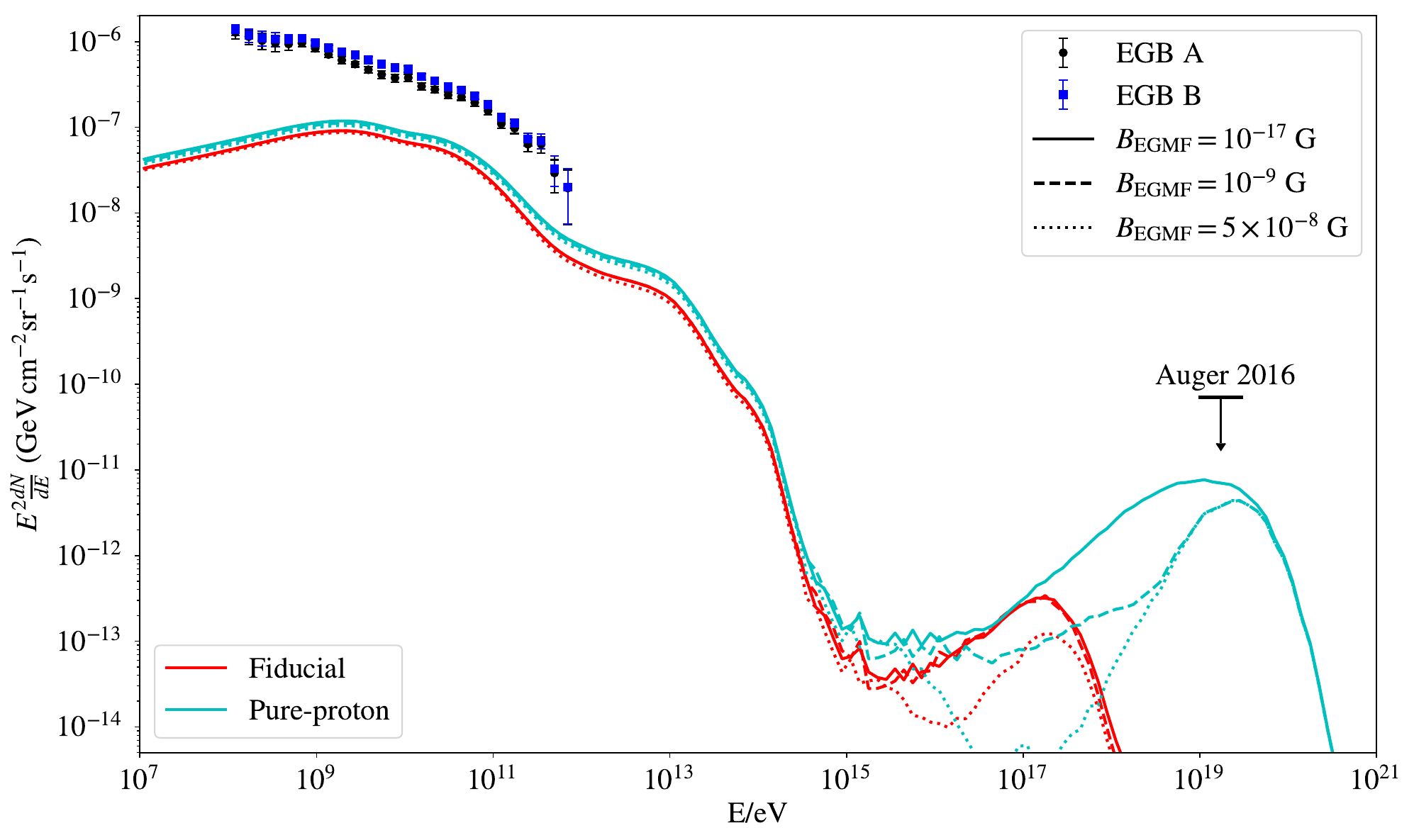}
	\caption{Effect of the extragalactic magnetic field on the predicted gamma-ray signal in representative UHECR source models, described in \S~\ref{sec:benchmarks}. For reference, the LAT EGB models A and B~\cite{Ackermann+14} are plotted. The Auger upper-bound on UHE gamma-rays~\cite{Aab+16c} is plotted in black.}
	\label{fig:EGMFsensitivity}
\end{figure*}

\par 
Having this in mind, we tabulate the average observed LAT gamma-ray flux given a UHECR of any species produced with any given energy and distance. Since the EM cascades we are considering are seeded by UHECR propagation, the energy and distance range relevant for EM cascades is driven by the range of energies and distances at which EM cascades are seeded by UHECRs and observed by LAT. We consider UHECRs with energies between $10^{17}$ eV and $10^{21}$ eV, initial comoving distances of $0-7.96$ Gpc ($z\lesssim 5$) from Earth, and mass numbers $A=1-56$. Pions produced by UHECR interactions also seed EM cascades and so we consider pions with energies between $10^{14}$ eV and $10^{21}$ eV and the same initial comoving distances as for UHECRs. LAT observes gamma-rays in the $100$ MeV to $820$ GeV energy range so the lower end of the energy range which needs to be tabulated is set by this band. Therefore, the parameter space of interest is EM cascades seeded by either electrons/positrons (hereafter, electrons) or photons with energies in the $10^{8}-10^{21}$ eV range initiated at $0-7.96$ Gpc in comoving distance from Earth. For UHECRs and EM cascades we tabulate the observed fluxes in bins $20$ Mpc wide in comoving distance and a tenth of a decade wide in energy. These bin sizes were chosen in order to ensure results were insensitive to details of the injection within a bin. 

\par
For EM cascades initiated by particles with energies less than $10^{12}$ eV, we simulate the cascades directly with \textsc{ELMAG}~\cite{Kachelriess+11}. However, for EM cascades more energetic than $10^{12}$ eV direct Monte Carlo simulation of the EM cascade becomes extremely computationally expensive. To overcome this limitation we perform abridged simulations of cascades above $10^{12}$ eV in energy, building higher energy-distance bin EM cascades out of pre-calculated EM cascades in lower energy-distance bins. Namely, we simulate the EM cascade with \textsc{ELMAG}, stopping the simulation of particles which have moved to a lower energy or distance bin than the bin in which the cascade was initiated. Once all particles' simulation has been stopped their energy-distance bin distribution is calculated. The observed flux can then be obtained by summing over the fluxes of the lower energy-distance bins weighted by the number of particles whose simulation terminated in that bin. Therefore, observed flux $f_{ij}^{\nu\mu}(E)$ of particles of type $\mu$ ($\gamma$ or $e^\pm$) from a cascade initiated in the $ij$th energy-distance bin by a particle of type $\nu$ is given by

\begin{align} \label{eq:EM2EM}
	f_{ij}^{\nu\mu}(E) = \displaystyle \sum_{m,n,\rho} N_{ijmn}^{\nu\rho} f_{mn}^{\rho\mu}(E) ~~,
\end{align}

\noindent
where $N_{ijmn}^{\mu\rho}$ is the number of particles of type $\rho$ produced in the $mn$th energy-distance bin due to an EM cascade initiated in the $ij$th energy-distance bin by a particle of type $\nu$ and $f_{mn}^{\rho\mu}(E)$ is the observed flux of particles of type $\mu$ due to an EM cascade initiated in the $mn$th energy-distance bin by a particle of type $\rho$. Following this procedure, the fluxes due to EM cascades at high energies can be much more efficiently calculated by building them up from those already calculated at lower energies. 

\par
Next, we would like to know the EM signal due a UHECR propagating from the $ij$th energy-distance bin. In order to calculate this, we simulate UHECR propagation from the $ij$th energy-distance bin using \textsc{CRPropa3}~\cite{Batista+16} and we tabulate the distribution of EM secondaries produced in propagation. These EM secondaries seed the EM cascades and so the total EM signal $f_{ij}^{A\mu}(E)$ produced by a CR of mass number $A$ propagating from the $ij$th energy-distance bin is given by

\begin{align} \label{eq:UHECR2EM}
	f_{ij}^{A\mu}(E) = \displaystyle \sum_{m,n,\nu} N^{A\nu}_{ijmn} f_{mn}^{\nu\mu}(E) ~~,
\end{align}

\noindent
where $N_{ijmn}^{A\nu}$ is the number of secondary particles of type $\nu$ produced in the $mn$th energy-distance bin by a CR of mass number $A$ propagating from the $ij$th energy-distance bin and $f_{mn}^{\nu\mu}(E)$ is the observed flux of particles of type $\mu$ due to an EM cascade initiated by a particle of type $\nu$ in the $mn$th energy-distance bin.

\par
In the case of EM cascades seeded by pions, we can use~\eqref{eq:UHECR2EM} to calculate the observed flux $f_{ijmn}^{\pi\nu}(E)$ by simply replacing $N^{A\nu}_{ijmn}$ in the sum by either $N^{\pi^0\nu}_{ijmn}$ or $N^{\pi^\pm\nu}_{ijmn}$ where

\begin{align}
	N^{\pi^0\nu}_{ijmn} &= 2\delta_{\gamma,\nu}\,, \quad \text{where $m$ is such that $E_m \leq \frac{1}{2} E_i < E_{m+1}$,} \\
	N^{\pi^\pm\nu}_{ijmn} &= \delta_{e^\pm,\nu}\,, \quad \text{where $m$ is such that $E_m \leq \frac{1}{4} E_i < E_{m+1}$.} 
\end{align}

\noindent
This approximation is well-motivated since even the longest lived pions decay promptly (within $10^{-2}$ pc) via nearly a single decay channel.

\par
To test the dependence of these results on the value of the EGMF strength assumed we calculated the EM cascades for three benchmark (RMS) values of the EGMF strength: $10^{-17}$ G, $10^{-9}$ G, and $5\times 10^{-8}$ G. These values span the range of observationally allowed values~\cite{Durrer+13}. Figure~\ref{fig:EGMFsensitivity} shows the gamma-ray flux in two different UHECR source models, the fiducial and pure-proton models from \S~\ref{sec:benchmarks}, plotted for each of these EGMF strengths. As can be seen, while there are significant differences at UHEs, due to the increasing importance of synchrotron losses as the EGMF strength increases, the flux in the LAT energy band is very insensitive to this parameter.

\section{Supplementary Figures} \label{app:supplfigs}

\subsection{Neutrino Constraints on Source Evolution}

\par
\begin{figure*}
	\centering
	\begin{minipage}{0.49\linewidth}
		\centering
		\subfloat[]{\label{fig:mz0_map_epos}\includegraphics[width=\textwidth]{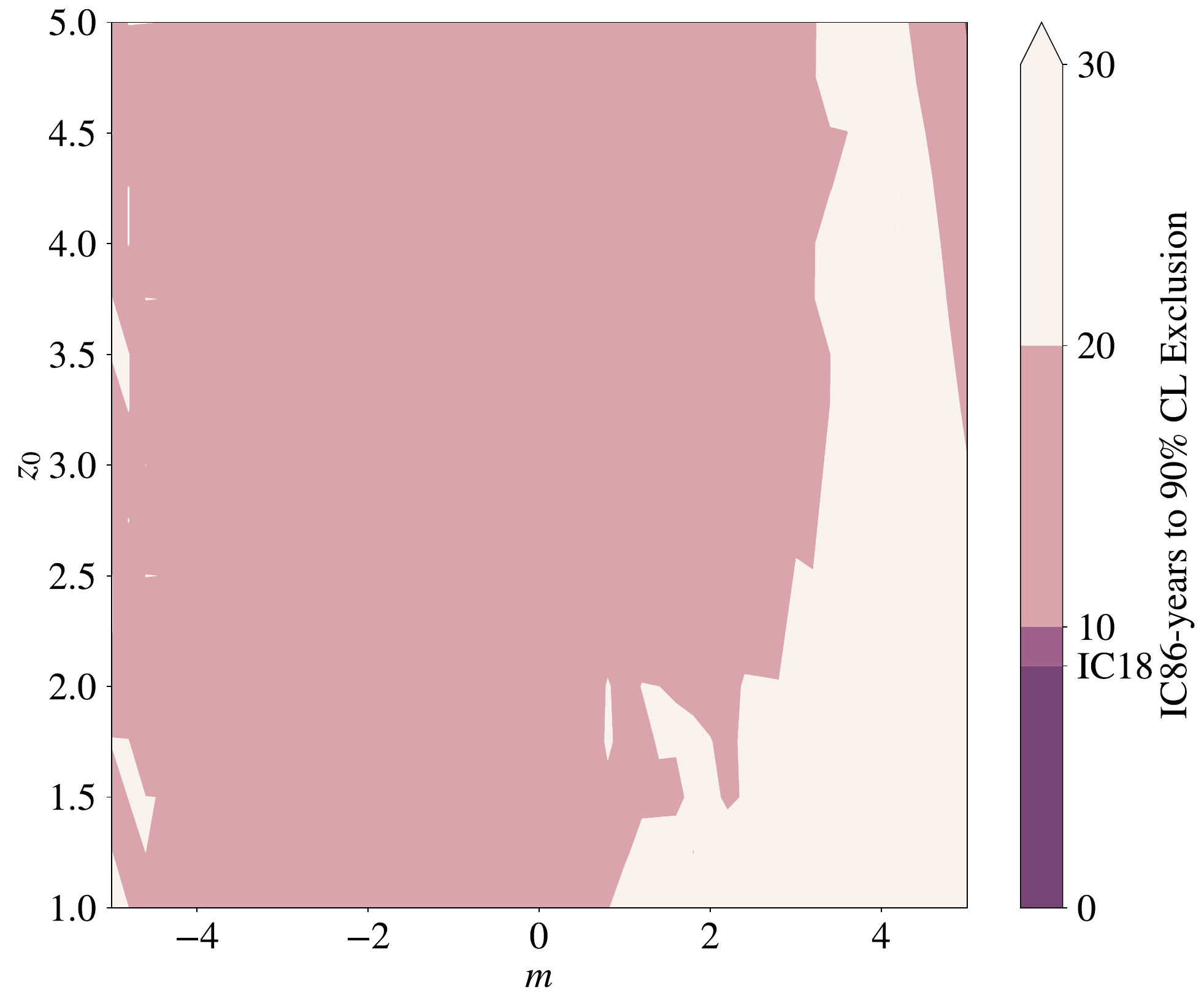}}
	\end{minipage}
	\begin{minipage}{0.49\linewidth}
		\centering
		\subfloat[]{\label{fig:mz0_map_sibyll}\includegraphics[width=\textwidth]{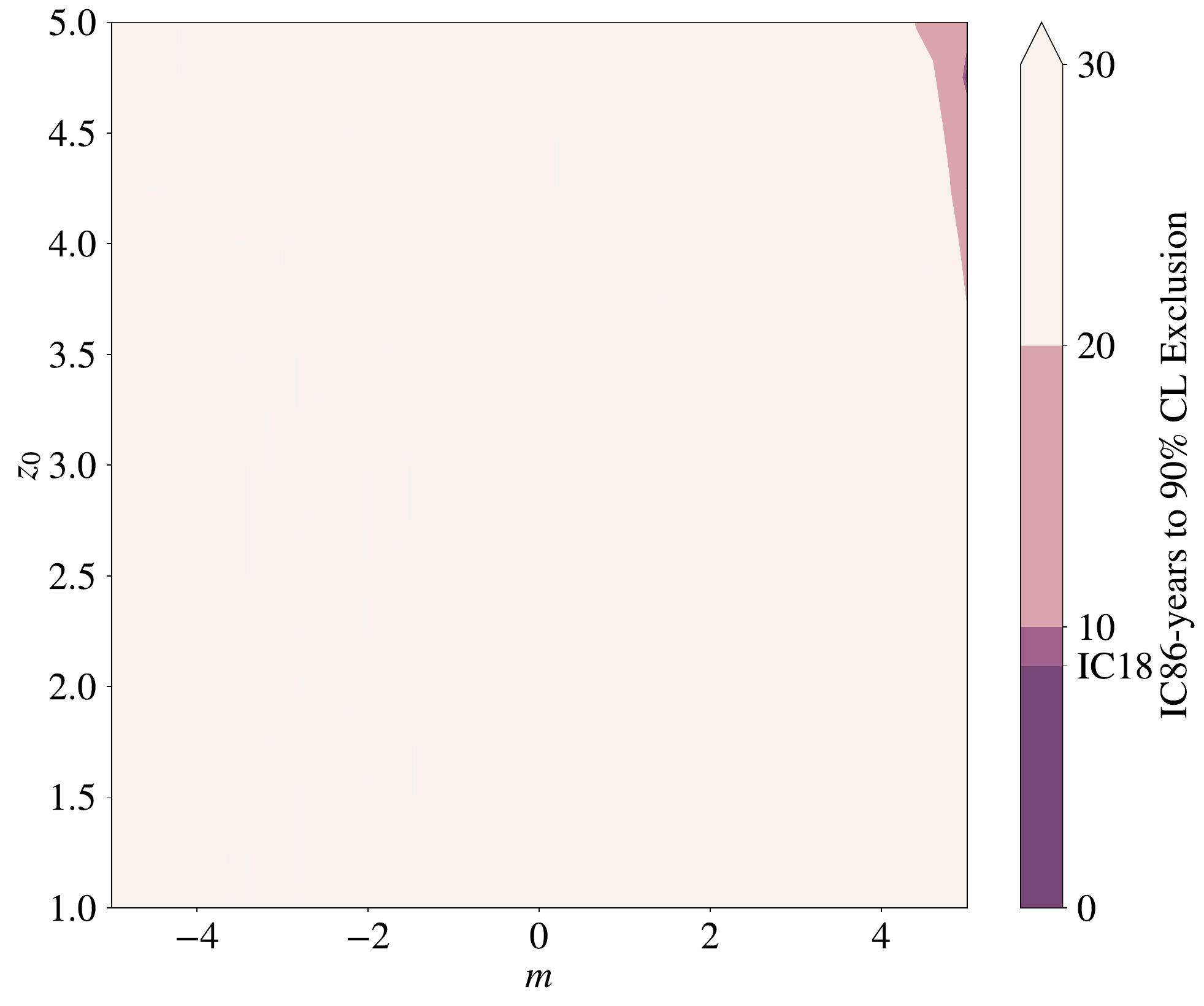}}
	\end{minipage}
	\begin{minipage}{0.49\linewidth}
		\centering
		\subfloat[]{\label{fig:mz0_map_nu}\includegraphics[width=\textwidth]{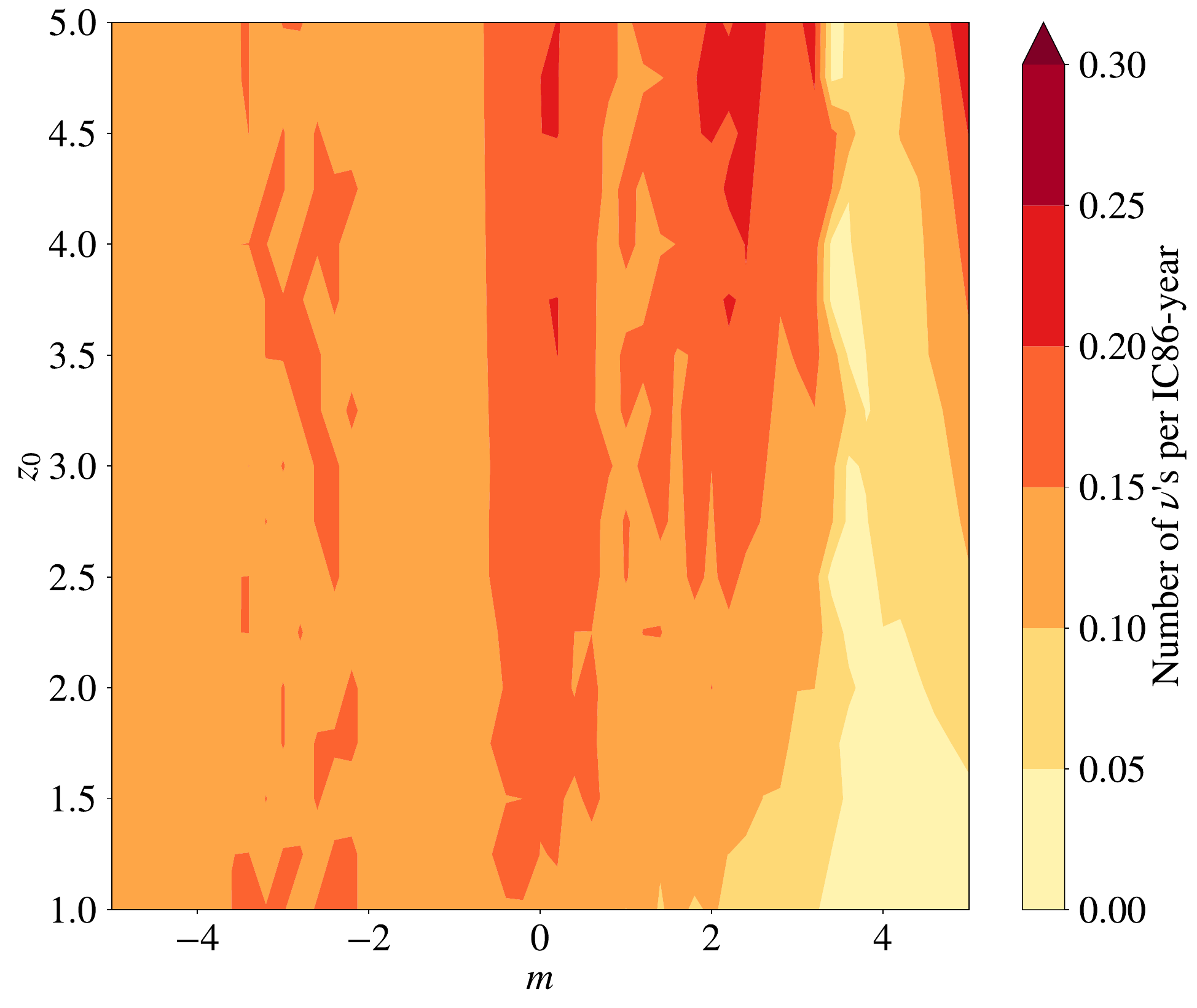}}
	\end{minipage}
	\begin{minipage}{0.49\linewidth}
		\centering
		\subfloat[]{\label{fig:mz0_map_nu_sibyll}\includegraphics[width=\textwidth]{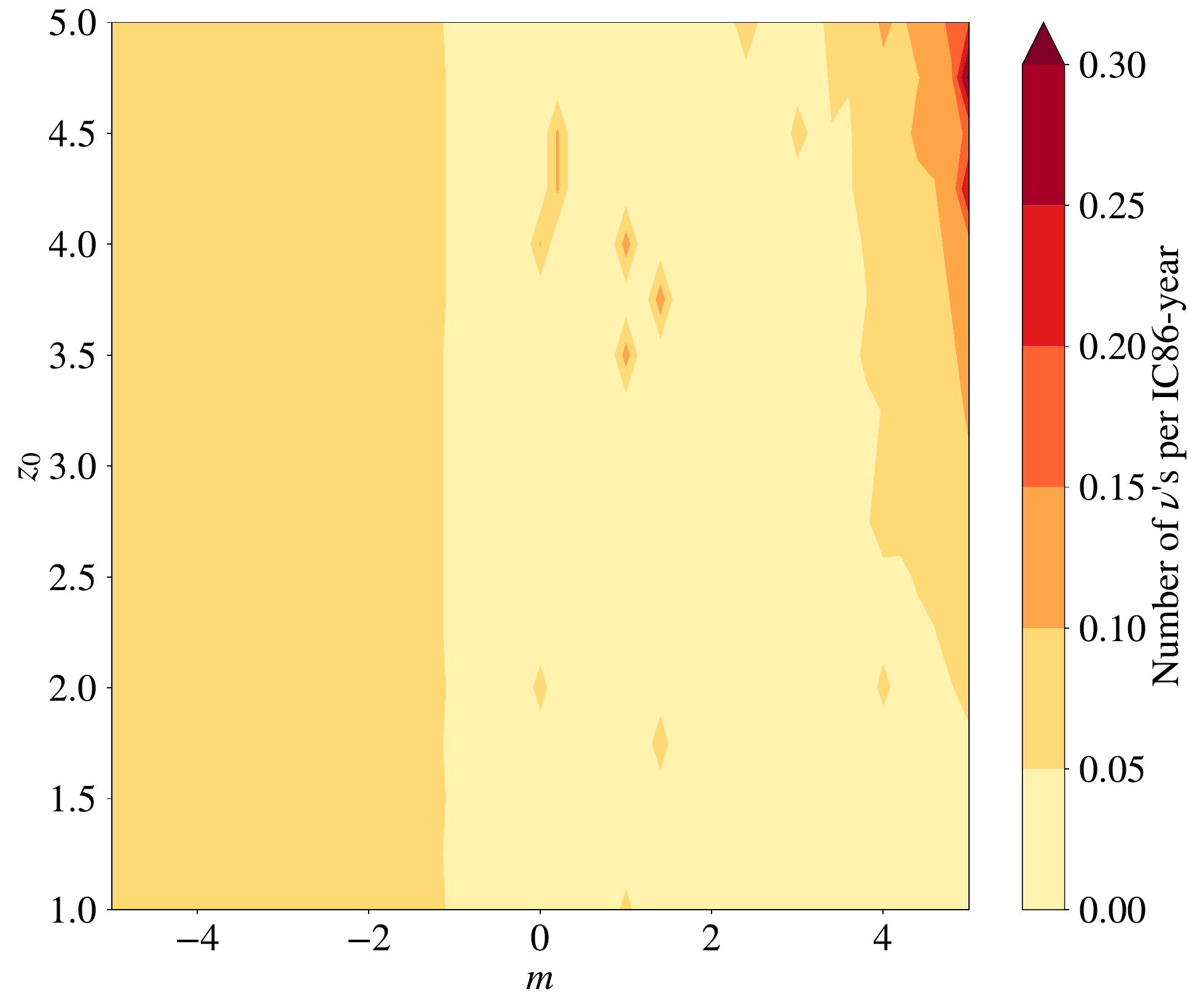}}
	\end{minipage}
	\caption{Neutrino constraints on source evolution. \textbf{Top:} Number of IC86-years to $90\%$ CL exclusion for various source evolutions. \textbf{Bottom:} Number of EHE cosmic neutrinos predicted to be observed per IC86-year for various source evolutions. All models inject a single-mass. Panels (a) and (c) use \textsc{EPOS-LHC} while panels (b) and (d) use \textsc{Sibyll2.3c} to interpret air shower data. Models were obtained by minimizing the neutrino flux by reducing the ambient photon field temperature in the source, while remaining within $3\sigma$ of the best-fit to UHECR data. The exposure corresponding to the latest IceCube bound~\cite{Aartsen+18} is labelled as ``IC18.'' Note the discontinuity the number of neutrinos as a function of $m$ in (d) at $m\simeq -1.5$ is due to the best-fit source temperature dropping suddenly as $m$ increases.}
	\label{fig:mz0_map}
\end{figure*}

Figure~\ref{fig:mz0_map} shows the number of IC86-years to $90\%$ CL exclusion for single-mass models throughout $m-z_0$ parameter space for the two considered HEGs. The models presented in Fig.~\ref{fig:mz0_map} were obtained by minimizing the predicted neutrino flux by lowering the temperature of the ambient photon field in the source while remaining within $3\sigma$ of the best-fit to the CR spectrum and composition. 

\par
Figure~\ref{fig:mz0_map} shows that the constraining power of neutrino data is strongly sensitive to the HEG used to interpret the Auger data, though neutrinos are not currently constraining in either HEG. Adopting \textsc{EPOS-LHC} leads to a model that produces many more neutrinos than if one adopts \textsc{Sibyll2.3c}. This is because \textsc{Sibyll2.3c} infers a heavier composition on Earth from the observed $X_\mathrm{max}$ data, allowing for relatively cooler source environments to describe the Auger data. 

\subsection{Spectral Dependence on Nearest-Source Distance in Single-Mass Scenarios} 

\par
\begin{figure*}
	\centering
	\includegraphics[width=\linewidth]{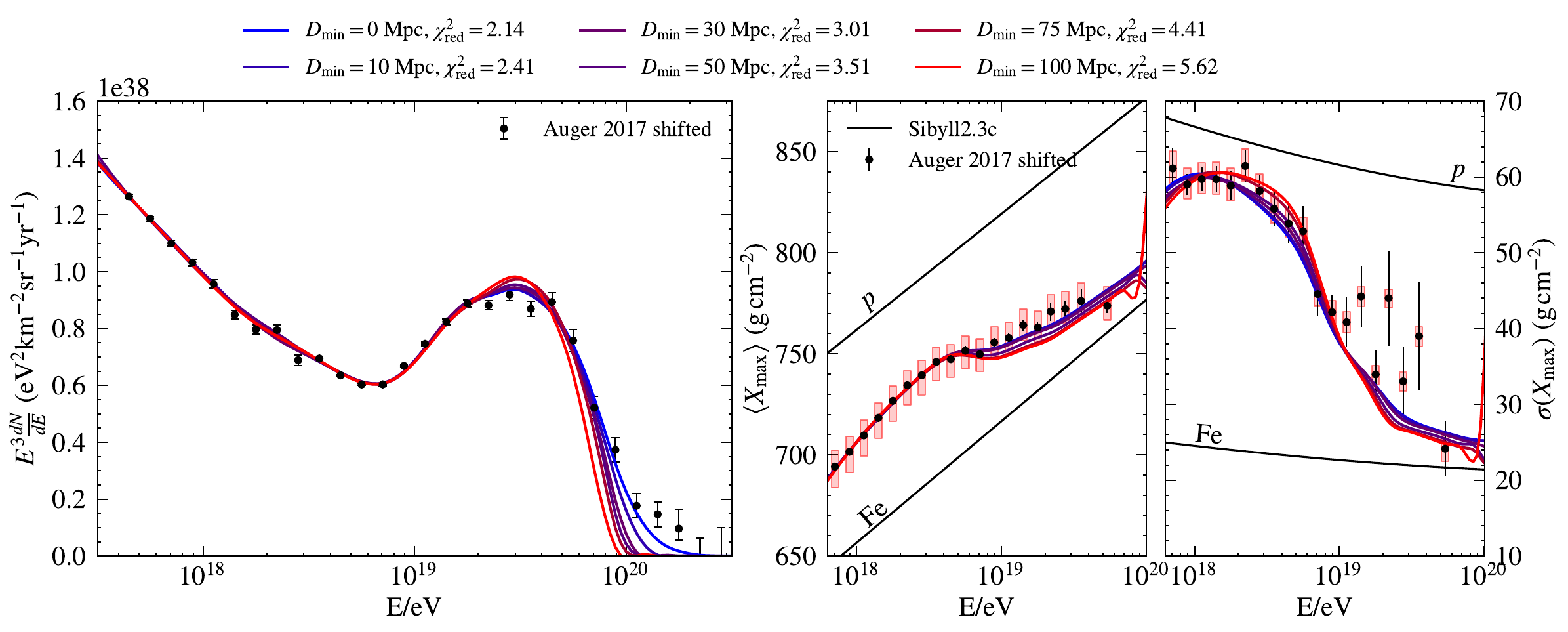}
	\caption{Impact of the nearest source distance $D_\mathrm{min}$ on the predicted CR spectrum (left) and composition (right) at Earth in single-mass models. Data points are the same as in Fig.~\ref{fig:UFA_evo_summary}.}
	\label{fig:UFA_mindist_singlemass}
\end{figure*}

Figure~\ref{fig:UFA_mindist_singlemass} shows the effect of a nonzero nearest source distance in single-mass models. As can be seen, these models generally prefer nearest source distances $\lesssim 10$ Mpc from Earth (though this is strongly sensitive to the choice of systematic shifts in the data) and prefer a heavier composition as $D_\mathrm{min}$ increases.

\subsection{Observed Proton Fraction with a Subdominant Pure-Proton Component}

\par
\begin{figure*}[!htpb]
	\centering
	\begin{minipage}{0.49\linewidth}
		\centering
		\subfloat[]{\label{fig:obsfp_detailed_map_epos}\includegraphics[width=\textwidth]{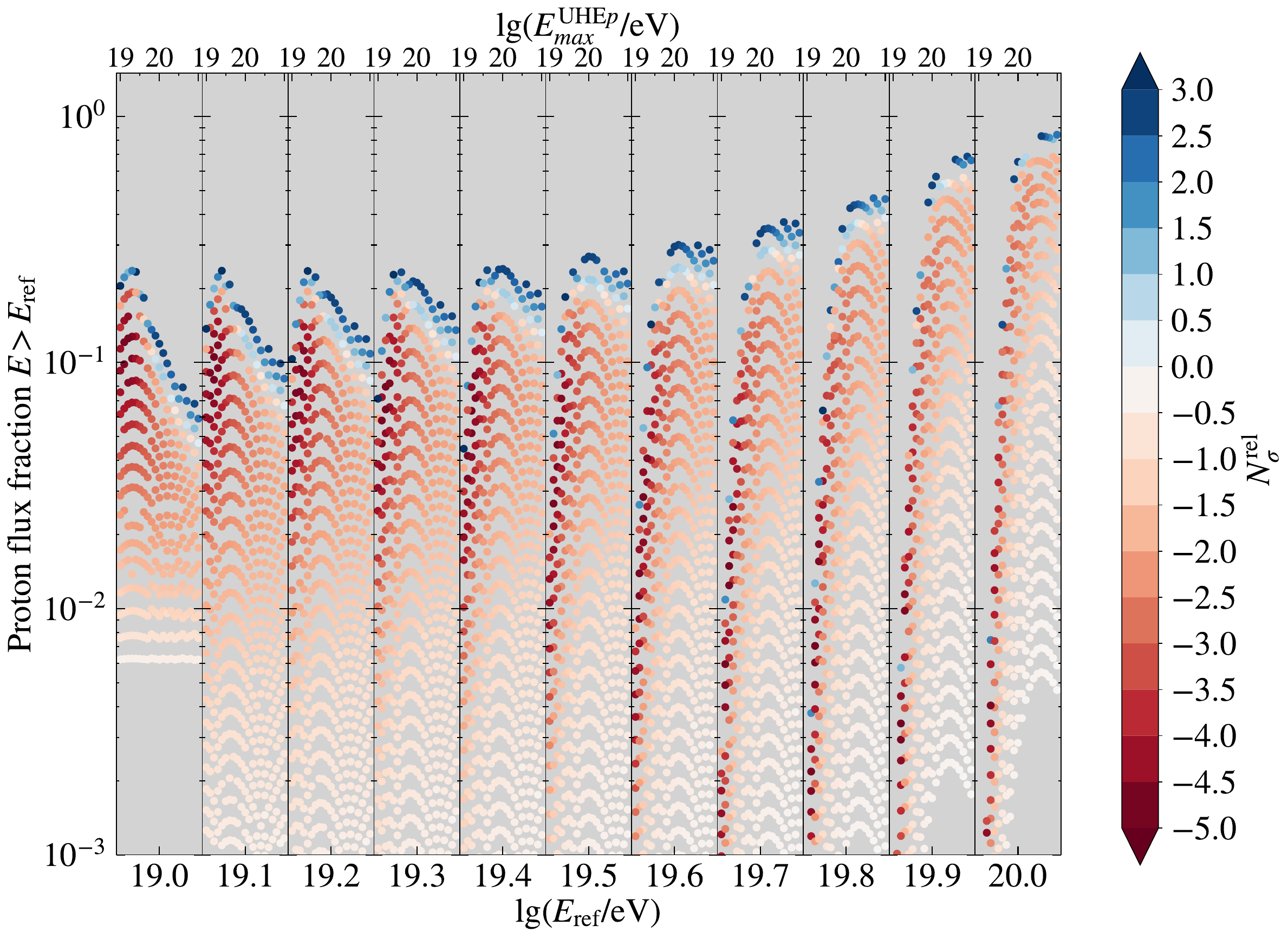}}
	\end{minipage}
	\begin{minipage}{0.49\linewidth}
		\centering
		\subfloat[]{\label{fig:obsfp_detailed_map_sibyll}\includegraphics[width=\textwidth]{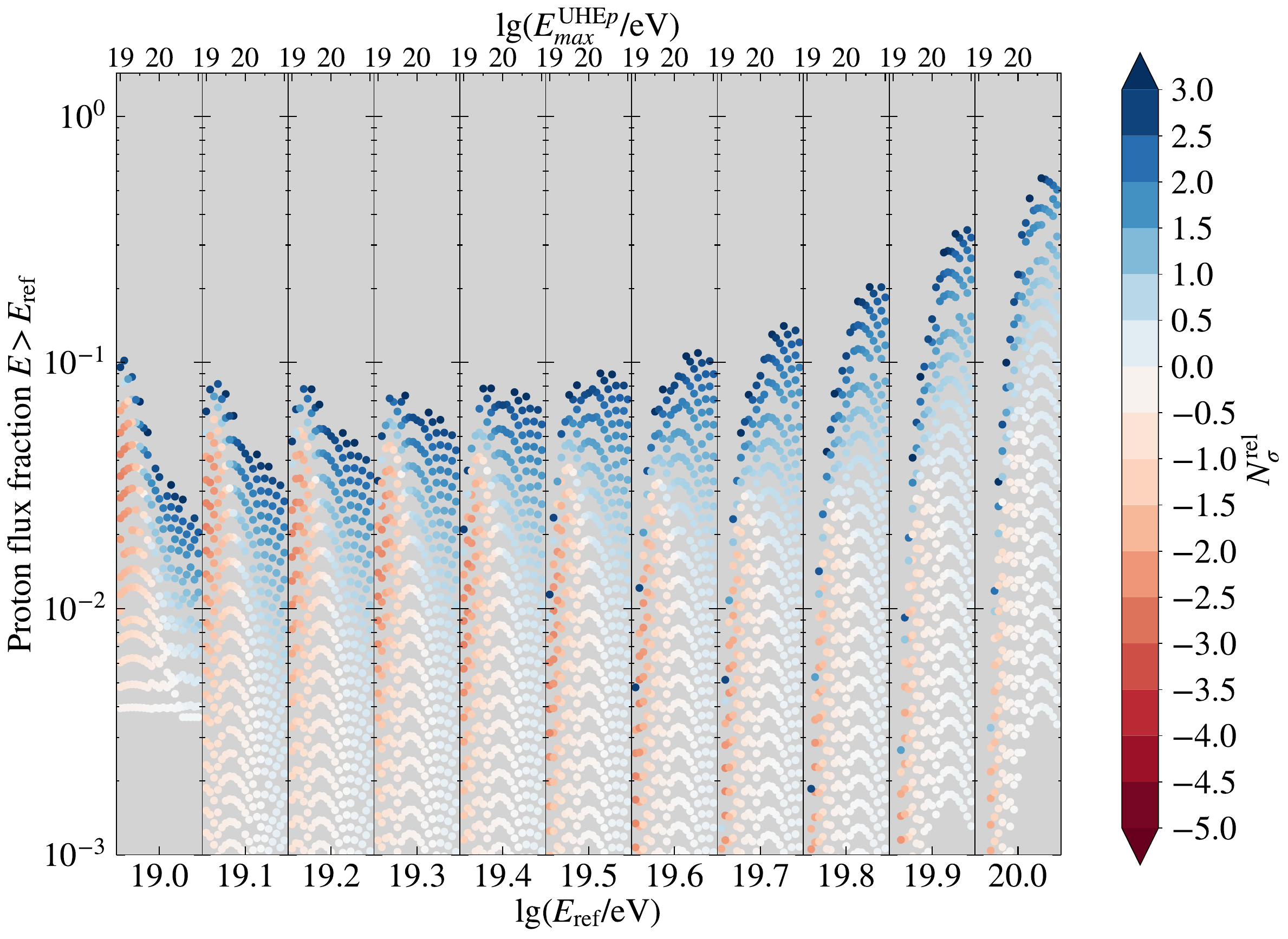}}
	\end{minipage}
	\begin{minipage}{0.49\linewidth}
		\centering
		\subfloat[]{\label{fig:obsfp_detailed_map_galmix_epos}\includegraphics[width=\textwidth]{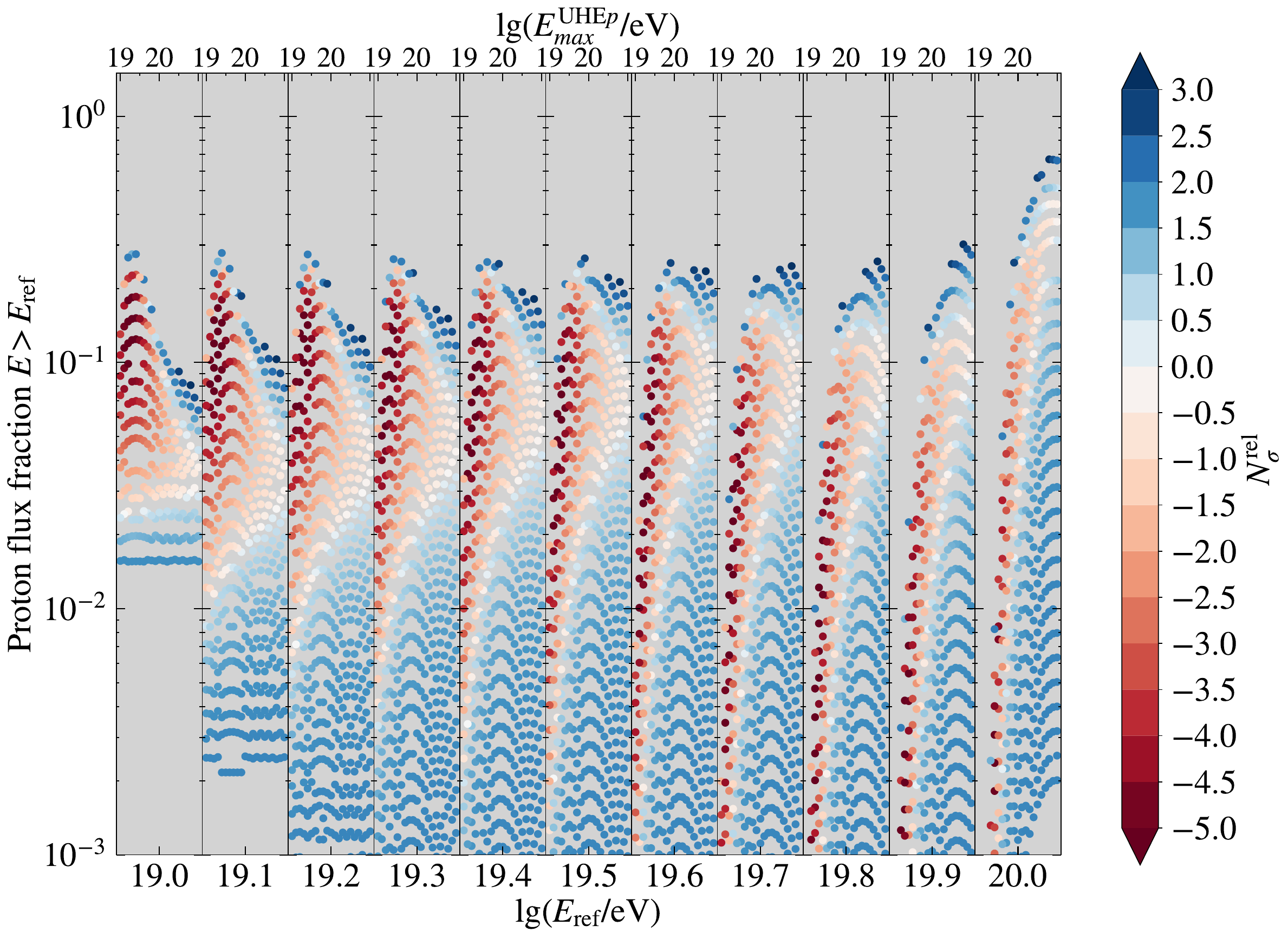}}
	\end{minipage}
	\begin{minipage}{0.49\linewidth}
		\centering
		\subfloat[]{\label{fig:obsfp_detailed_map_galmix_sibyll}\includegraphics[width=\textwidth]{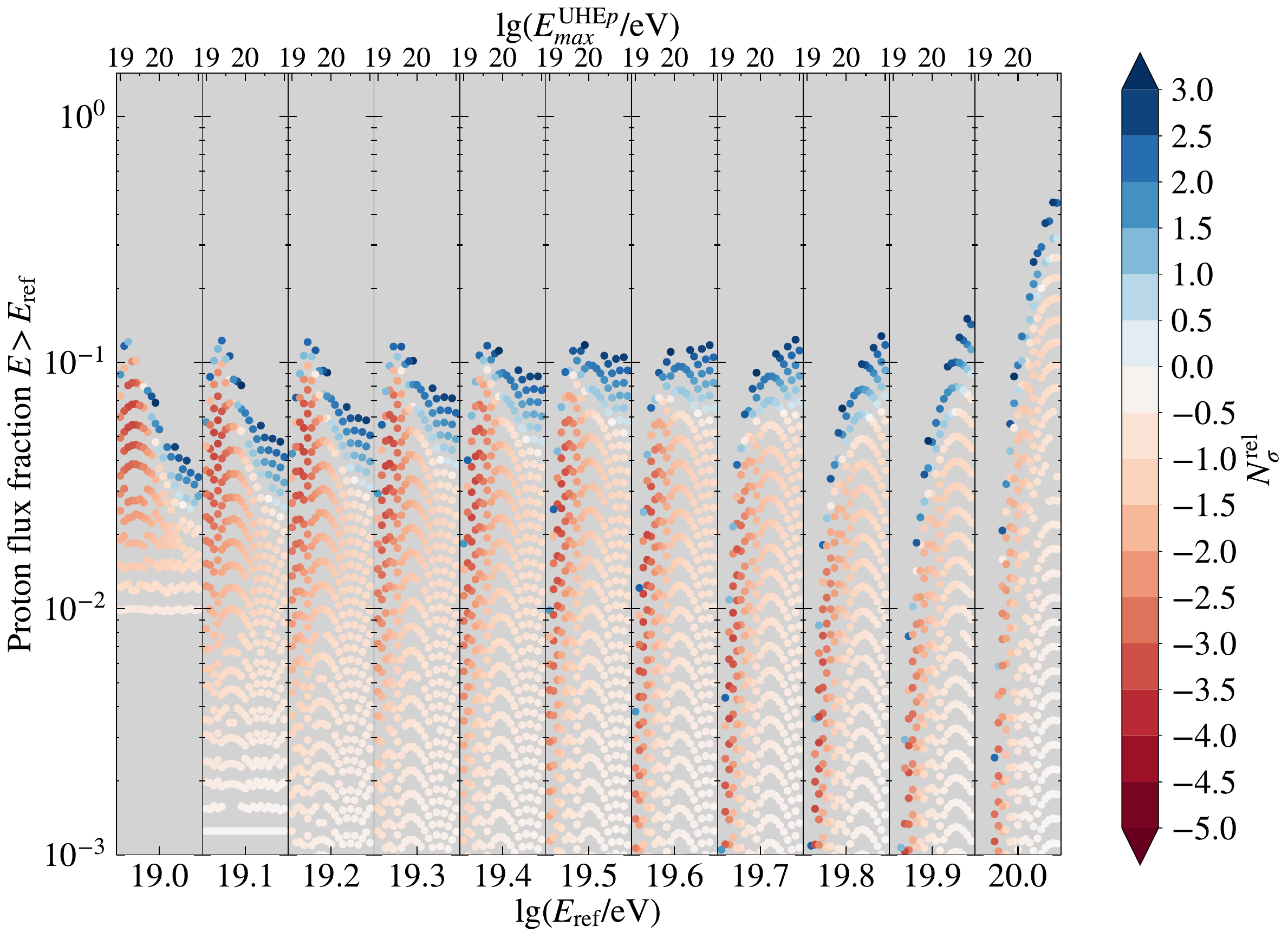}}
	\end{minipage}
	\caption{Observed proton fraction above the reference energy $E_\mathrm{ref}$ with a subdominant pure-proton component whose cutoff energy is $E^{\mathrm{UHE}p}_{max}$. Each column corresponds to a given lg$E_\mathrm{ref}$, as indicated by the label on the lower x-axis; within each column lg$E^{\mathrm{UHE}p}_{max}$ runs from 19-21 as indicated by the upper x-axis. The four panels are for a subdominant pure-proton component in addition to (a) single-mass \textsc{EPOS-LHC}, (b) single-mass \textsc{Sibyll2.3c}, (c) galactic mix \textsc{EPOS-LHC}, and (d) galactic mix \textsc{Sibyll2.3c} models. Color indicates the number of standard deviations the best-fitting UFA15 model with the additional subdominant pure-proton component is from the corresponding best-fit UFA15 model without this component. Grey indicates regions such that the specified proton flux fraction cannot be realized by the models we considered. Only models consistent with current Auger data are plotted.}
	\label{fig:obsfp_detailed_maps}
\end{figure*}

Figure~\ref{fig:obsfp_detailed_maps} shows the observed proton flux fraction above a given reference energy as a function of the cutoff energy of the subdominant pure-proton component. Considerable observed proton fractions in models consistent with current Auger data are possible, regardless of the cutoff energy of the additional component. 

\subsection{Details of Subdominant Pure-Proton Component Fit Improvement}

\begin{figure*}[!htpb]
	\centering
	\begin{minipage}{0.49\linewidth}
		\centering
		\subfloat[]{\label{fig:deltachi2spec_map_epos}\includegraphics[width=\textwidth]{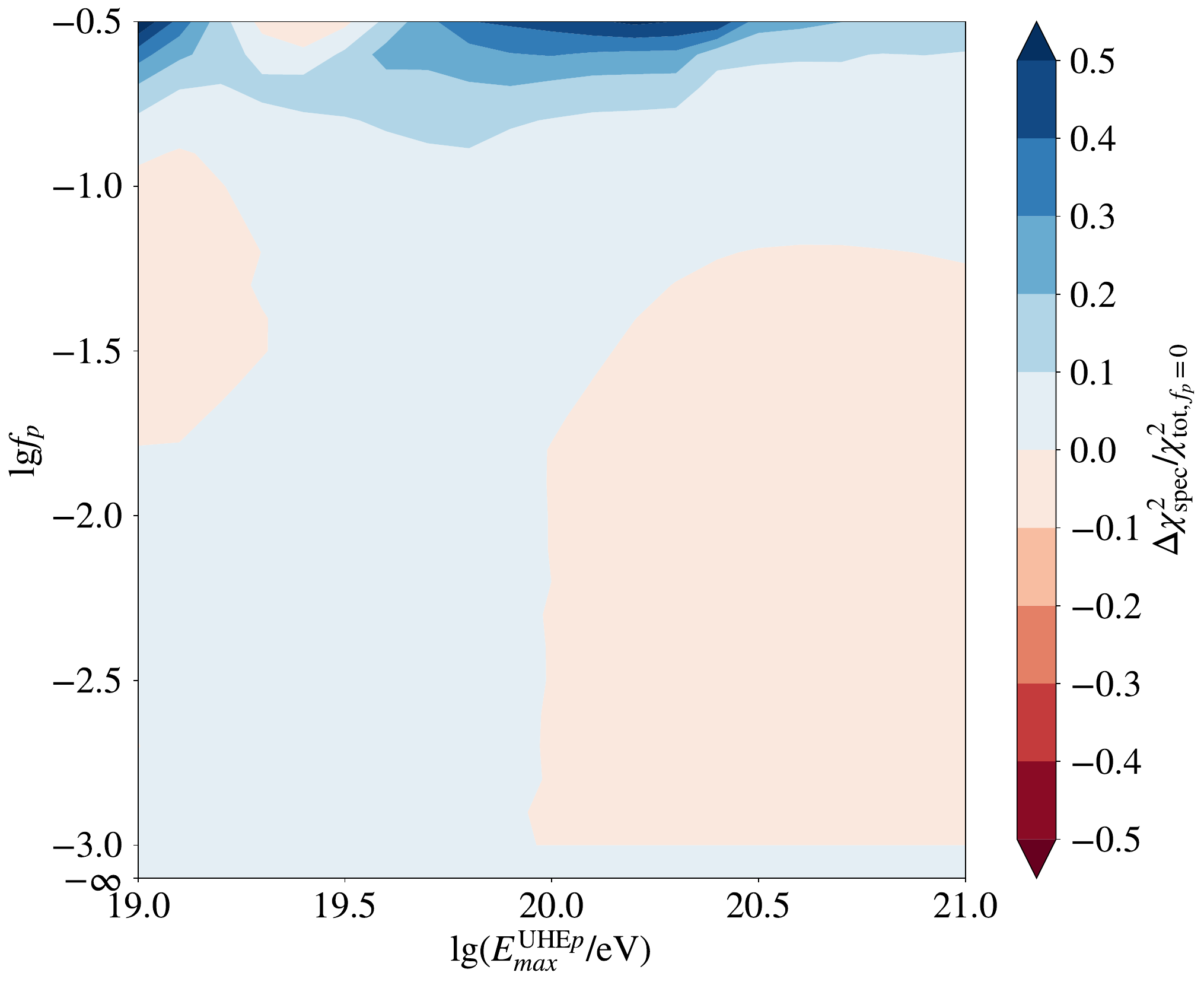}}
	\end{minipage}
	\begin{minipage}{0.49\linewidth}
		\centering
		\subfloat[]{\label{fig:deltachi2spec_map_sibyll}\includegraphics[width=\textwidth]{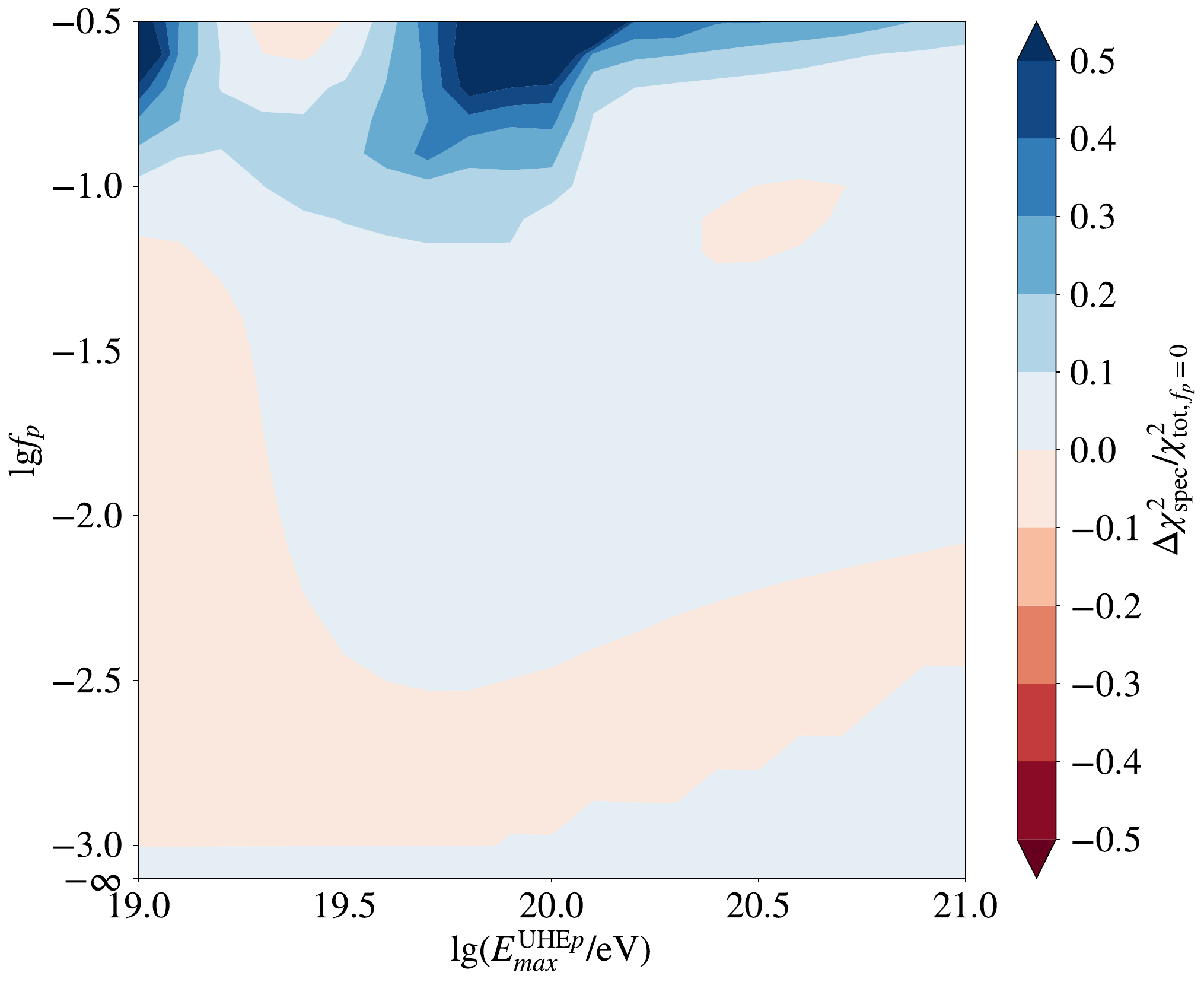}}
	\end{minipage}
	\begin{minipage}{0.49\linewidth}
		\centering
		\subfloat[]{\label{fig:deltachi2spec_map_galmix_epos}\includegraphics[width=\textwidth]{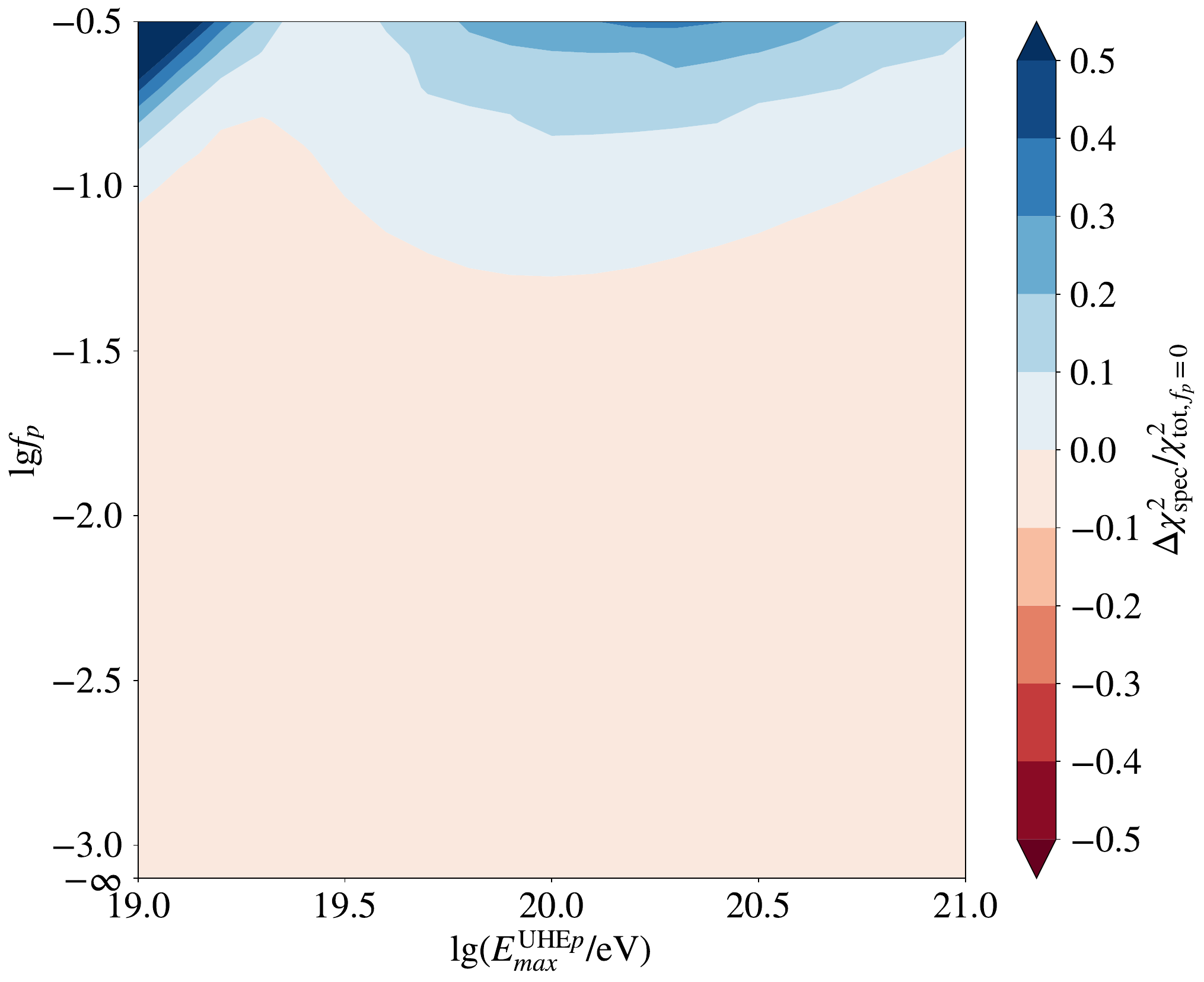}}
	\end{minipage}
	\begin{minipage}{0.49\linewidth}
		\centering
		\subfloat[]{\label{fig:deltachi2spec_map_galmix_sibyll}\includegraphics[width=\textwidth]{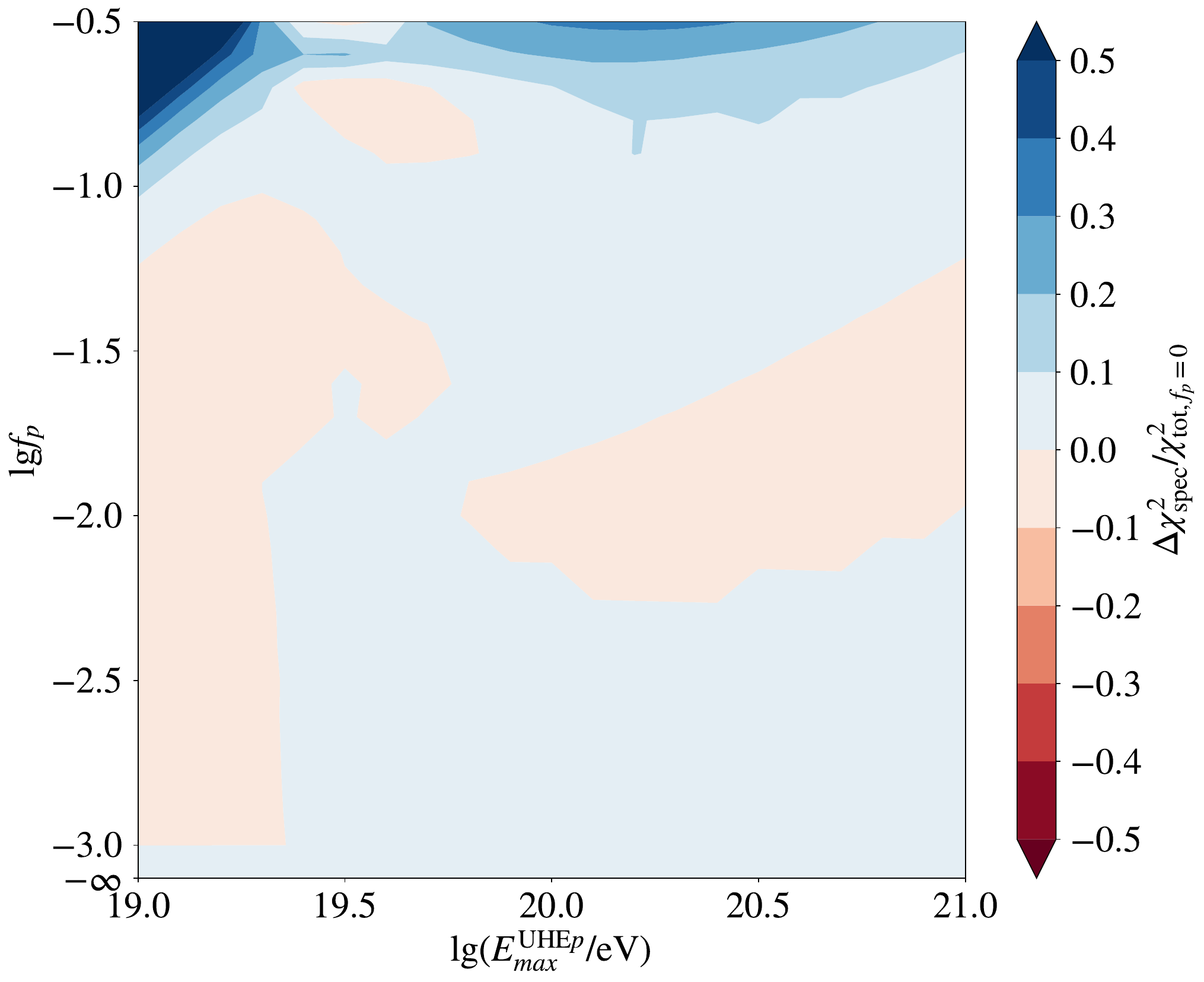}}
	\end{minipage}
	\caption{Effect of an additional pure-proton component on the spectrum contribution to the total $\chi^2$ relative to the best-fit in a single-component model. Additional pure-proton component with (a) single-mass \textsc{EPOS-LHC}, (b) single-mass \textsc{Sibyll2.3c}, (c) galactic mix \textsc{EPOS-LHC}, and (d) galactic mix \textsc{Sibyll2.3c} models.}
	\label{fig:deltachi2spec_maps}
\end{figure*}

\begin{figure*}[!htpb]
	\centering
	\begin{minipage}{0.49\linewidth}
		\centering
		\subfloat[]{\label{fig:deltachi2lna_map_epos}\includegraphics[width=\textwidth]{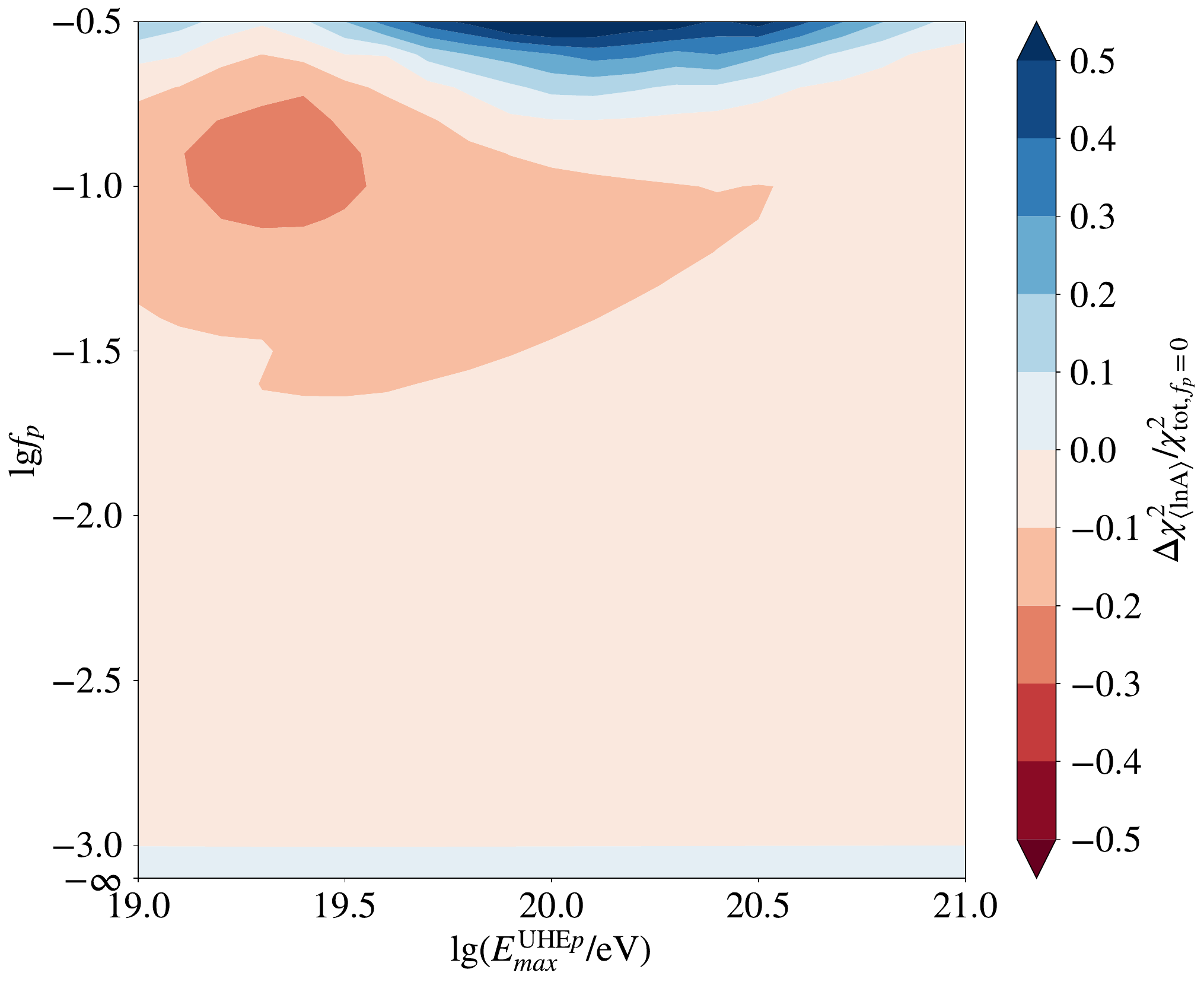}}
	\end{minipage}
	\begin{minipage}{0.49\linewidth}
		\centering
		\subfloat[]{\label{fig:deltachi2lna_map_sibyll}\includegraphics[width=\textwidth]{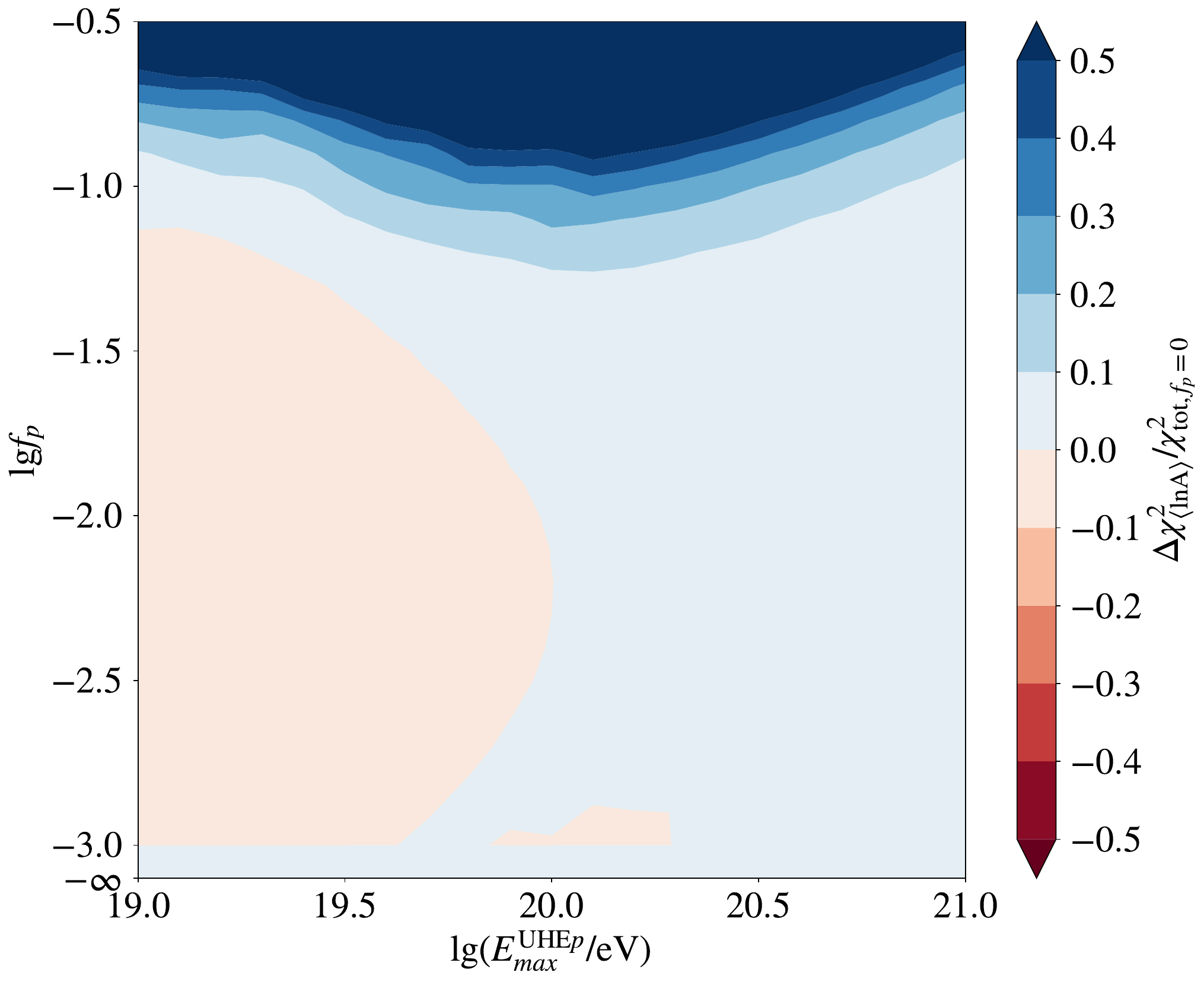}}
	\end{minipage}
	\begin{minipage}{0.49\linewidth}
		\centering
		\subfloat[]{\label{fig:deltachi2lna_map_galmix_epos}\includegraphics[width=\textwidth]{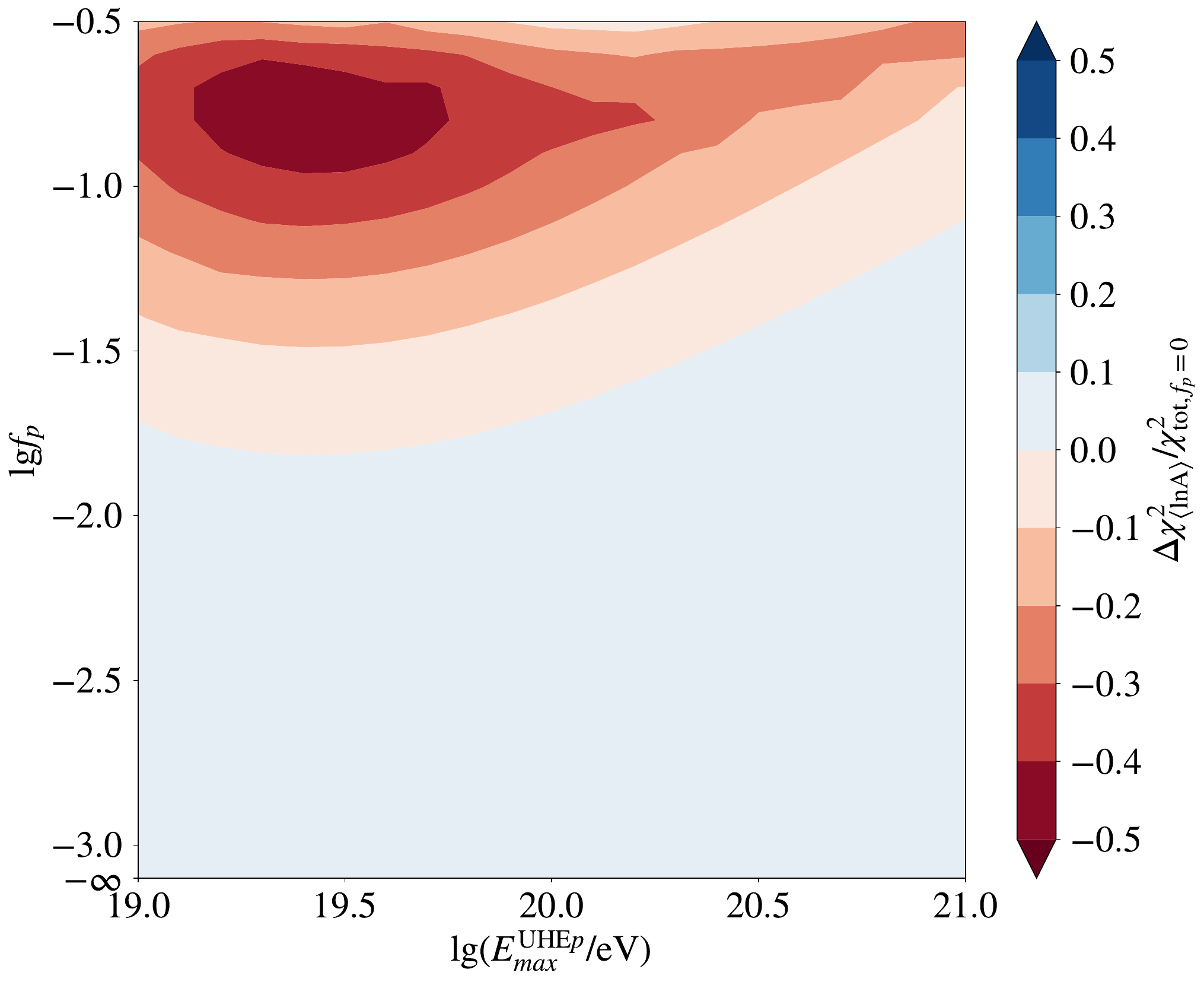}}
	\end{minipage}
	\begin{minipage}{0.49\linewidth}
		\centering
		\subfloat[]{\label{fig:deltachi2lna_map_galmix_sibyll}\includegraphics[width=\textwidth]{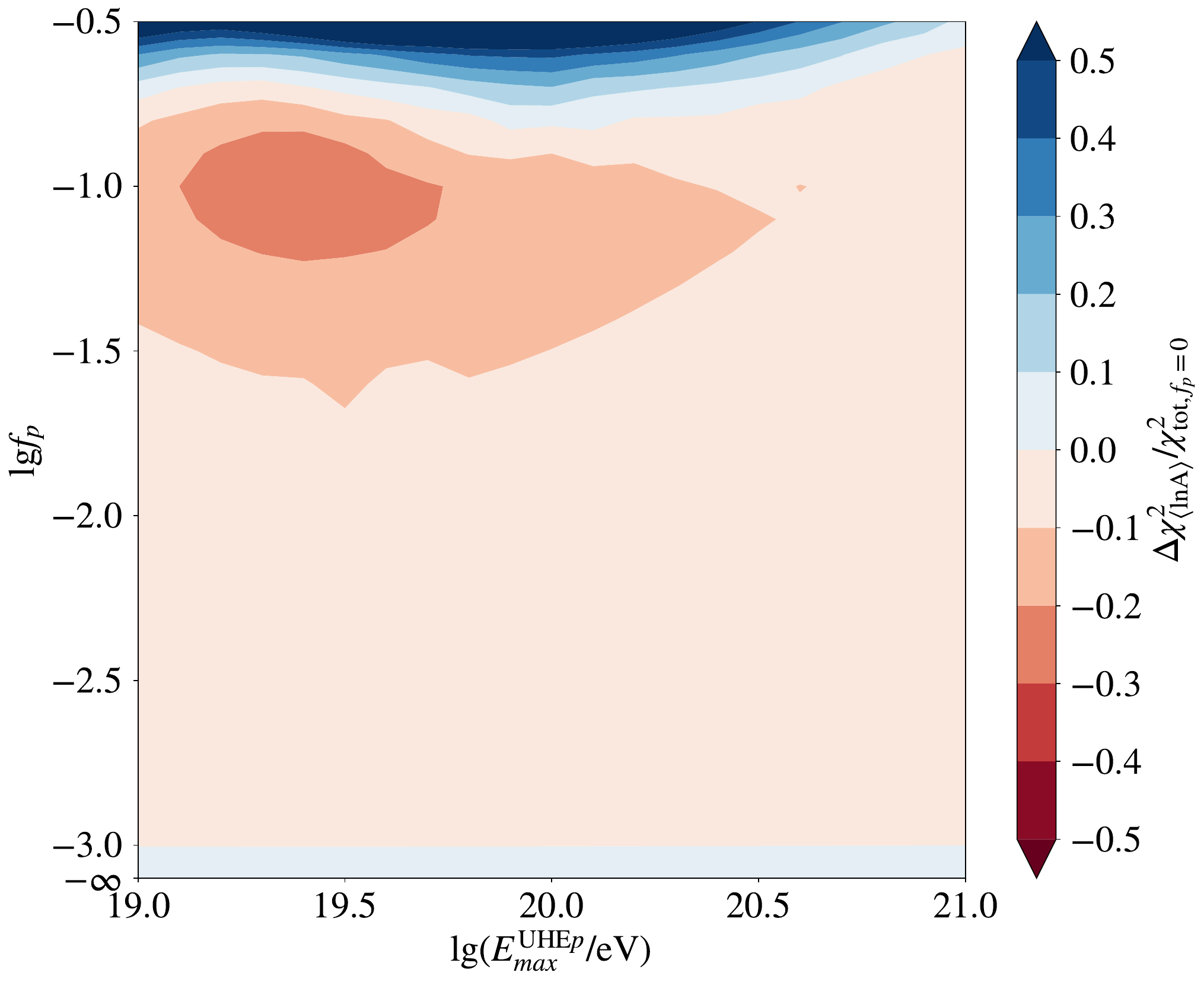}}
	\end{minipage}
	\caption{Effect of an additional pure-proton component on the $\langle \ln{A} \rangle$ contribution to the total $\chi^2$ relative to the best-fit in a single-component model. Additional pure-proton component with (a) single-mass \textsc{EPOS-LHC}, (b) single-mass \textsc{Sibyll2.3c}, (c) galactic mix \textsc{EPOS-LHC}, and (d) galactic mix \textsc{Sibyll2.3c} models.}
	\label{fig:deltachi2lna_maps}
\end{figure*}

\begin{figure*}[!htpb]
	\centering
	\begin{minipage}{0.49\linewidth}
		\centering
		\subfloat[]{\label{fig:deltachi2vlna_map_epos}\includegraphics[width=\textwidth]{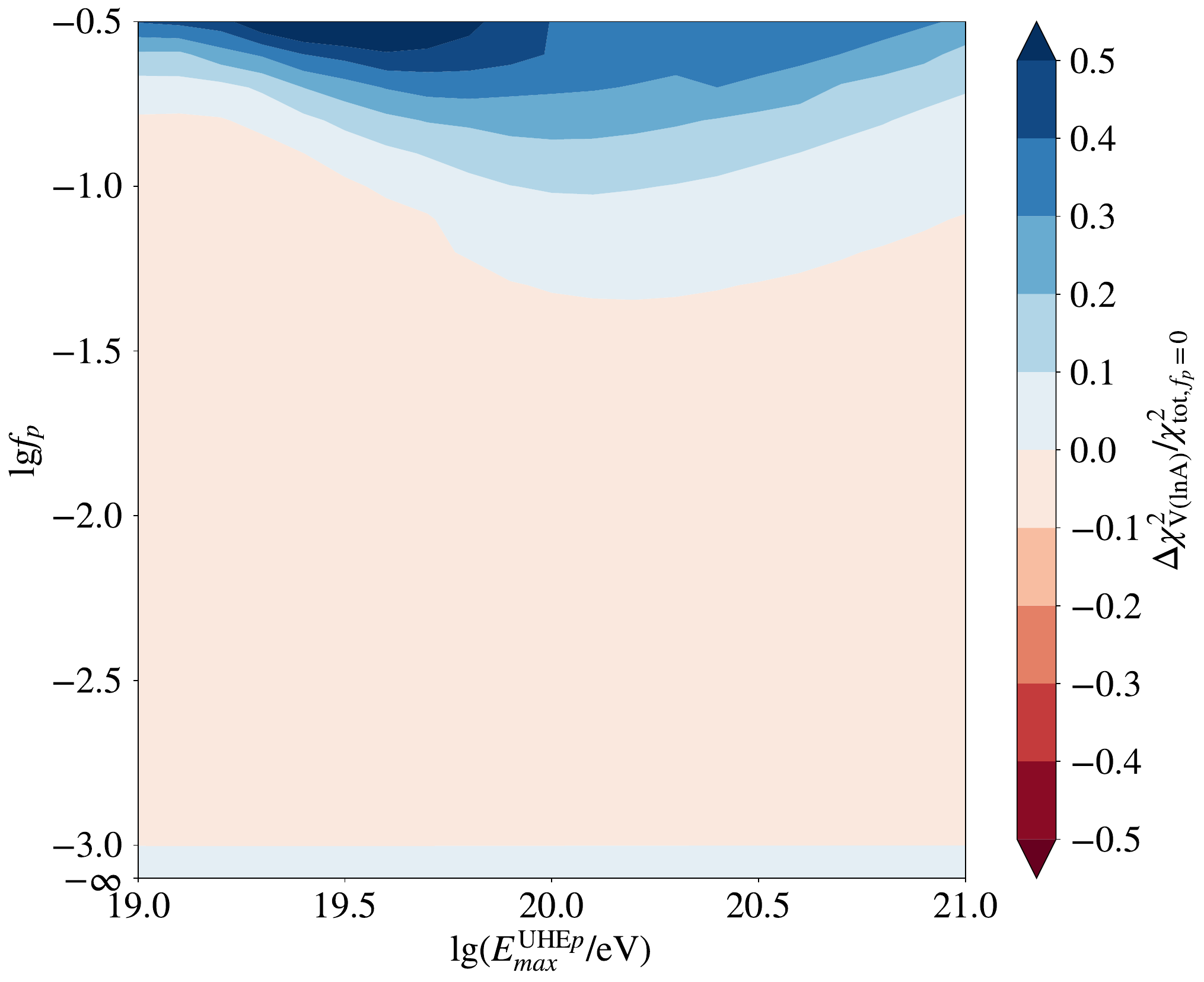}}
	\end{minipage}
	\begin{minipage}{0.49\linewidth}
		\centering
		\subfloat[]{\label{fig:deltachi2vlna_map_sibyll}\includegraphics[width=\textwidth]{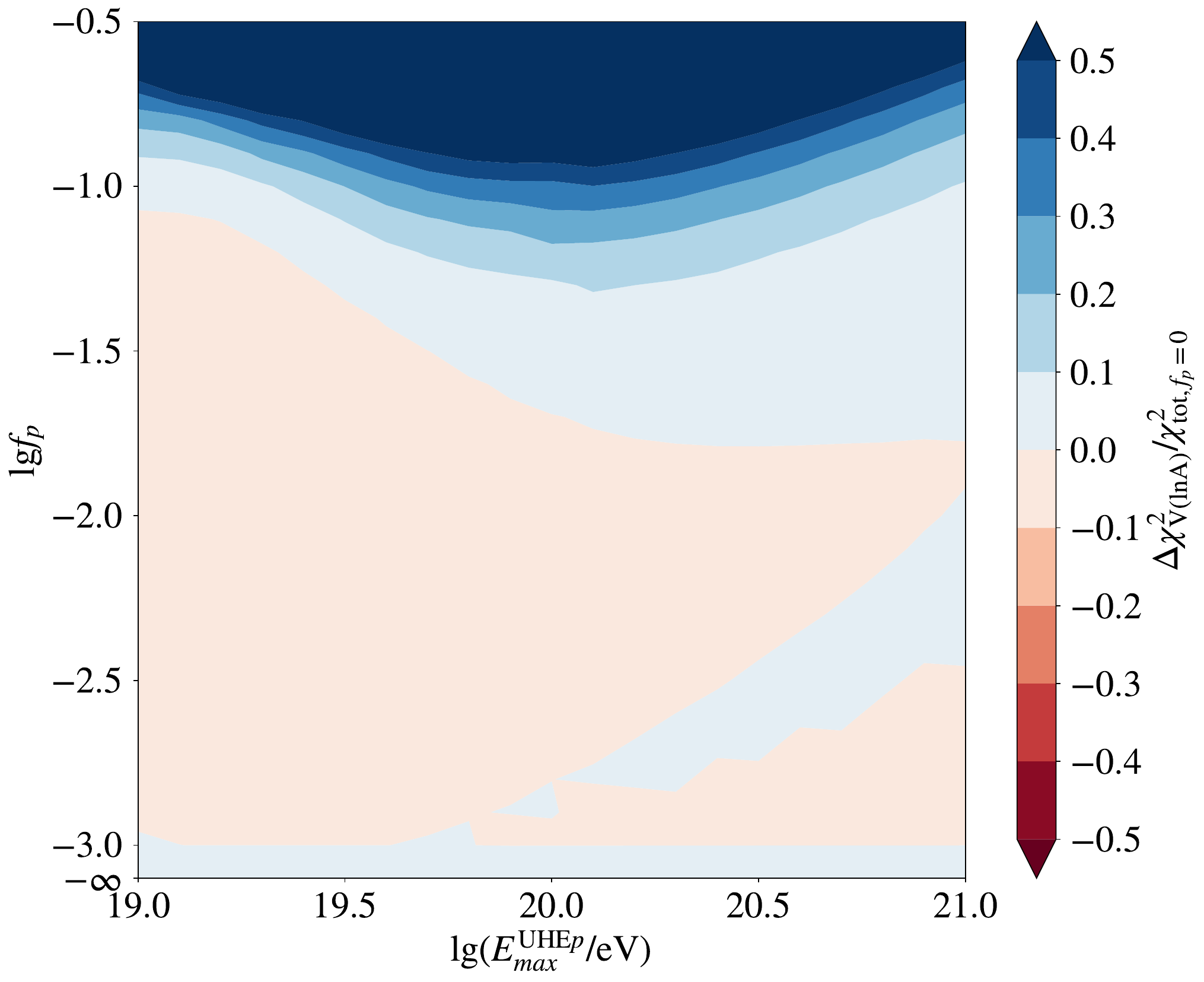}}
	\end{minipage}
	\begin{minipage}{0.49\linewidth}
		\centering
		\subfloat[]{\label{fig:deltachi2vlna_map_galmix_epos}\includegraphics[width=\textwidth]{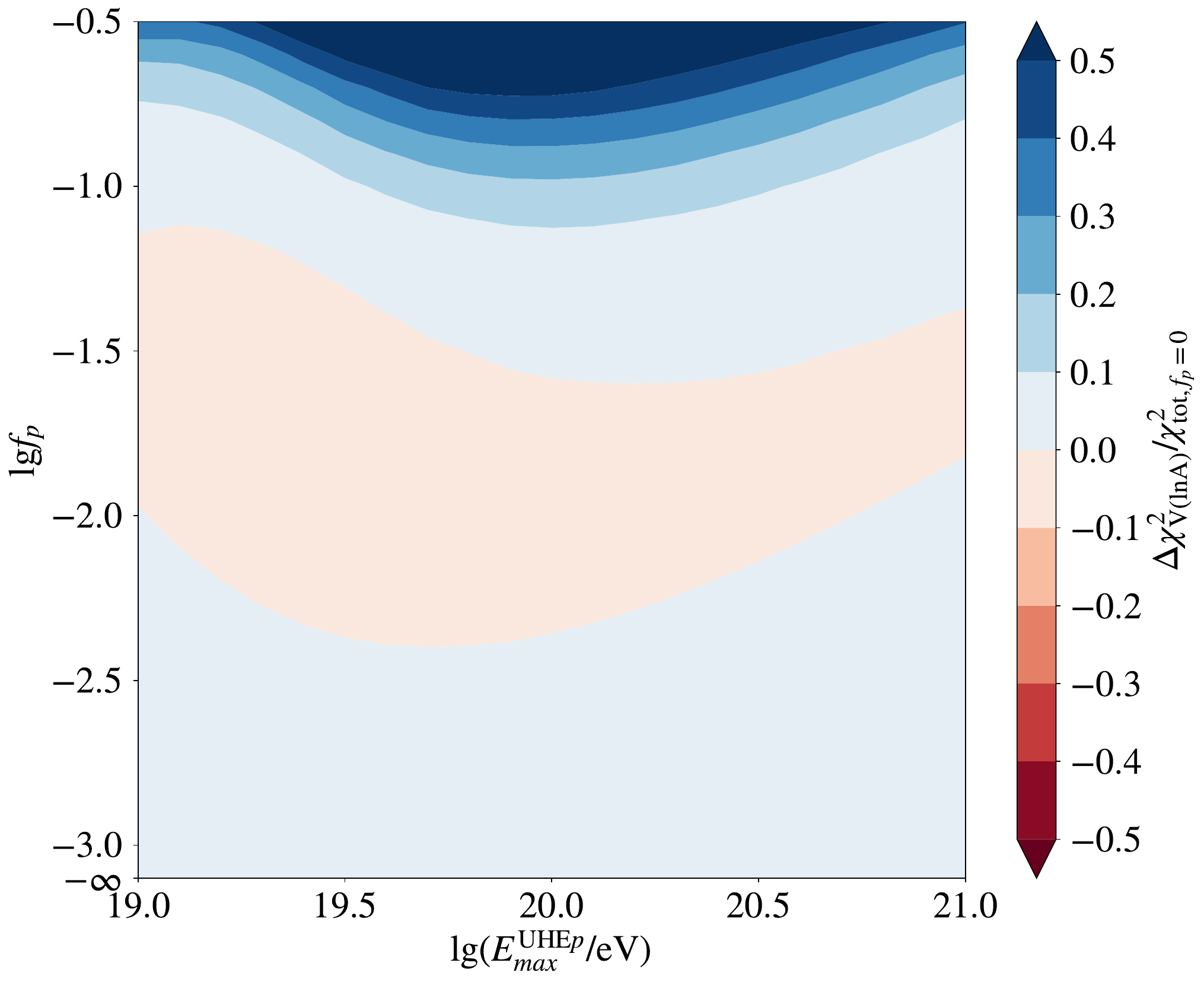}}
	\end{minipage}
	\begin{minipage}{0.49\linewidth}
		\centering
		\subfloat[]{\label{fig:deltachi2vlna_map_galmix_sibyll}\includegraphics[width=\textwidth]{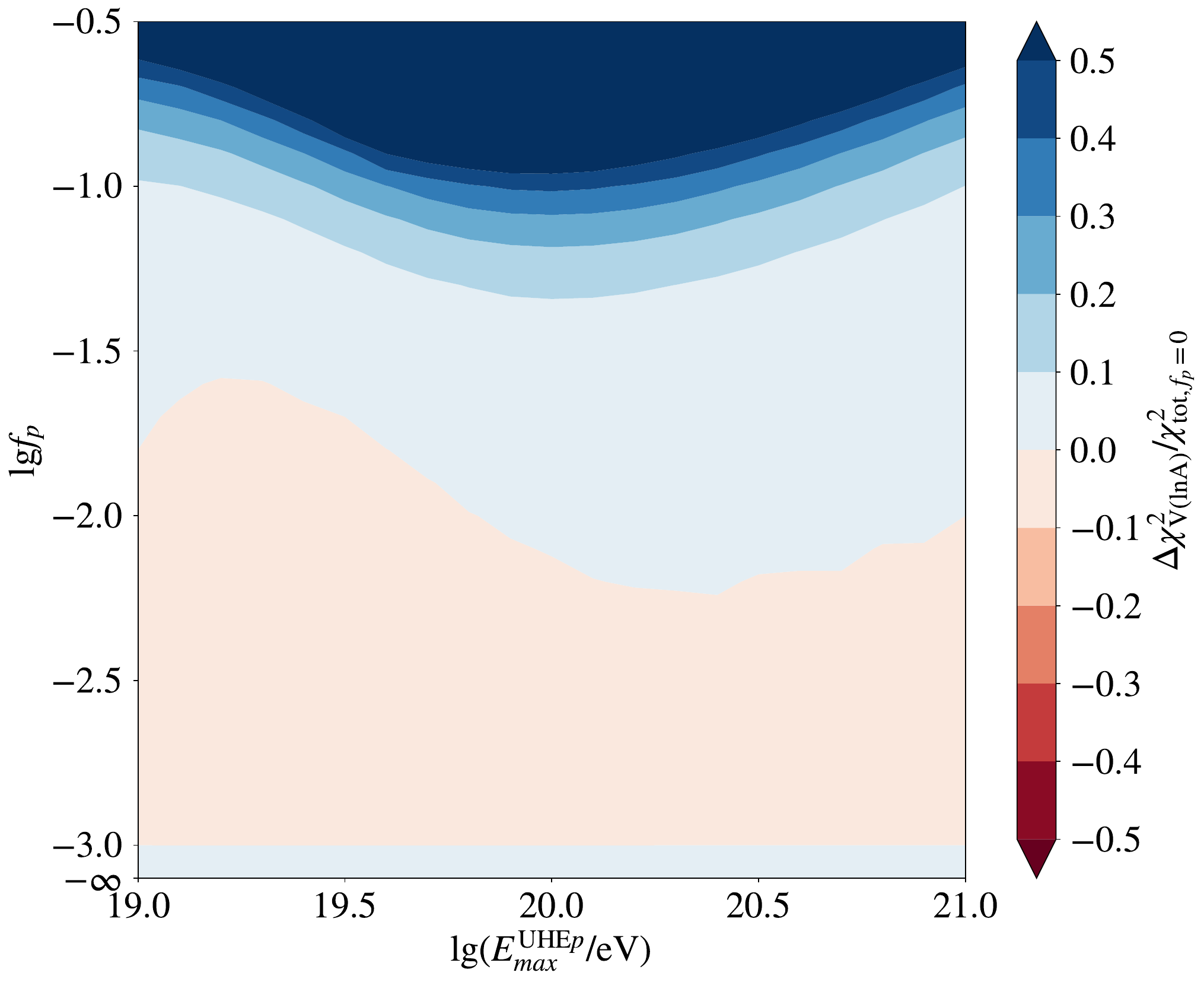}}
	\end{minipage}
	\caption{Effect of an additional pure-proton component on the $V(\ln{A})$ contribution to the total $\chi^2$ relative to the best-fit in a single-component model. Additional pure-proton component with (a) single-mass \textsc{EPOS-LHC}, (b) single-mass \textsc{Sibyll2.3c}, (c) galactic mix \textsc{EPOS-LHC}, and (d) galactic mix \textsc{Sibyll2.3c} models.}
	\label{fig:deltachi2vlna_maps}
\end{figure*}

Figures~\ref{fig:deltachi2spec_maps},~\ref{fig:deltachi2lna_maps}, and~\ref{fig:deltachi2vlna_maps} show the individual effects of the subdominant pure-proton component on the $\chi^2$ from the spectrum, $\langle \ln{A} \rangle$, and $V(\ln{A})$ relative to the total $\chi^2$ in the best-fit model with a single component. From these figures one can see that all components of the $\chi^2$ --- spectrum, $\langle \ln{A} \rangle$, and $V(\ln{A})$ --- are improved by the addition of a subdominant pure-proton component in some part of the parameter space we explored. 

\par
The most significant improvement comes from an improved fit to $\langle \ln{A} \rangle$, while the improvement to the other observables is more mild. Generally, high values of $f_p$ are excluded due to a quickly degrading ability to fit $V(\ln{A})$. This is often in spite of the fact that the description of $\langle \ln{A} \rangle$ may be improved over this range of values of $f_p$. This is a generic feature we have found in fitting both the spectrum and composition observed by Auger: often models fitting $\langle \ln{A} \rangle$ are at odds with those fitting $V(\ln{A})$. This puts very strong constraints on the types of models which are able to describe both of these observables simultaneously. 

\par
The fact that a subdominant pure-proton component most strongly improves the fit to $\langle \ln{A} \rangle$ is due to the difficulty photodisintegration models have simultaneously reproducing the ankle and spectral cutoff (which controls the composition at the highest energies in photodisintegration models independent of HEG), the composition at the ankle, and the elongation rate between the ankle and the end of the UHECR spectrum. Single-component photodisintegration models are able to reproduce the first two of these features well but are challenged to some degree to achieve the correct elongation rate. Namely, photodisintegration models predict the spectrum getting heavier too quickly compared to observations between the ankle and the end of the spectrum. The addition of a light second component can then improve the description of the composition in this region.

\par
\textsc{EPOS-LHC} benefits most from the addition of a subdominant light component (see Fig.~\ref{fig:UHEproton_maps}), since \textsc{EPOS-LHC} infers a lighter composition compared to \textsc{Sibyll2.3c}. This induces a more dramatic decrease in the predicted elongation rate in the ankle-to-spectral-cutoff region compared to \textsc{Sibyll2.3c} (e.g. compare the predicted elongation rates in Fig.~\ref{fig:UFA_benchmark_CR} with those in Fig.~\ref{fig:UFA_evo_CR}). In other words, \textsc{EPOS-LHC} requires the spectrum get heavier even faster than \textsc{Sibyll2.3c} due to the larger discrepancy in the mass it infers from $X_\mathrm{max}$ data at the ankle and the mass at the spectral cutoff set by spectral data, which is HEG independent. 

\par
Similarly, galactic mix models generically benefit from a second light component more than single-mass models because the composition in galactic models is heavier at the highest energies than that in single-mass models (whose injected mass is set by the relative positions of the ankle and the spectral cutoff independent of HEG). This also acts to decrease the elongation rate above the ankle. 

\par
Clearly, advances in the fidelity of HEGs will be highly beneficial for exploiting the full information in the UHECR $\langle X_\mathrm{max} \rangle$ and $\sigma(X_\mathrm{max})$ observations.

\subsection{Best-Fit Parameters of Various Models}

\par
Table \ref{tab:params} gives the best-fit parameters for some models presented throughout this paper. The parameters in this table are as follows: $\gamma$, the spectral index at injection into the source environment; the composition indicates either the mass number injected into the source or a galactic mix; $R_\mathrm{max}$, the rigidity at which the injected spectrum is exponentially cutoff; $\dot{\varepsilon}_{17.5}$, the comoving CR power density above $10^{17.5}$ eV of CR sources at $z=0$; $D_\mathrm{min}$, the comoving distance to the nearest extragalactic UHECR source; $T$, the temperature of the ambient photon field surrounding the source; $\delta$, the power law index of the escape length as a function of rigidity; $R_{19}^\mathrm{Fe}$, the escape-to-interaction time ratio for an $10^{19}$ eV iron nucleus; $\gamma_\mathrm{gal}$, the spectral index of the observed galactic CR spectrum; $A_\mathrm{gal}$, the mass number of the galactic CRs; $f_\mathrm{gal}$, the galactic CR (number) flux fraction at $10^{17.5}$ eV; $E_\mathrm{max}^\mathrm{gal}$, the energy at which the galactic CR spectrum is exponentially cutoff; $f_p$, the fraction of energy carried by the additional pure-proton component relative to the total energy of all CRs escaping their source above $10^{19}$ eV [see~\eqref{eq:fp_def}]; and, $E_\mathrm{max}^{\mathrm{UHE}p}$, the energy at which the additional pure-proton component is exponentially cutoff. 

\begin{table*}[!htpb]
\begin{tabular}{l c c c c}

\hline \hline
\textbf{Model} & Evolution & $D_\mathrm{min}$ & Extra Proton Component I & Extra Proton Component II \\
\hline \hline
\textit{Source Parameters} & & & & \\
\hline
$\gamma$ & \textbf{0.09 (-0.72)} & \textbf{-1.07 (-1.05)} & -1 (-1) & -1 (-1) \\
Composition & \textbf{25 (31)} & Gal. Mix & \textbf{32.9 (31.4)} & Gal. Mix \\
$\log_{10}(R_\mathrm{max}/\mathrm{V})$ & \textbf{18.4 (18.5)} & \textbf{18.6 (18.6)} & \textbf{18.6 (18.5)} & \textbf{18.5 (18.5)} \\
$\dot{\varepsilon}_{17.5}$ (erg$\,$Mpc$^{-3}\,$yr$^{-1}$) & \textbf{5.2 (6.2)$\times 10^{44}$} & \textbf{10.3 (7.5)$\times 10^{44}$} & \textbf{8.7 (7.0)$\times 10^{44}$} & \textbf{8.7 (7.1)$\times 10^{44}$} \\ 
Evolution & $\mathbf{m=4.2,\, z_0=2}$ \textbf{(SFR)} & SFR & SFR & SFR \\
$D_\mathrm{min}/$Mpc & 0 (0) & \textbf{40 (40)} & 0 (0) & \textbf{40 (40)} \\
\hline \hline
\textit{Source Environment} & & & & \\
\hline
Photon Field & MBB $\sigma=2$ & BB & BB & BB \\
$T$ (K) & \textbf{450 (90)} & \textbf{750 (500)} & \textbf{400 (350)} & \textbf{450 (400)} \\
$\delta$ & \textbf{-1.01 (-0.60)} & \textbf{-1.01 (-1.01)} & \textbf{-0.90 (-1.01)} & \textbf{-1.01 (-1.01)} \\
$\log_{10}R_{19}^\mathrm{Fe}$ & \textbf{1.89 (1.47)} & \textbf{2.45 (2.21)} & \textbf{2.48 (2.25)} & \textbf{2.53 (2.25)} \\
\hline \hline
\textit{Galactic CR Spectrum} & & & & \\
\hline
$\gamma_\mathrm{gal}$ & \textbf{-3.52 (-3.35)} & \textbf{-3.50 (-3.45)} & \textbf{-3.54 (-3.44)} & \textbf{-3.56 (-3.47)} \\
$A_\mathrm{gal}$ & \textbf{33.5 (28.1)} & \textbf{33.0 (29.4)} & \textbf{30.4 (27.6)} & \textbf{30.9 (27.9)} \\
$f_\mathrm{gal}$ (\%) & \textbf{73 (84)} & \textbf{75 (81)} & \textbf{75 (81)} & \textbf{75 (81)} \\
$\log_{10}(E_\mathrm{max}^\mathrm{gal}/\mathrm{eV})$ & \textbf{18.5 (18.2)} & \textbf{18.5 (18.6)} & \textbf{18.6 (18.5)} & \textbf{18.5 (18.7)} \\
\hline \hline
\textit{Extra Proton Parameters} & & & & \\
\hline
$f_p$ (\%) & 0 (0) & 0 (0) & \textbf{10 (3.2)} & \textbf{12.6 (6.3)} \\
$\log_{10}(E^{\mathrm{UHE}p}_\mathrm{max}/\mathrm{eV})$ & --- & --- & \textbf{19.2 (19.0)} & \textbf{19.3 (19.3)} \\
\hline \hline
$\chi^2/$ndf & 185.8/55 (109.9/55) & 297.7/55 (168.8/55) & 159.9/54 (108.2/54) & 183.7/54 (141.5/54) \\
Corresponding Figure & Fig.~\ref{fig:evochi2} & Fig.~\ref{fig:UFA_mindist_summary} & Fig.~\ref{fig:UHEproton_summary} & Fig.~\ref{fig:UHEproton_summary} \\
\hline \hline

\end{tabular}
\caption{\label{tab:params}Best-fit parameters in various models presented throughout this paper. Bolded values indicate the parameters were free during the minimization procedure. Values with parentheses are for models using \textsc{Sibyll2.3c}, while those without parentheses are for models using \textsc{EPOS-LHC}. All fits were performed to shifted Auger data as explained in \S~\ref{sec:benchmarks}. The models explored are as follows: ``Evolution'' is the best-fit source evolution in a single-mass model; ``$D_\mathrm{min}$'' is the best-fit minimum source distance in a galactic mix model; and, ``Extra Proton Component I \& II'' additional subdominant proton component models with single-mass and galactic mix models, respectively. SFR is the star formation rate evolution of~\cite{Robertson+15} presented in~\eqref{eq:SFR}. BB and MBB are blackbody and modified blackbody photon fields, respectively. See text for definitions of other parameters.}
\end{table*}

\end{document}